\documentclass[12pt]{article}
\usepackage[english]{babel}
\usepackage[utf8]{inputenc}
\usepackage{amsmath}
\usepackage{amsthm}
\usepackage{amsfonts}
\usepackage{graphicx} 
\usepackage{subcaption}
\usepackage[round]{natbib}   

\usepackage{physics}
\usepackage{csvsimple}
\usepackage{mdframed}
\usepackage{bbm}
\usepackage{soul}
\usepackage{booktabs}
\usepackage{threeparttable}
\usepackage{eqparbox}
\usepackage{wrapfig}
\usepackage{bm}
\usepackage{mathtools}
\usepackage{pgf,tikz}
\usepackage{float}
\usepackage[margin=1.in]{geometry}
\usepackage{tocloft}
\usepackage{parskip}
\usepackage[%
    colorlinks=true,
    pdfborder={0 0 0},
    linkcolor=blue,
    citecolor={blue}
]{hyperref}
\usepackage{multirow, makecell}
\usepackage{courier}
\usepackage[skip=1ex]{caption}
\usepackage{hyperref}
\usepackage{color, colortbl}
\definecolor{Gray}{gray}{0.9}
\definecolor{LightCyan}{rgb}{0.88,1,1}
\usepackage{setspace}

\usepackage{titlesec}

\newcommand{\blind}{1}
\newcommand{\rev}[1]{{\color{black}#1}}
\newcommand{\re}[1]{{\color{black}#1}}
\newcommand{\revi}[1]{{\color{black}#1}}

\titleformat{\paragraph}
{\normalfont\normalsize\bfseries}{\theparagraph}{1em}{}
\titlespacing*{\paragraph}
{0pt}{3.25ex plus 1ex minus .2ex}{1.5ex plus .2ex}

\usepackage{authblk}
\title{Powerful Partial Conjunction Hypothesis Testing\\ via Conditioning}
\author[]{Biyonka Liang}
\author[]{Lu Zhang}
\author[]{Lucas Janson}
\affil[]{Department of Statistics, Harvard University}
\date{May 15, 2024}

\newcommand{\btheta}{\bm \theta}
\newcommand{\bT}{\bm T}

\newcommand{\PP}{\mathbb{P}}
\newcommand{\inp}{\stackrel{p}{\to}}
\newcommand{\RR}{\mathbb{R}}
\newcommand{\EE}[1]{\mathbb{E}[{#1}]}
\newcommand{\Ep}[2]{\mathbb{E}_{#1}[{#2}]}

\newcommand{\indic}[1]{\mathbbm{1}\left\{{#1}\right\}}
\newcommand{\given}{\;\middle|\;}
\newcommand{\iid}{\stackrel{i.i.d.}{\sim}}

\newcommand{\hattheta}{\hat{\bm{\theta}}^{(n)}_{(m-r+2:m)}}
\newcommand{\hatthetaelem}{\hat{\bm{\theta}}^{(n)}}
\newcommand{\Tseq}{\bm{T}^{(n)}}
\newcommand{\LFNseq}{\bm{\theta}^{(n)}}
\newcommand{\tildeg}{\tilde{G}\left(x, \Tseq\right)}
\newcommand{\copies}{X^{(k)}_\ell}
\newcommand{\Ib}{\mathbf{I}}
\newcommand{\bZ}{\bm{Z}}
\newcommand{\cN}{\mathcal{N}}

\usepackage{algorithm}
\usepackage{algorithmic}

\newcommand{\NA}{---}
\newcommand{\repl}{n}

\newcommand{\nspace}{\Theta_0^{r/m}}
\newcommand{\ccone}{\Theta_0^{\text{cone}}}

\newcommand{\mytest}{f(\bm{T}_{(1:m-r+1)}) > c_\alpha\left(\bT_{(m-r+2:m)}, \bT_{(m-r+2:m)}\right)}

\newtheorem{innercustomgeneric}{\customgenericname}
\providecommand{\customgenericname}{}

\newcommand{\newcustomtheorem}[2]{%
  \newenvironment{#1}[1]
  {%
  \renewcommand\customgenericname{#2}%
  \renewcommand\theinnercustomgeneric{##1}%
  \innercustomgeneric
  }
  {\endinnercustomgeneric}
}

 \newtheorem{theorem}{Theorem}
 \newtheorem{definition}{Definition}
  \newtheorem{lemma}{Lemma}
 \newcustomtheorem{customthm}{Theorem}
\newcustomtheorem{customlemma}{Lemma}
\newcustomtheorem{customassumption}{Assumption}
\newcustomtheorem{customdef}{Definition}

\begin{document}
    
\pagenumbering{arabic}

\maketitle

\begin{abstract}
A Partial Conjunction Hypothesis (PCH) test combines information across a set of base hypotheses to determine whether some subset is non-null. PCH tests arise in a diverse array of fields, but standard PCH testing methods 
can be highly conservative\rev{, leading to low power especially in low signal settings commonly encountered in applications.} In this paper, we introduce the conditional PCH (cPCH) test, a new method for testing a single PCH that directly corrects the conservativeness of standard approaches by conditioning on certain order statistics of the base p-values. Under distributional assumptions commonly encountered in PCH testing, the cPCH test \rev{is valid} and \rev{produces nearly uniformly distributed p-values under the null (i.e., cPCH p-values are only very slightly conservative)}.  We demonstrate that the cPCH test \rev{matches or outperforms existing} single PCH tests \rev{with particular power gains in low signal settings,} maintains Type I error control even under model misspecification, and can be used to outperform state-of-the-art multiple PCH testing procedures \rev{in certain settings, particularly when side information is present}. Finally, we illustrate an application of the cPCH test through a replicability analysis across DNA microarray studies.
\end{abstract}

\noindent%

\section{Introduction}
\subsection{Motivation}
Partial Conjunction Hypothesis (PCH) tests are necessary to address important statistical questions in a diverse array of fields. For example, in areas such as genomics \cite[]{huang, hdmt, dact}, psychology \cite[]{baronkenny}, and social policy evaluation \cite[]{Karmakar}, researchers are interested in understanding complex causal relationships between a cause and an effect. Methods such as causal mediation analysis and causal factor analysis formulate questions about such relationships as the conjunction of links in a causal graph, where each link in the graph represents a hypothesis relating an effect to an outcome. The primary goal of these methods is to identify true causal relationships, which requires procedures for testing whether the hypotheses representing the conjunction of certain links are individually non-null. Another major application area is replicability analysis, where the partial conjunction of hypotheses represents the corroboration of scientific results across independent studies. In recent years, scientific replicability has become a major topic of interest in the analysis of observational studies, where subtle biases in population, data processing, and diagnostic measures can significantly influence conclusions. ``Replicability crises'' have been observed across a wide range of domains, including genetic epidemiology \cite[]{hirsch, nci, kraft}, economics \cite[]{camerer}, psychology \cite[]{sanford}, and medicine \cite[]{ioannidis}, underscoring the need for replicability analysis to convincingly validate scientific claims.

\subsection{Problem Statement}\label{section:problemstatement}
Partial Conjunction Hypothesis (PCH) testing provides a statistical framework for evaluating the partial conjunction of a set of $m$ hypotheses. Formally, it tests whether or not at least $r$ out of $m$ \emph{base hypotheses}, $H_{0,i}$, $i = 1, ..., m$, are individually non-null, i.e., letting $r^\star$ be the true number of non-null base hypotheses: 
\begin{definition}[Partial Conjunction Hypothesis (PCH)]
$$H_{0}^{r/m}: r^\star<r$$
\end{definition}
where $r \in \{2, 3, ..., m\}$.
$H_{0}^{r/m}$ can equivalently be interpreted as stating that at least $m-r+1$ of the $H_{0,i}$ are \emph{true}, and its complement as stating that at least $r$ of the $H_{0,i}$ are \emph{false}. For example, in replicability analysis, where $H_{0, i}$ is that no effect was present in study $i$, the rejection of $H_{0}^{2/m}$ provides explicit evidence for scientific replicability (i.e., that an effect was present in two or more studies). In causal mediation and causal factor analysis, where  $H_{0, i}$ represents the absence of the $i$th link in a causal chain, the rejection of $H_0^{m/m}$ provides explicit evidence for a causal relationship between the beginning and end of the chain. 

Notably, $H_0^{r/m}$ is composite, so the null space consists of $r$ disjoint \emph{null configurations} corresponding to $r^\star
= 0, ..., r-1$. For instance, in the $r=m=2$ case, the null configurations are:
 \[   H_0^{2/2} = 
 \begin{cases} 
 \text{Both } H_{0,i} \text{ are true ($r^\star=0$), or}\\
 \text{exactly one of } H_{0, i} \text{ is false ($r^\star=1$).}
 \end{cases}\]
 
In this paper, we consider the setting where the base hypotheses $H_{0,i}$ have independent \emph{base p-values}, $p_1,\dots,p_m$. Although we have multiple \emph{base} null hypotheses (with corresponding \emph{base} p-values), this paper primarily focuses on the problem of testing a single PCH (at a time), i.e., we aim to test $H_{0}^{r/m}$ for a \emph{single} set of $m$ base null hypotheses. Our improved PCH test can then be separately applied to multiple PCH's and the resulting p-values plugged into existing multiple testing procedures, and we will also explore this empirically in Section~\ref{section:mt} and \ref{section:dmd}.
  
\subsection{Testing a Single PCH} \label{section:classical_intro}
Standard methods for testing a single PCH apply global null tests to the largest $m-r+1$ base p-values. Letting $p_{(1)}\leq ...\leq p_{(m)}$ be the sorted base p-values, the resulting \emph{PCH p-value}, $p^{r/m}$, is of the form 
$$p^{r/m}(p_1, ..., p_m) = g\left(p_{(r)}, ..., p_{(m)}\right),$$
where $g: \mathbb{R}^{m-r+1} \to \mathbb{R}$ corresponds to the combining function of the global null test that is used. These standard single PCH tests are generally referred to by the same name as the global null test that they employ, with Simes' and Fisher's global null tests being the most common, \rev{under the assumption that the individual base p-values are independent} \cite[]{adafilt}.\rev{\footnote{Another common approach uses Bonferroni's correction, which is valid under any dependence structure of the base p-values. In this paper, we focus on Fisher's and Simes' tests as they have been shown to dominate Bonferroni's when the base p-values are independent \cite[]{dact}, which is our setting of interest.}}
\rev{For instance, the combining function for Fisher's test is $g_{\text{F}}\left(p_{(r)}, ..., p_{(m)}\right) = -2 \sum_{i=r}^{m} \log p_{(i)} $, which we can recognize as the standard Fisher test applied to the largest $m-r+1$ p-values.} 
When $r=m$, Simes' and Fisher's tests are equivalent and specifically when $r=m=2$, they are both referred to as the \emph{Max-P test} \cite[]{dact}.

Intuitively, applying a global null test to the $m-r+1$ ``least promising'' (i.e., largest) base p-values tests $H_0^{r/m}$ because rejecting the global null test provides evidence that strictly fewer than $m-r+1$ base hypotheses are null (equivalently, that at least $r$ base hypotheses are non-null) and therefore, $H_0^{r/m}$ should be rejected. Intuitively, PCH tests of this form are valid because the largest base p-values stochastically dominate the Unif(0,1), so the resulting PCH p-value will also stochastically dominate the Unif(0,1) \cite[]{BH2008}. 

Despite their widespread usage in various applied and methodological studies such as \cite{voxels, zuo, Rietveld, Small, Karmakar}, standard single PCH tests are highly conservative even when the tests for the individual base hypotheses are non-conservative \cite[]{BH2008,zhang, dact}. For example, under the global null (where all $H_{0,i}$ are true), even when the base p-values are uniformly distributed under the null, $p_{(r)}, ..., p_{(m)}$ are highly superuniform (i.e., $\mathbb{P}\left(p_{(i)} \leq t\right)\ll t$ for all $i = r, ..., m$). Thus, the resulting PCH p-value will also be superuniform. The only null case where standard single PCH tests will produce uniform p-values is when there are exactly $r-1$ non-null base hypotheses each having \emph{infinite} signal strength, which we refer to as the \emph{least favorable null} case as it is the parameter configuration in the null that maximizes the probability of rejecting $H_0^{r/m}$ \cite[]{BH2008, dickhaus}. Under this setting, $p_{(r)}, ..., p_{(m)}$ are guaranteed to correspond perfectly with the true null base p-values (since the $r-1$ non-null base p-values will all be 0). Thus, as long as the null base p-values are uniformly distributed and the global null test used is not conservative, the resulting PCH test will produce uniformly distributed PCH p-values. However, the least favorable null case (and in particular the infinite signal size of the non-nulls) is generally unrealistic in real-world data settings, and any other null case constitutes a situation where standard methods would be conservative to some degree.
 
 This conservativeness is concerning primarily because it extends to the alternative space. For example, in alternative configurations where the signal strengths associated with the non-null base hypotheses are low, standard single PCH tests can have especially low power since many of the $p_{(r)}, ..., p_{(m)}$ will likely correspond to null base p-values. Thus, the low power of standard single PCH tests in applied settings is fundamentally linked to their conservativeness under the null. However, alternative configurations with low signals are especially prevalent in applications, where effects are often subtle, such as genetic epidemiology \cite[]{sesia, adafilt}. Therefore, a primary challenge in PCH testing research has been to develop methods that correct the conservativeness of standard PCH tests.

\subsection{\re{Existing Work Adjusting for PCH Conservativeness}}\label{section:mt_intro}

\re{To our knowledge, there is only one prior work, \cite{miles2021optimal}, which directly corrects the conservativeness of single PCH testing. \cite{miles2021optimal} defines a test that achieves a minimax optimality criterion and has Type I error exactly $\alpha$ for unit fraction values of $\alpha$.  As their test is exact for $r=m=2$ for unit fraction values of $\alpha$, it performs similarly to ours in that specific setting; see Figure~\ref{fig:mmcomp} in Supplementary Materials~\ref{section:singlepch_additionalresults}. However, their test does not have a natural inversion to p-values, thereby limiting its utility as reporting p-values is standard practice for hypothesis testing in nearly all application areas. Additionally, without p-values, it is unclear how one could use this test with most multiple testing procedures for multiple PCH testing. Though \cite{miles2021optimal} propose a way to convert their test to a p-value, they do not prove its validity and empirically find it is conservative under the null. 
In comparison, the test we propose admits a natural inversion to p-values and is far less conservative under the null.
As we show in Section~\ref{section:mt}, our test outperforms using \cite{miles2021optimal}'s p-values for multiple PCH testing. Additionally, their test ``does not have an obvious unique natural extension to the setting with [$m \geq 3$]'' \citep{miles2021optimal}. Our test extends naturally to larger $m$ that are common across various PCH testing applications. 
} 

Several works correct the conservativeness of standard single PCH tests by sharing information across \emph{multiple} PCH tests to infer which PCH's are most likely to be under different null configurations. Naturally, this approach can only be applied when there are multiple (ideally very many) different PCH's being tested at once. Importantly, \re{methods which rely on there being multiple PCH tests }
\emph{cannot} be applied to testing a single PCH. Broadly, we can categorize these methods into two types: empirical Bayes and filtering methods.

Empirical Bayes methods such as \cite{HY2014, huang, hdmt, dact, dreyfuss} aim to predict the proportion of PCH's belonging to each null configuration, often adapting existing methods for estimating the proportion of nulls in the multiple testing literature \cite[]{efron2001, storey2002, storey, jinandcai, twogroup}. These approaches usually produce asymptotically valid PCH p-values under certain regularity conditions. However, because they rely on the consistency of their estimation method for their asymptotic validity guarantees, they can experience high Type I Error inflation when the number of hypotheses is small or when violations of regularity conditions make estimation unreliable.

Filtering methods filter out unpromising hypotheses to facilitate an analysis of the remaining ones, which tend to be less conservative \cite[]{adafilt, dickhaus}. For example, the AdaFilter method in \cite{adafilt} uses a data-adaptive threshold based on a Bonferroni correction to reduce the set of PCH's to the ones closest to the least favorable null case. \cite{dickhaus} filters and re-scales PCH p-values based on a user-provided threshold such that any multiple testing procedure like Benjamini--Hochberg can be applied to the reduced set while controlling FDR on the entire set.
These methods are highly effective in settings where there is a large number of PCH's being simultaneously tested and the global null is expected to be the overwhelmingly dominant null configuration. In these settings, most of the null PCH p-values will be highly conservative, thus allowing filtering to effectively exclude unpromising hypotheses. However, outside of this particular, albeit important, setting, the performance of these methods can suffer. For instance, we show that in situations where the proportion of global nulls is small relative to other null configurations, these filtering methods can have lower power than just using standard methods for single PCH testing to compute individual PCH p-values and then applying Benjamini--Hochberg; see Supplementary Materials~\ref{section:mt_m_4} for more details.

Overall, since multiple testing approaches to PCH testing must correct for both the multiplicity of the hypotheses being tested and the conservativeness of the individual p-values, they tend to be tailored to certain multiple testing settings (i.e., when the global null is the predominant null configuration) and can be less powerful outside of those settings. Alternatively, a generic way to generate multiple PCH testing procedures would be to develop a powerful and non-conservative \emph{single} PCH test \re{which admits (nearly) non-conservative} p-values (as we do in this paper),
and then the p-values from such a test could be fed into any existing (non-PCH-specific) multiple testing procedure.  By leveraging the vast literature on multiple testing procedures, this generic multiple PCH testing procedure has the potential to be powerfully applied in almost any setting, such as those with side information, as we explore further in Section~\ref{section:mt}.

Hence, to our knowledge, no method before ours allows one to compute \re{nearly} non-conservative p-values for PCH testing, thus enabling both powerful single PCH testing and the aforementioned generic multiple PCH testing procedure. 


\subsection{Our Contributions}
In this paper, we propose the conditional PCH (cPCH) test, a new approach to correcting the conservativeness of a standard (single) PCH test by conditioning on a function of the data. 
In the commonly encountered situation when the underlying test statistics associated with each of the $m$ hypotheses within a PCH test are independent and Gaussian, we show that the cPCH test \rev{is valid and only slightly conservative} under \emph{any} null configuration for any $m$ and $r$. 
\re{In contrast, all other existing single PCH tests are either highly conservative under certain null configurations, or have other limitations such only being applicable when $r=m=2$ and lacking a natural inversion to p-values.}
We demonstrate via simulation that the cPCH test is \re{almost always} more powerful than standard single PCH tests \rev{with particular power gains in low signal size settings}, maintains Type I error control even under model misspecification, and, in combination with different \rev{multiple testing procedures, can outperform existing multiple PCH testing methods.
} \re{
In particular, because cPCH p-values can be directly used as input to different multiple testing procedures, it enables the user to perform multiple PCH testing in settings that existing multiple PCH testing approaches are not designed to accommodate, such as settings with side information. While other single PCH tests can also be used as input, the power of many multiple testing procedures can be especially sensitive to conservative p-values. Because the cPCH test produces approximately uniform null p-values, it is almost always more powerful than using a different single PCH test with the same multiple testing procedure.}


\subsection{Reproducibility} 
All code, along with a tutorial for its use, is provided at 
\if1\blind
{\texttt{https://github.com/biyonka/cpch}.
} \fi
\if0\blind
{[link removed for anonymization].
} \fi

\subsection{Notation}
For a distribution $P_{\bm\theta}$ in a parametric model $\{P_{\bm{\theta}}:\bm\theta\in\Theta\}$, let $\mathbb{P}_{\bm \theta}$ denote a probability taken with respect to $P_{\bm \theta}$. We write $\Phi$ and $\phi$ to denote the cumulative distribution function (CDF) and probability density function (PDF) of the standard normal random variable, respectively. Let $I_m$ be the $m \times m$ identity matrix. 

\section{The Conditional PCH Test}\label{section:cpchtesting}

\subsection{Preliminaries}\label{section:preliminaries}
Although we will ultimately argue that our method applies more broadly, we will begin by assuming that each of the base null hypotheses $H_{0,i}$ tests whether a scalar parameter $\theta_i=0$, and that the data for such a test can be summarized into a single unit-variance Gaussian test statistic $T_i$ with mean $\theta_i$, for $i = 1, ..., m$:
\begin{equation}\label{eq:distassump}
    H_{0, i}: \theta_i = 0 \text{ vs. } H_{1, i}: \theta_i \neq 0,\text{ \hspace{0.7cm}}T_i \sim \mathcal{N}(\theta_i, 1).
\end{equation}
 \rev{Though our test can be adapted to one-sided testing, we focus on two-sided testing, as this is standard in the literature \citep{adafilt,dact,hdmt, Karmakar, Small, dickhaus}.}
As in the problem statement in Section~\ref{section:problemstatement}, we assume the test statistics are independent across $i$. \rev{
 Independence across $i$ is commonly assumed in PCH settings and is often reasonable. For instance, when the PCH's represent links in a causal graph such as in causal mediation analysis, independence across the links is commonly required for results to be interpreted causally \cite[]{Karmakar, Small}. In replicability analysis, the studies considered are often conducted independently from one another \cite[]{dact, dickhaus}.} We can write the PCH compactly as $H_0^{r/m}: \bm\theta \in \Theta_0^{r/m}$, where $\Theta_0^{r/m}=\{\btheta \in \RR^m: \norm{\btheta}_0 < r \}$ is the partial conjunction null space and $\norm{\cdot}_0$ is the $\ell_0$ norm. 
It may seem constraining to assume a parametric form of both the null and non-null base test statistic distributions. However, this distributional assumption approximates a wide array of applications, as the data for the base hypotheses are commonly summarized as asymptotically normal parameter estimators such as maximum likelihood estimators, method of moments estimators, and most causal estimators for average treatment effects \cite[]{voxels,  Rietveld, zhang, Barfield, adafilt,dact}. The replicability analysis of microarray studies presented in Section~\ref{section:dmd} provides a real data example that adheres to the above setting according to \cite{adafilt}.

We introduce some key notation for this setting. Recall we originally ordered the base p-values $\{p_i\}_{i=1}^m$ as $p_{(1)}\le \cdots \le  p_{(m)}$. Since we are in a two-sided testing setting, we will analogously order the $T_i$ by their \emph{magnitudes}, i.e., we order $\{T_{(i)}\}_{i=1}^m$ by $|T_{(1)}|\leq ...\leq|T_{(m)}|$ and let $\bm{T}_{(i:j)} = (T_{(i)}, ..., T_{(j)})$ for $i \leq j$. Note that the indices are ordered in reverse for the $T_{(i)}$ as they are for the $p_{(i)}$, e.g., the most significant and hence \emph{smallest} p-value $p_{(1)}$ corresponds to the \emph{largest}-magnitude 
test statistic $T_{(m)}$. From now on, we also present the combining functions of global null tests as functions of the base test statistics $T_i$ instead of the base p-values $p_i$. In this new formulation, a PCH test with some test-statistic-combining function $f$ will reject $H_{0}^{r/m}$ when $f(\bm{T}_{(1:m-r+1)}) \geq c_\alpha$ where $c_\alpha$ is the rejection threshold. \rev{For instance, the Fisher test would have $f_{\text{F}}(\bm{T}_{(1:m-r+1)}) = -2\sum_{i=1}^{m-r+1} \log \left(2(1-\Phi(|T_{(i)}|))\right)$ and $c_\alpha = x_{2(m-r+1),1-\alpha}$
where $x_{2(m-r+1),1-\alpha}$ is the $1-\alpha$ quantile of the $\chi^2_{2(m-r+1)}$ distribution.}

\subsection{Methodological Motivation and Intuition}\label{section:motivation}

Given a test statistic $f\left(\bm{T}_{(1:m-r+1)}\right)$, the primary challenge of PCH testing is to specify a rejection threshold $c_\alpha$ such that the test which rejects when $f\left(\bm{T}_{(1:m-r+1)}\right) \geq c_\alpha$ is valid, in that it controls Type I error at the desired level $\alpha$, and powerful, in that it rejects as often as possible when the PCH is false. In particular, our design objective is to generate a test that produces uniform PCH p-values $p^{r/m}$ under \emph{any} null configuration, i.e., $\mathbb{P}_{\bm\theta}(p^{r/m} \leq \alpha) = \alpha$ for all $\bm\theta \in \Theta_0^{r/m}$, since conservativeness in the p-value distribution at some point in the null space will imply a loss of power \rev{for alternative configurations with low signals, which
are especially prevalent in applications such as genetics \cite[]{BH2008, BHY2009, dact}.} 

A common way to generate new tests and further understanding of existing methods in comparison is to first consider an oracle test that has the desired properties (in our case, one that produces uniform p-values under every null) and admits a plug-in version with (hopefully) similar statistical behavior. For illustrative purposes, we focus in this subsection on the $r=m=2$ case where the test statistic for all standard combining functions (from Fisher's and Simes' global null tests) reduces to $f(T_{(1)}) = |T_{(1)}|$. By the symmetry of the problem (i.e., the rejection threshold is invariant to relabelling of the indices of $T_1$ and $T_2$), it is sufficient to represent all $\bm\theta \in \Theta_0^{2/2}$ by $\theta_{(2)}$ (where once again, we order the $\theta_i$'s by magnitude, i.e., $|\theta_{(1)}| \leq |\theta_{(2)}|$) since $\theta_{(1)} = 0$ by definition of $\Theta_0^{2/2}$. Thus, a natural choice of extra information to provide to our oracle is $\theta_{(2)}$. So, we define the rejection threshold of our PCH Oracle test $c_\alpha(\theta_{(2)})$ as the value satisfying
$$\mathbb{P}_{\theta_{(2)}}\left(\left|T_{(1)}\right| > c_\alpha\left(\theta_{(2)}\right) \right) = \alpha,$$ i.e., $c_\alpha(\theta_{(2)})$ is the $1-\alpha$ quantile of $|T_{(1)}|$'s distribution when $\{\theta_1, \theta_2\} = \{0, \theta_{(2)}\}$. By definition of $c_\alpha(\theta_{(2)})$, this oracle test produces uniform PCH p-values for \emph{every} $\btheta\in\Theta_0^{2/2}$, as desired.

Since PCH tests do not have access to the true $\theta_{(2)}$ in practice, they must use a rejection threshold that does not require knowledge of the true $\theta_{(2)}$. One valid choice is $c_\alpha = \sup_{\theta_{(2)}}c_\alpha\left(\theta_{(2)}\right) = c_\alpha(\infty)$, which is the rejection threshold of the Max-P test: $c_\alpha(\infty) = \Phi^{-1}\left(1-\alpha/2\right)$. Thus, we can think of the Max-P test as a \emph{plug-in} version of the oracle test where $\hat{\theta}_{(2)}=\infty$ is a (worst-case) plug-in estimator of $\theta_{(2)}$.

At first glance, the choice of $\hat{\theta}_{(2)} = \arg\sup_{\theta_{(2)}}c_\alpha\left(\theta_{(2)}\right) = \infty$ seems to be the primary cause for the conservativeness of standard methods (recall all standard methods reduce to the Max-P test when $r=m=2$). A promising possibility for resolving the conservativeness of the Max-P test is to choose a different estimator for $\theta_{(2)}$ that is likely to be closer to the true $\theta_{(2)}$. A natural choice would be $\hat{\theta}_{(2)} = T_{(2)}$, the maximum likelihood estimator (MLE) of $\theta_{(2)}$. This choice defines a new PCH test, which we call the \rev{marginal (plug-in) PCH} test, with rejection threshold $c_\alpha\left(T_{(2)}\right)$, the $1-\alpha$ quantile of $|T_{(1)}|$'s distribution when $\theta_{(2)} = T_{(2)}$. Though this choice of plug-in estimator does not come with any obvious Type I error guarantees, we find that the Type I error inflation of the marginal PCH test is remarkably small, as shown in Figure~\ref{fig:t1error}. However, it is \emph{still highly conservative} near the global null and, like the Max-P test, this conservativeness extends to the alternative space as well; see Figure~\ref{fig:power_mPCH}. These observations lead to our primary motivating insight for the conditional PCH (cPCH) test:

\begin{figure}[htb!]
\centering
\begin{subfigure}{0.5\textwidth}
  \centering  \includegraphics[width=\linewidth]{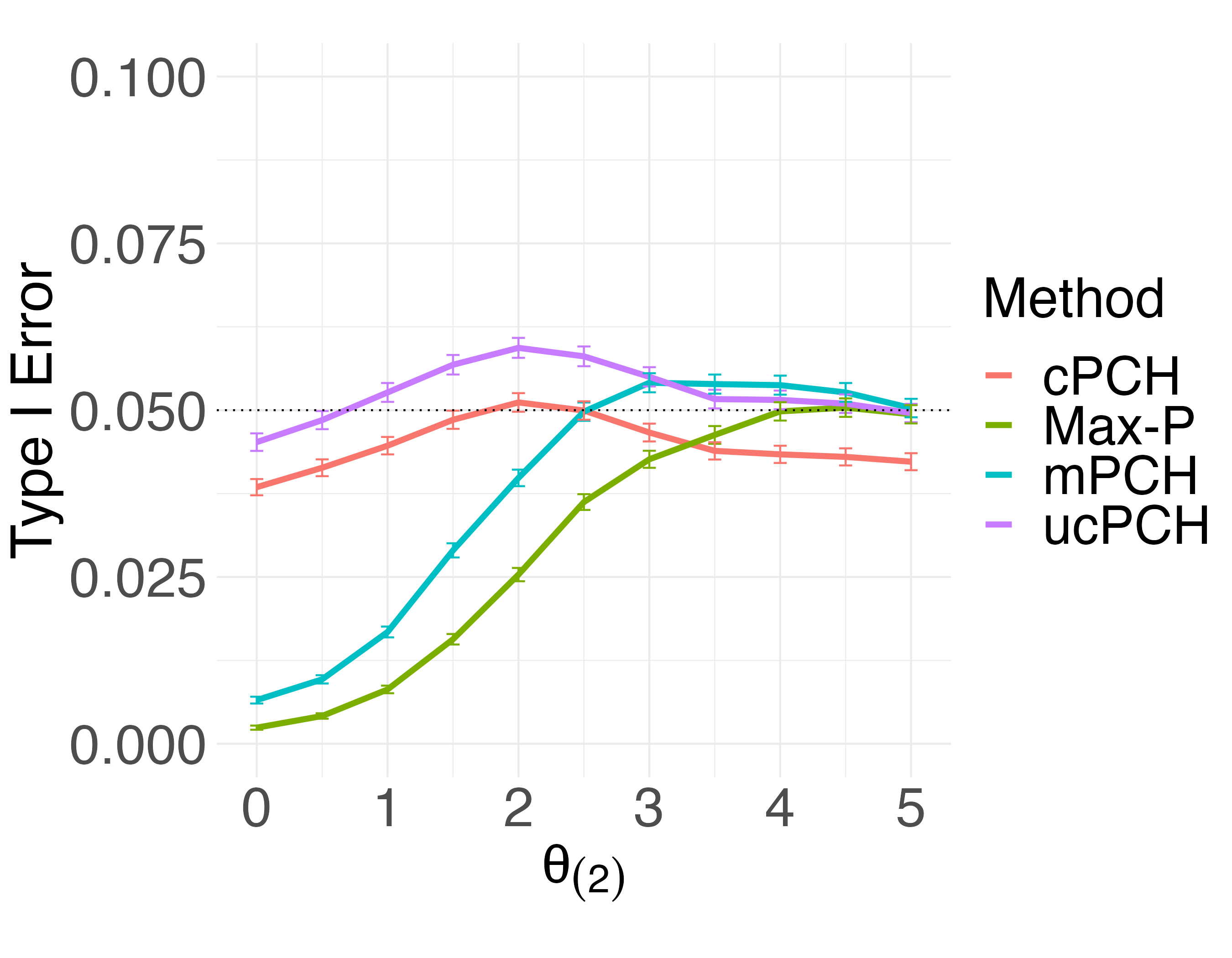}
  \caption{Type I error}
  \label{fig:t1error}
\end{subfigure}%
\begin{subfigure}{0.5\textwidth}
  \centering
  \includegraphics[width=\linewidth]{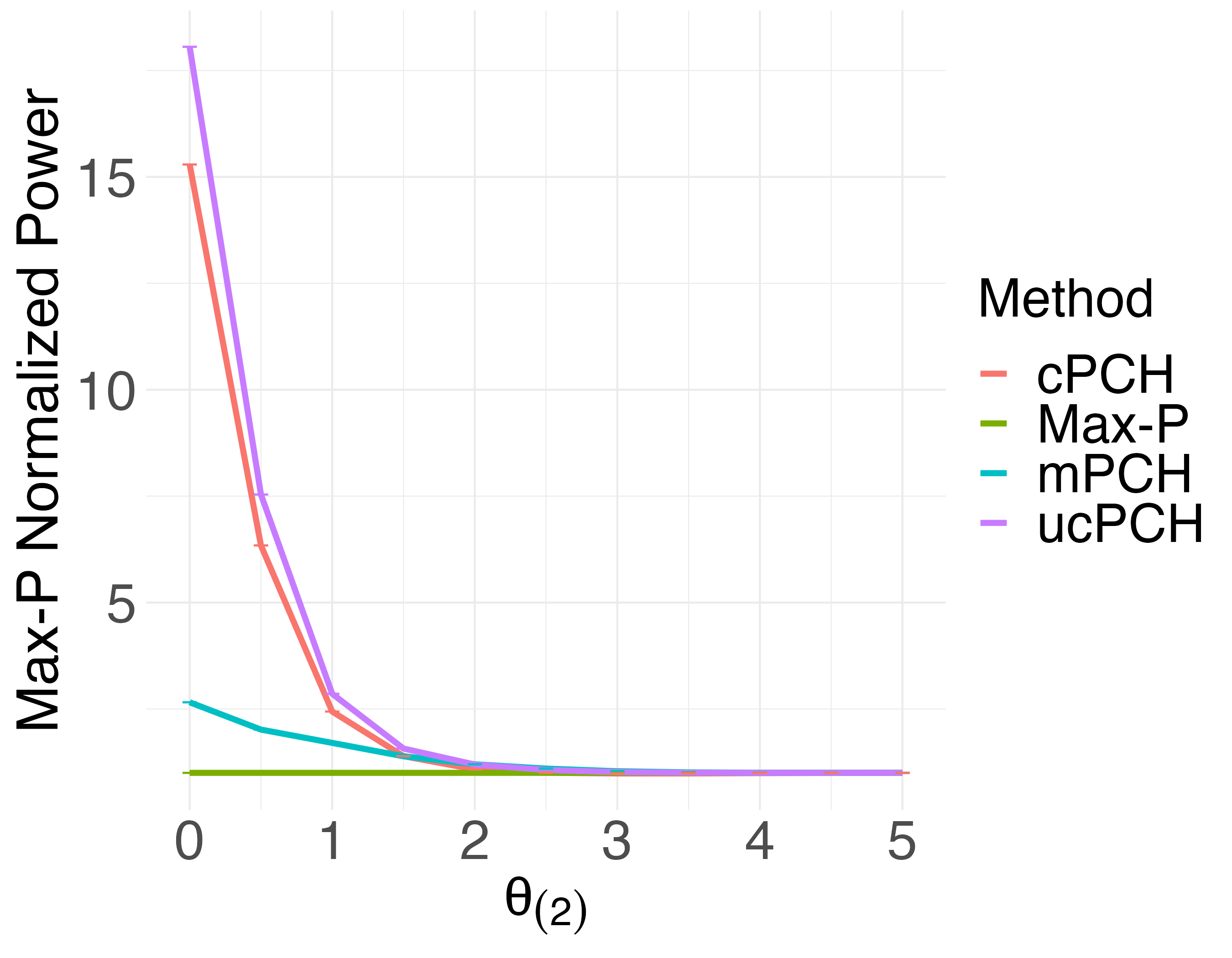}
  \caption{Max-P Normalized Power}
  \label{fig:power_mPCH}
\end{subfigure}
\caption{(a) Type I error and (b) \re{Max-P Normalized Power (i.e., the power of each test divided by the power of the Max-P test)} of the Max-P, marginal PCH (mPCH), \re{unadjusted cPCH} (ucPCH), and cPCH tests for $H_0^{2/2}$ at level $\alpha=0.05$ (dotted black line). In (a), each point represents $100{,}000$ independent simulations with $T_i \sim \mathcal{N}(\theta_i, 1)$ where $T_1$ and $T_2$ are independent and $\left(\theta_1, \theta_2\right) = \left(0, \theta_{(2)}\right)$. \revi{In (b), $T_1$ and $T_2$ are independent and $\left(\theta_1, \theta_2\right) = \left(\frac{\theta_{(2)}}{2}, \theta_{(2)}\right)$. Error bars depict $\pm 2$ standard errors.}}
\label{fig:test}
\end{figure}


The conservativeness of existing methods is not primarily caused by the choice of the estimator for the unknown parameters, but rather the sensitivity of the test statistic's distribution to the unknown parameters.

In particular, if $|T_{(1)}|$'s distribution did not depend very much on $\theta_{(2)}$, the marginal PCH test, which uses a reasonable plug-in estimator for $\theta_{(2)}$, should have similar Type I error to the oracle test which uses $\theta_{(2)}$'s exact value and produces exactly uniform p-values under every null $\btheta$.

One way to make the distribution of a test statistic less sensitive to unknown parameters is by conditioning on a function of the data $h(\bm{T})$. For example, an extreme case would be if the function of the data we conditioned on were sufficient for $\btheta$. Then, the conditional distribution of the test statistic would not depend on the unknown parameters at all.

Thus, we propose a new PCH test based on the distribution of the test statistic $|T_{(1)}|$ \emph{conditional} on a function of the data. In the $r=m=2$ case, a natural choice for conditioning is $h(\bm{T}) = T_{(2)}$, since we must preclude
conditioning on $T_{(1)}$, which uniquely determines the value of the test statistic $f(T_{(1)})$, and $T_{(2)}$ is all that remains.
So, we define the new cPCH Oracle test by the rejection threshold $c_\alpha(\theta_{(2)},   T_{(2)})$ satisfying $$\mathbb{P}_{\theta_{(2)}}\left(|T_{(1)}| > c_\alpha\left(\theta_{(2)},T_{(2)}\right) \given  T_{(2)} \right) = \alpha,$$ i.e., $c_\alpha\left(\theta_{(2)},  T_{(2)}\right)$ is the $1-\alpha$ quantile of the conditional distribution of $|T_{(1)}| \mid T_{(2)}$ when  $\left\{\theta_1, \theta_2\} = \{0, \theta_{(2)}\right\}$. As with the PCH Oracle test defined above, the cPCH Oracle test produces exactly uniform p-values under every null $\btheta$ by construction. From the cPCH Oracle test, we now define a new plug-in test in the hopes that it will resolve the conservativeness of the approaches described above. We call this test the \emph{\rev{unadjusted} cPCH test}, which uses the MLE $\hat{\theta}_{(2)} = T_{(2)}$ as an estimator for $\theta_{(2)}$ in the cPCH Oracle test's rejection cutoff.

\rev{Though the unadjusted cPCH test produces nearly uniform p-values under the null and is more powerful than the other approaches, it has slight Type I error inflation. We can correct for this by computing the maximum Type I error of the unadjusted cPCH test and adjusting the level so that the Type I error is upper bounded by $\alpha$. We can do this adjustment for any $\alpha, m,$ and $r$, which we detail fully in Section~\ref{section:validity}. 
We call this adjusted version of the unadjusted cPCH test the \emph{cPCH test}.
}

\rev{As shown in Figure~\ref{fig:t1error}, 
the cPCH test controls Type I error at the nominal level $\alpha$ across the entire null space (unlike the marginal PCH and unadjusted cPCH tests) while being far less conservative than the Max-P test.} 
Figure~\ref{fig:power_mPCH} shows that under the alternative, the cPCH test \rev{achieves particular power gains in the low signal} setting (i.e., when $\theta_1$ and $\theta_2$ are both small) \rev{while having similar power to Max-P \re{(i.e., the Max-P Normalized Power of the cPCH test asymptotes to $1$)}, the only other valid test, in the high signal setting.} 


Thus, via conditioning, we have developed a valid method for PCH testing that is more robust to plug-in estimation and, as we show further in Section~\ref{section:singlepower}, is empirically more powerful than existing methods for single PCH testing.

\subsection{Formal Definition}\label{section:formal_def}

We now formally define the cPCH test for general $m \geq 2$.
For a function $f: \mathbb{R}^{m-r+1} \to \mathbb{R}$, define $c_{\alpha}\left(\btheta_{(m-r+2:m)}, \bT_{(m-r+2:m)}\right)$ as the $1-\alpha$ quantile of the distribution of $f\left(\bT_{(1:m-r+1)}\right) \mid \bT_{(m-r+2:m)}$ where $\bT \sim \mathcal{N}(\btheta, I_m)$ with $\btheta \in \mathbb{R}^m$ such that $\btheta_{(1:m-r+1)} = \bm{0}$ and $\btheta_{(m-r+2:m)}$ comes from the first argument to $c_\alpha$. Note $c_{\alpha}\left(\btheta_{(m-r+2:m)}, \bT_{(m-r+2:m)}\right)$ is permutation invariant to the elements of the underlying $\btheta$, i.e., the resulting quantile would be the same if $\bT \sim \mathcal{N}(\sigma(\btheta), I_m)$  where $\sigma(\btheta)$ is some permutation of the elements of $\btheta$. Therefore, we represent any $\btheta \in \Theta_0^{r/m}$ by $\btheta_{(m-r+2:m)}$. We condition on $\bT_{(m-r+2:m)}$ because, intuitively, we want to condition on as much information as possible to make the Type I error as insensitive as possible to the error in the estimation of $\btheta_{(m-r+2:m)}$. 

\begin{definition}[Conditional PCH (cPCH) test]
The level-$\alpha$ cPCH test rejects $H_{0}^{r/m}$ when $f\left(\bm{T}_{(1:m-r+1)}\right) > c_{a(\alpha)}\left(\bT_{(m-r+2:m)}, \bT_{(m-r+2:m)}\right)$\rev{, where $a(\alpha) \in [0, \alpha]$ is the value such that $$
    \max_{\bm{\theta} \in \Theta_0^{r/m}}  \mathbb{P}_{\bm{\theta}_{(m-r+2:m)}} \left(\bm{T}_{(1:m-r+1)} > c_{a(\alpha)}\left(\bT_{(m-r+2:m)}, \bT_{(m-r+2:m)}\right)\right) = \alpha.
$$}
\label{cpchdef}
\end{definition}
\rev{In words, $a(\alpha)$ is the level at which the maximum Type I error of the unadjusted cPCH test is equal to $\alpha$.} \re{Note, we suppress the dependence on $m$ and $r$.}
In the definition above, $\bm{T}_{(m-r+2:m)}$ in the first argument of \rev{$c_{a(\alpha)}$} serves as the estimator for $\btheta_{(m-r+2:m)}$, while the same quantity in the second argument denotes what is conditioned on. Analogous to our construction of the cPCH test when $r=m=2$ in Section~\ref{section:motivation}, $\bm{T}_{(m-r+2:m)}$ is the MLE for $\btheta_{(m-r+2:m)}$.


For all results in this paper, we focus on the combining functions of the Simes and Fisher global null tests. Note, however, that the definition of the cPCH test is general, as it allows the analyst to in principle specify any $f$ of her choice. Our code allows the analyst to specify $f$ and provides implementations of the cPCH test using Simes' and Fisher's combining functions as used in this paper.

While the cPCH test has valid Type I error control by definition, to perform it, we must be able to compute \re{the unadjusted cPCH test and find the appropriate adjustment level $a(\alpha)$} 
for any $m$, $r$, and nominal level $\alpha$. We discuss such computational details in the following section.

\section{Computing the cPCH Test}\label{section:computing_cpch}

\subsection{Computing Unadjusted cPCH p-values}\label{section:computing}
\re{
In this section, we discuss the challenges of computing unadjusted cPCH p-values} and describe our novel procedure for computing these p-values efficiently.

Given the observed $f(\bT_{(1:m-r+1)}) = f_{\text{obs}}$, the \rev{unadjusted} cPCH p-value can be written as
$$ \mathbb{P}_{\hat{\btheta}}\left(f(\bm{T}_{(1:m-r+1)}) \geq f_{\text{obs}} \,\mid\, \bT_{(m-r+2:m)}\right),
$$
where $\hat{\btheta}$ is comprised of $m-r+1$ zeroes and the elements of $\bT_{(m-r+2:m)}$. Thus, $\hat{\btheta}$ is a shorthand for the MLE of $\btheta_{(m-r+2:m)}$ as in Definition~\ref{cpchdef}. Because the value of 

$\mathbb{P}_{\hat{\btheta}}\left( f(\bm{T}_{(1:m-r+1)}) \geq f_{\text{obs}} \given \bT_{(m-r+2:m)} \right)$ is invariant to re-indexing of the elements of $\hat{\btheta}$, throughout this paper, we default to $\hat{\btheta} = \left(0, ..., 0, T_{(m-r+2)}, ..., T_{(m)}\right)$. 

Computing this probability exactly requires us to derive the density of $\bT_{(1:m-r+1)} \mid \bT_{(m-r+2:m)}$ where $\bT_{(1:m-r+1)}, \bT_{(m-r+2:m)}$ are order statistics of independent but non-identically distributed (i.n.i.d.) random variables.
Though the conditional density functions of order statistics of i.n.i.d. random variables has been studied in previous works \cite[]{beg, ozbey2019distributions}, calculating the conditional densities based on the techniques in such works would 
involve enumerating all $m!$ possible permutations of the order statistics, \revi{which scales exponentially in $m$}. 
Alternatively, if we can generate samples from $f\left(\bT_{(1:m-r+1)}\right) \mid \bT_{(m-r+2:m)}$, we can estimate this conditional distribution by taking many independent samples and computing the empirical distribution. Thus, even if the analytic form of the conditional density is intractable, we can compute \rev{unadjusted} cPCH p-values with high accuracy as long as we can efficiently sample from the conditional density. In particular, our proposed sampling procedure is computationally more efficient than computing the exact conditional densities when $r$ is small, e.g., when $r=2$ as in replicability analysis settings \cite[]{ followup, HY2014, adafilt}, the complexity of computing a cPCH p-value is linear, i.e., $O(m)$.  

Now we briefly describe our strategy for the above sampling problem \revi{with a complete outline of the procedure provided in Supplementary Materials~\ref{section:computing_further_details}}. First, when $r=m=2$, we derive the analytic form of the conditional density $T_{(1)} \mid T_{(2)}$, thus allowing us to obtain exact \rev{unadjusted} cPCH p-values without sampling. When $m > 2$, we develop a new, efficient procedure for sampling from $f\left(\bT_{(1:m-r+1)}\right) \mid \bT_{(m-r+2:m)}$. On a high level, our approach involves conditioning on extra events about which of the $T_1, ..., T_m$ correspond to the order statistics $\bm{T}_{(1:m-r+1)}$ and $\bT_{(m-r+2:m)}$. By doing so, we can express the conditional density of $f\left(\bT_{(1:m-r+1)}\right) \mid \bT_{(m-r+2:m)}$ as a mixture distribution such that \rev{(a) the involved mixture weights can be computed analytically, and (b) the mixture components can be estimated by sampling from rather simple probability distributions.}

\revi{Using $10{,}000$ samples,} we found that the average computation time of a single cPCH p-value is $< 30$ milliseconds for any $2\leq m \leq 4$ and $2\leq r \leq m$. When $r=m=2$, we can compute the \rev{unadjusted} cPCH p-value exactly (and hence, no sampling is required); in this case, the computation time is $\approx 1$ millisecond. See Figure~\ref{fig:comptimes} in the Supplementary Materials~\ref{section:singlepch_additionalresults} for further details on computation time.
  
Notably, our sampling scheme does not rely on any specific properties of the normal distribution; the only assumption necessary is that the base test statistics are independent. Though we have generally assumed that the $T_i$ are normally distributed, we can compute \rev{unadjusted} cPCH p-values \rev{(and hence, cPCH p-values)} assuming the base test statistics are distributed under \emph{any} one-parameter location family, such as a t-distribution with fixed degrees of freedom. Our code contains implementations of the cPCH test for both normal and t-distributed base test statistics.

\subsection{Computing $a(\alpha)$}\label{section:validity}
\re{To find $a(\alpha)$, one must be able to efficiently traverse the Type I error curve (as a function of the nominal level) of the unadjusted cPCH test to identify the nominal level at which the unadjusted cPCH test has maximum Type I error equal to $\alpha$. The efficiency of computing unadjusted cPCH p-values enables us to estimate the Type I error efficiently at any specific $\btheta \in \Theta_0^{r/m}$, so the primary challenge of computing $a(\alpha)$ is efficiently searching over $\btheta \in \Theta_0^{r/m}$ across a range of nominal levels.
Fortunately, for most common PCH testing scenarios, such as causal mediation analysis, where $m = 2$, and replicability analysis, where $m$ is often $\leq 5$ \cite[]{followup, HY2014, adafilt}, the dimension of $\Theta_0^{r/m}$ is generally small. Additionally Figure~\ref{fig:t1error} suggests that the Type I error curve is smooth and unimodal. Because of these properties, finding the maximum of the Type I error curve is amenable to computational approaches such as grid search and stochastic gradient descent (SGD). Further implementational details on these approaches, including justification of the use of stochastic gradient descent for finding $a(\alpha)$, are included in the Supplementary Materials~\ref{sm:a_alpha} and~\ref{sec:sgd_comp}.



 We emphasize that $a(\alpha)$ is a universal constant that only depends on $m$, $r$, and $\alpha$.
We pre-computed $a(\alpha)$ across $m\leq 5$, $2\leq r \leq m$, and $\alpha \in [0, 1]$ and incorporated these pre-computed values into our implementation provided at \if1\blind
{\texttt{https://github.com/biyonka/cpch}} \fi
\if0\blind
{[link removed for anonymization]} \fi. Hence, there is no computational or implementational cost to the user to find $a(\alpha)$, as long as they are considering an $m \leq 5$, which is common in PCH testing. Furthermore, the adjustment table (provided as a CSV in our Github, with a subset provided in Table~\ref{tab:lookup} of Supplementary Materials~\ref{sec:lookup}) can be used to immediately look up the corresponding $a(\alpha)$ for various common values $\alpha$; note that the adjustment is often small. 
In our Github repository, we also provide all scripts necessary to find $a(\alpha)$ for $m > 5$ and a CSV file of the full adjustment table for various $m, r,$ and $\alpha$. }


Additionally, we provide an exhaustive characterization of the unadjusted cPCH p-value distribution at effectively \emph{every} $\btheta \in \Theta^{r/m}_0$ for realistic values of $m$ and $r$, and show that it is approximately uniform and non-conservative; see Supplementary Materials~\ref{appendix:approx_valid}. Hence, a user can generally proceed by using the unadjusted cPCH test for larger $m$ without incurring much Type I error inflation. In fact, we show that under the least favorable null case, the limiting Type I error of the unadjusted cPCH test (for any $m$) is exactly $\alpha$ as the sample size of the data used for computing the base test statistics approaches infinity; see Supplementary Materials~\ref{appendix:lfn_valid} for formal theorem statement and proof. Such a result is consistent with the intuition of the unadjusted cPCH test being a finite-sample correction for an exactly valid oracle test, and hence with sufficiently large data, should have similar performance to the oracle test.

\section{Simulations} \label{section:simulations} 
 \subsection{Single PCH Testing}\label{section:singlepower}
In this section, we empirically evaluate the power and Type I error of the cPCH test in comparison with existing approaches (Simes' and Fisher's tests) and the cPCH Oracle test presented in Section~\ref{section:motivation}. Recall, we define $r^\star$ as the true number of non-null hypotheses. We generate data by sampling $\bm{T}$ independently from the model:
$$T_h \iid \mathcal{N}(\theta, 1),\; h = 1, ..., r^\star, \text{ \hspace{0.5cm}} T_l  \iid \mathcal{N}(0, 1),\; l = r^\star+1, ..., m.$$
We generate p-values for testing $H_0^{r/m}$ using the cPCH test, the cPCH Oracle test, and the standard PCH tests (Fisher and Simes).

\begin{figure}[ht]
        \centering
        \includegraphics[width = \textwidth]{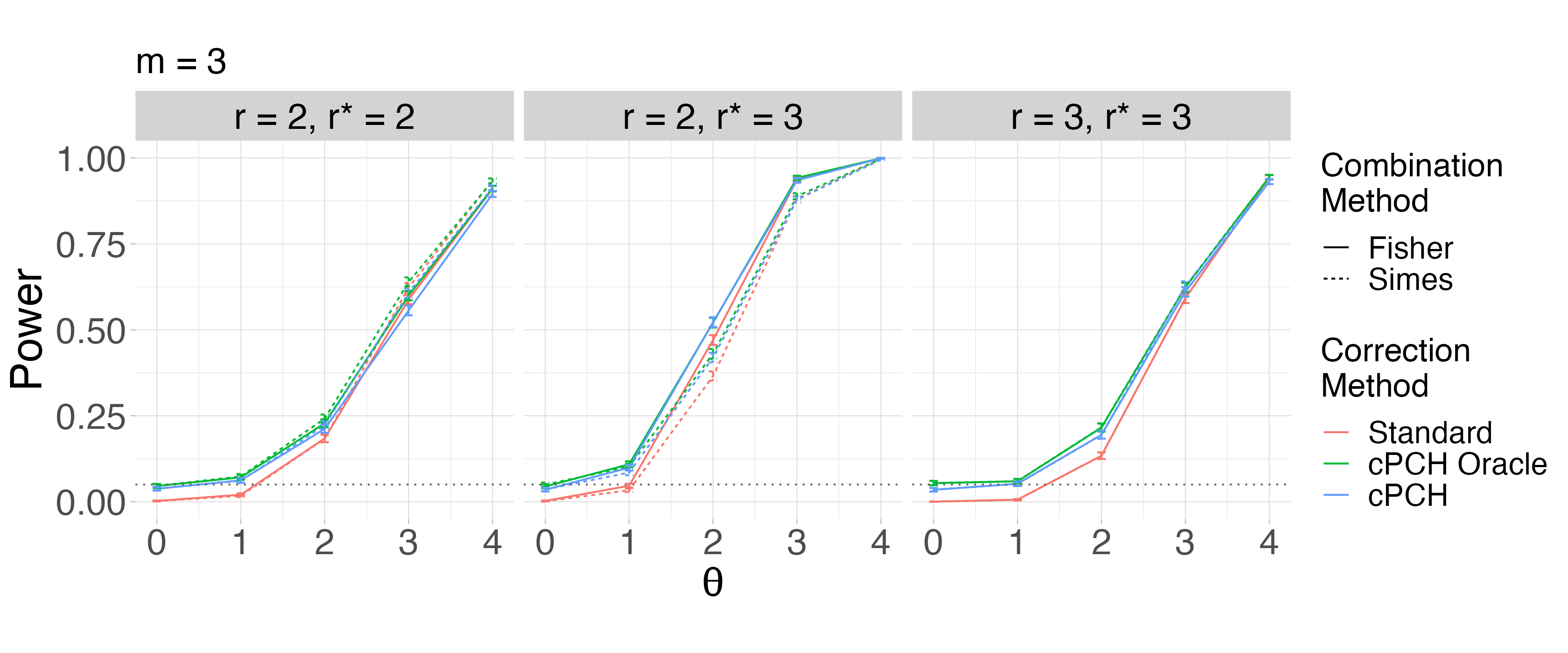}
        \vspace{-22pt}
 \caption{Power 
 across all alternative cases $(r^\star \geq r)$ for testing $H_0^{r/3}$ at level $\alpha = 0.05$ (dotted grey line). Each point represents the proportion of p-values below $\alpha$ over $5000$ independent replicates of the data generating procedure described in Section~\ref{section:singlepower}. Error bars depict $\pm 2$ standard errors.} 
        \label{fig:single_pc_power}
    \end{figure}

As shown in Figure~\ref{fig:single_pc_power}, for $m=3$, the cPCH test has \emph{nearly identical power to the cPCH Oracle test} and in particular, is more powerful than standard single PCH tests \revi{in low-signal regimes while matching or nearly matching the power of standard single PCH tests for larger signal strengths}. Figure~\ref{fig:adjusted_single_pc_null_plot_m_3} in the Supplementary Materials~\ref{section:singlepch_additionalresults}, the analogous plot for null values of $r^\star$, shows that the cPCH test controls Type I error for all null configurations and is far less conservative under the null than the standard single PCH tests. We perform this simulation for various $m$ and find similar results; see the Supplementary Materials~\ref{section:singlepch_additionalresults} for further details. 

As the cPCH test assumes a known model for the base test statistics, it is important to evaluate its robustness to model misspecification. Since we can compute cPCH p-values for $\bm{T}$'s distributed under any one-parameter location family, we can assess the performance of the cPCH test (which assumes $\bm{T}$ is normally distributed) when $\bm{T}$ is generated from a different one-parameter location family, such as the t-distribution with a fixed scale and degrees of freedom. Figure~\ref{fig:robust} in the Supplementary Materials~\ref{section:robustness} shows that the cPCH test remains powerful and maintains Type I error control for most null configurations under t-distributed model misspecification, even in some settings where the base test statistics have Cauchy distribution, i.e., the degrees of freedom is $1$. Notably, both the properly specified and misspecified cPCH test maintains Type I error control for all null configurations when the degrees of freedom is $\geq 5$. In most applied settings, the degrees of freedom of a t-distributed base test statistic reflects the underlying sample size used to compute the base test statistic. Thus, excluding settings where the sample size is extremely small, we expect the cPCH test to be robust to model misspecification. See the Supplementary Materials~\ref{section:robustness} for further details.

\subsection{Multiple PCH Testing}\label{section:mt}

Much of the existing literature on PCH testing addresses the conservativeness of standard PCH testing by sharing information across multiple PCH tests. Although the main focus of this paper is on testing a \emph{single} PCH, we conduct an extensive simulation study to show how cPCH testing can be used for multiple PCH testing and, in some cases, improve upon existing multiple PCH testing procedures. From now on, we assume there are $M$ PCH's being simultaneously tested. We denote $T_{ij}$ and $p_{ij}$ as the $i$th base test statistic and p-value, respectively, for the $j$th PCH being tested, $i = 1, ..., m$, $j = 1, ..., M$. \re{Recall that the cPCH test only requires that the test statistics for each individual PCH has independence across $i$, and hence, does not impose any limitations the dependence structure across the PCH's for multiple testing. The users can simply select the multiple testing procedure to be used with the cPCH p-values to accommodate the expected dependence across the PCH's.} 
\re{Our simulation study explores the $r=m=2$ setting. } 
We choose $r=m=2$ because it is the simplest setting for PCH testing and the one that many state-of-the-art methods are designed for, thus allowing us to compare the cPCH test with the most methods. 



\re{We consider the following single PCH testing approaches (which we combine with different multiple testing procedures): our cPCH test, the test of \cite{miles2021optimal} (with their proposed conversion to a p-value, which is necessary for its use in multiple testing), and the standard (i.e., Fisher's, Simes') PCH test. We consider the following multiple PCH testing procedures (which require there being multiple PCH's being tested simultaneously): HDMT \citep{hdmt}, AdaFilter 
 \citep{adafilt}, DACT \citep{dact}, and the method of \cite{dickhaus}. A detailed overview of all methods under comparison can be found in the Supplementary Materials~\ref{section:methodsundercomparision_appendix}. 
 
  Along with the base test statistics $T_{ij}, i=1,2$, $j=1,..., M$, all methods also have access to a univariate covariate $X_j, j=1,..., M$ that is informative of the PCH being non-null. \revi{Specifically, we generate the data $(X_j, T_{ij})$, $i = 1, 2$, $j = 1, ..., 10,000$ by sampling independently from the following model:
\begin{align*}
X_j &\sim \text{Unif}[0, 1]\\
\gamma_{ij} \mid X_j & = \text{Bern}(\pi_1(X_j)), \text{ where } \pi_1(X_j) = \begin{cases} 
                                              1 & \text{if } X_j \geq 0.95 \\
                                              0.1 & \text{otherwise} 
                                          \end{cases}\\
T_{ij} \mid \gamma_{ij}, X_j &\sim  \mathcal{N}(\theta \gamma_{ij} , 1)
\end{align*}
Here, $X_j$ represents a (univariate) covariate 
 that, when large, is highly informative that the PCH is non-null.
 }
Multiple PCH testing with covariates can arise in several important settings such as fMRI analysis, where naturally occurring covariate information like the $(x, y, z)$ coordinates of the brain voxel locations can be highly informative of brain activation regions, and genetic microarray studies, where the large amount of redundancy across microarray studies enables prior information, such as discoveries from an earlier study, to inform hypothesis testing in later studies \citep{li2017accumulation}. We explore such a microarray example using real data in Section~\ref{section:dmd}. 
While it is unclear how to leverage covariate information in many state-of-the-art multiple PCH testing approaches (such as those mentioned above), all single PCH testing procedures, including the cPCH test, can be combined with a covariate-assisted multiple testing procedure such as AdaPT--GMM \cite[]{adaptgmm} to utilize available side information. Hence,} we combine the single PCH tests with three different FDR controlling procedures: Benjamini--Hochberg \cite[]{Benjamini1995}, Storey's procedure \cite[]{storey}, and AdaPT--GMM \cite[]{adaptgmm}, a covariate-assisted multiple testing procedure. \re{We set the FDR level $q=0.1$ and $M=10{,}000$.}

   \begin{figure}[ht]
        \centering
        \includegraphics[width = \textwidth]{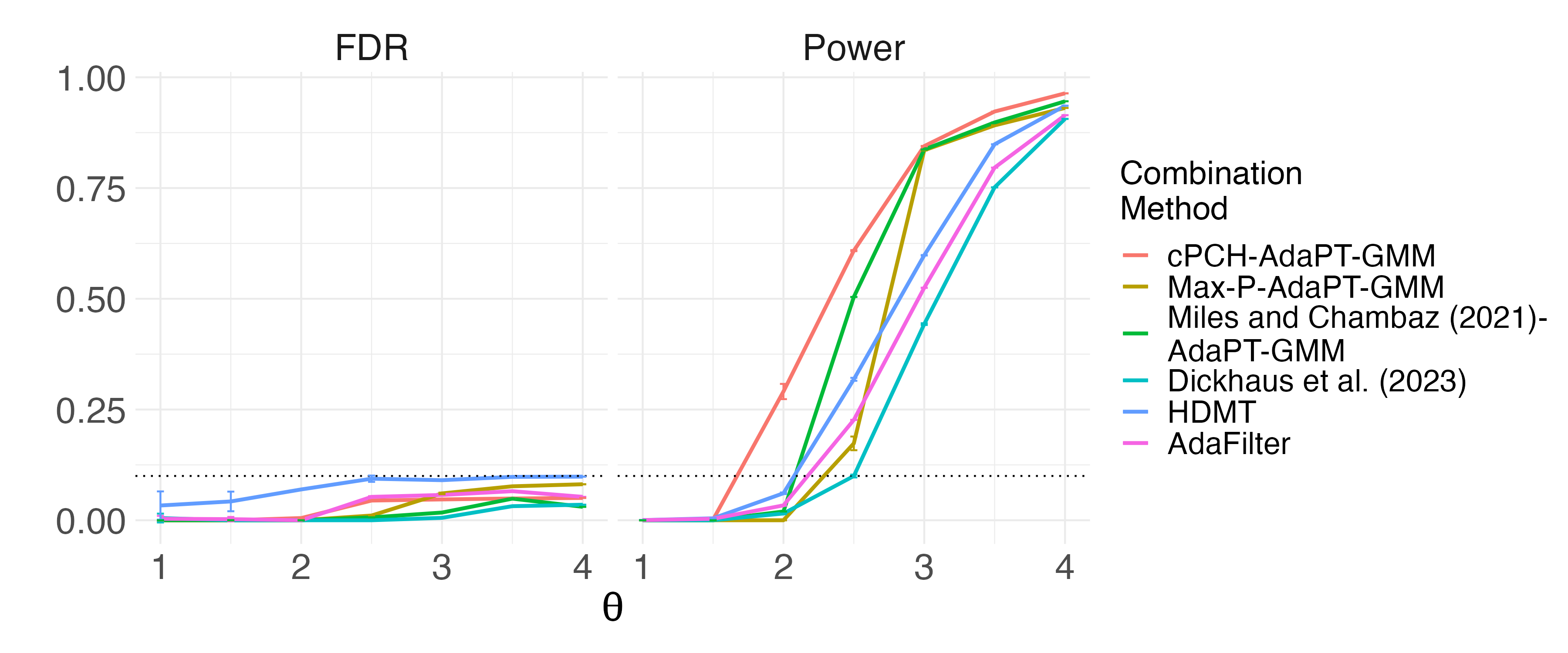}
        \vspace{-22pt}
 \caption{FDR and power of various PCH multiple testing methods at nominal FDR level $q = 0.1$ (dotted black line) for testing $H_0^{2/2}$. Each point represents $100$ independent replicates of the data generating procedure described in Section~\ref{section:mt} for a given $\theta$. Error bars depict $\pm 2$ standard errors.} 
        \label{fig:mt_cov}
    \end{figure}

\re{For presentational purposes, Figure~\ref{fig:mt_cov} only shows the single PCH testing approaches with AdaPT--GMM, as ADaPT--GMM provided the highest power gains while still controlling FDR for all single PCH tests. We also removed the DACT method as it had significant FDR inflation; see Figure~\ref{fig:full_adapt} in the Supplementary Materials~\ref{section:sm_mt} for the full plot depicting all methods under comparison.
As shown in Figure~\ref{fig:mt_cov},} the cPCH test with AdaPT--GMM is more powerful than all other approaches, while still maintaining FDR control.   Additionally, the cPCH test with AdaPT--GMM is much more powerful than every other method in the low signal regime, which, as we discussed in Section~\ref{section:classical_intro}, is an important regime for various application areas such as genetic epidemiology. 
Hence, the primary advantage of the cPCH test for multiple PCH testing is its flexibility to be combined with nearly any (non-PCH-specific) multiple testing procedure, allowing users to choose a procedure based on their problem specifics. \re{While the other single PCH tests are also able to utilize covariate information, they are often still significantly less powerful than cPCH with AdaPT--GMM, as the power of many multiple testing procedures can be especially sensitive to conservative p-values. Because the cPCH test produces approximately uniform null p-values, it is almost always more powerful than using a different single PCH test with the same multiple testing procedure.} \re{We also performed an additional simulation with $m=4$ and no covariate information and showed various settings in which cPCH can outperform state-of-the-art approaches; see Supplementary Materials~\ref{section:pch_mt_sim_study} for details. } 

\section{Differential Gene Expression Analysis}
\label{section:dmd}
Duchenne Muscular Dystrophy is a genetic disorder characterized by progressive muscle degeneration in young children. Understanding the genetic markers of Duchenne muscular dystrophy enables clinicians to link the effects of the disease with their associated genes, thus providing a pathway for targeted drug therapies. Analyzing the average power in both a single PCH testing and covariate-assisted setting, we demonstrate that the cPCH test has improved power for detecting replicating genetic markers that may be associated with Duchenne muscular dystrophy progression over existing methods in these settings.

We utilize four independent Duchenne muscular dystrophy-related microarray datasets (GDS214, GDS563, GDS1956, and GDS3027) from the Gene Expression Omnibus (GEO) database, a public functional genomics data repository. The datasets are first pre-processed using \texttt{limma}, the standard package for processing microarray data. For each gene, \texttt{limma} outputs the test statistic from a two-sample t-test for testing whether that gene is differentially expressed in Duchenne muscular dystrophy patients compared to healthy subjects; the test statistics are post-processed as in \cite{adafilt} to account for some data artifacts. Therefore, we expect the test statistics to follow an approximately normal location family with unit variance, as assumed in Condition~\eqref{eq:distassump}. In total, $M = 1871$ unique genes are shared among the four studies.


First, we can treat this data as a set of p-values and assess the empirical power, i.e., the total positive rate, calculated as the proportion of p-values below the nominal level $\alpha$ across the $M$ genes, of the cPCH test compared to that of its standard counterparts for single PCH testing. As shown in Table~\ref{tab:avgpower_dmd}, the cPCH test makes substantially more rejections than its standard counterparts (note that the method of \cite{miles2021optimal} is not applicable because $m=4>2$) at nominal level $\alpha = 0.05$.
      \begin{table}[!htb]
      \centering
     \begin{tabular}{lccc}
     & $r=2$  & $r=3$ & $r=4$  \\
     \midrule
 cPCH-Fisher & 27.0 (+15.4\%) & 17.1 (+34.4\%) & ~8.3 (+84.0\%) \\
 cPCH-Simes & 23.7 (+17.9\%)   & 16.2 (+38.5\%) & ~8.3 (+66.0\%) \\
     \end{tabular}
      \caption{Total positive rate of the cPCH test compared to its standard counterparts. The first value in each entry represents the proportion of cPCH p-values below $\alpha$ and the second value represents the percent increase compared to the corresponding standard single PCH test. Treating the p-values as independent, the implied standard errors are all $< 0.9$. 
      }
      \label{tab:avgpower_dmd}
  \end{table}

Next, we emulate a covariate-assisted setting. Microarray datasets often contain a large amount of redundancy, as assays run across various independent labs aim to profile many of the same genes potentially associated with a disease \citep{li2017accumulation}. Hence, we can use discoveries from an earlier study as covariate information to assist multiple PCH testing on later studies. To emulate this setting, we first apply the Benjamini--Hochberg procedure to the data from the earliest performed microarray, and then use the decision vector corresponding to the discovered genes as covariate information for performing multiple PCH testing for $r= 2, 3$ on the remaining three studies ($m=3$). We use nominal FDR level $q = 0.1$ throughout. We plug the cPCH and standard Fisher p-values generated from the remaining three studies into the AdaPT--GMM multiple testing procedure \citep{adaptgmm}, which enables us to utilize the covariate information from the first study. We also used each of the other studies as holdout and obtained generally similar results, see Table~\ref{tab:additional_dmd}.

\begin{table}[!htb]
      \centering
     \begin{tabular}{lrrrr}
      $r$ & cPCH-AdaPT--GMM &  AdaFilter & Dickhaus et al. (2023) & Fisher-AdaPT--GMM\\
  \midrule
2 & 335 & 268 & 172 & 307 \\
3 & 80 & 79 & 13 &  69\\
     \end{tabular}
      \caption{Rejections made in the follow-up study testing scenario described in Section~\ref{section:dmd}. Fisher's combining function is used throughout as we found outperforms Simes in this example. See Supplementary Materials~\ref{section:genes} for analogous results using the Simes combining function. \re{Note, we only compare the above methods because the other approaches (e.g., HDMT, DACT, and the test of \cite{miles2021optimal}) are only designed for the $r=m=2$ case.}
      }
      \label{tab:follow_dmd}
  \end{table}

\re{Table~\ref{tab:follow_dmd} shows that cPCH with AdaPT--GMM discovers more differentially expressed genes than alternative methods. Table~\ref{tab:selected_genes_3} in the Supplementary Materials~\ref{section:genes} shows $20$ of the total genes discovered by cPCH--AdaPT--GMM using $r=3$. As desired, many of these genes have biological functions associated with muscle maintenance and cell growth regulation. In particular, MYH3, MYH8, MYL4, and MYL5 are known genetic markers for Duchenne muscular dystrophy.
}

\if1\blind
{
\section{Acknowledgements}
The authors would like to thank Nathan Cheng for his help with Lemma 4, Jingshu Wang for sharing her pre-processing code for the Duchenne muscular dystrophy data, James Dai and Frank Wang for providing further information on the HDMT package, and Xihong Lin for helpful discussions regarding this work. 
} \fi
\if0\blind
{} \fi


\if1\blind
{
\section{Funding}
 B.L. is partially supported by the National Science Foundation via the Graduate Research Fellowship Program. L.Z. and L.J. are partially supported by a grant from the National Science Foundation (Grant \#DMS-2134157). B.L., L.Z., and L.J. are partially supported by a CAREER grant from the National Science Foundation (Grant \#DMS-2045981).

\section{Disclosure Statement}
The authors report there are no competing interests to declare.
} \fi
\if0\blind
{} \fi

\addcontentsline{toc}{section}{References}
\bibliographystyle{agsm}
\bibliography{bibtex}

@article{ozbey2019distributions,
  title={{On Distributions of Order Statistics for Nonidentically Distributed Variables}},
  author={Ozbey, Fahrettin and G{\"u}ng{\"o}r, Mehmet and Bulut, Yunus},
  journal={Appl. Math},
  volume={13},
  number={1},
  pages={11--16},
  year={2019}
}

@article{BH2008,
 ISSN = {0006341X, 15410420},
 author = {Yoav Benjamini and Ruth Heller},
 journal = {Biometrics},
 number = {4},
 pages = {1215--1222},
 publisher = {[Wiley, International Biometric Society]},
 title = {{Screening for Partial Conjunction Hypotheses}},
 volume = {64},
 year = {2008}
}

@article{dact,
author = {Zhonghua Liu and Jincheng Shen and Richard Barfield and Joel Schwartz and Andrea A. Baccarelli and Xihong Lin},
title = {Large-Scale Hypothesis Testing for Causal Mediation Effects with Applications in Genome-wide Epigenetic Studies},
journal = {Journal of the American Statistical Association},
volume = {117},
number = {537},
pages = {67-81},
year  = {2022},
publisher = {Taylor & Francis},
doi = {DOI: 10.1080/01621459.2021.1914634},
eprint = { 
    
        https://doi.org/10.1080/01621459.2021.1914634
    
    

}

}

@misc{adafilt,
      title={{Detecting Multiple Replicating Signals using Adaptive Filtering Procedures}}, 
      author={Jingshu Wang and Lin Gui and Weijie J. Su and Chiara Sabatti and Art B. Owen},
      year={2021},
      eprint={1610.03330},
      archivePrefix={arXiv},
      primaryClass={stat.ME}
}

@article{BHY2009,
author = { Yoav Benjamini and Ruth Heller and Daniel Yekutieli},
address = {England},
copyright = {Copyright 2009 The Royal Society},
issn = {1364-503X},
journal = {Philosophical Transactions of the Royal Society of London. Series A: Mathematical, Physical, and Engineering Sciences},
publisher = {The Royal Society},
pages = {4255-4271},
title = {{Selective Inference in Complex Research}},
volume = {367},
year = {2009},
language = {eng},
number = {1906},
}

@article{HY2014,
author = {Ruth Heller and Daniel Yekutieli},
title = {{Replicability Analysis for Genome-Wide Association Studies}},
volume = {8},
journal = {The Annals of Applied Statistics},
number = {1},
publisher = {Institute of Mathematical Statistics},
pages = {481 -- 498},
year = {2014},
doi = {10.1214/13-AOAS697},
URL = {https://doi.org/10.1214/13-AOAS697}
}

@article{followup,
author = { Marina   Bogomolov  and  Ruth   Heller },
title = {{Discovering Findings That Replicate From a Primary Study of High Dimension to a Follow-Up Study}},
journal = {Journal of the American Statistical Association},
volume = {108},
number = {504},
pages = {1480-1492},
year  = {2013},
publisher = {Taylor & Francis},
DOI = {10.1080/01621459.2013.829002},
eprint = { 
        https://doi.org/10.1080/01621459.2013.829002
}
}

@Article{huang,
  author =   {Yen-Tsung Huang},
  title =    {{Genome-wide Analyses of Sparse Mediation Effects under Composite Null Hypotheses}},
  journal = {Annals of Applied Statistics},
  volume = {13},
  pages = {60 - 84},
  year =     2019,
   link =  {https://doi.org/10.1214/18-AOAS1181},
}

@article{efron2001,
author = {Bradley Efron and Robert Tibshirani and John D Storey and Virginia Tusher},
title = {{Empirical Bayes Analysis of a Microarray Experiment}},
journal = {Journal of the American Statistical Association},
volume = {96},
number = {456},
pages = {1151-1160},
year  = {2001},
publisher = {Taylor & Francis},
doi = {10.1198/016214501753382129},

URL = { 
        https://doi.org/10.1198/016214501753382129
    
},
eprint = { 
        https://doi.org/10.1198/016214501753382129
    
}

}

@article{twogroup,
author = {Bradley Efron},
title = {{Microarrays, Empirical Bayes and the Two-Groups Model}},
volume = {23},
journal = {Statistical Science},
number = {1},
publisher = {Institute of Mathematical Statistics},
pages = {1 -- 22},
year = {2008},
doi = {10.1214/07-STS236},
URL = {https://doi.org/10.1214/07-STS236}
}

@article{Small,
author = {Bikram Karmakar and Dylan S. Small},
title = {{Assessment of the Extent of Corroboration of an Elaborate Theory of a Causal Hypothesis Using Partial Conjunctions of Evidence Factors}},
volume = {48},
journal = {The Annals of Statistics},
number = {6},
publisher = {Institute of Mathematical Statistics},
pages = {3283 -- 3311},
year = {2020},
doi = {10.1214/19-AOS1929},
URL = {https://doi.org/10.1214/19-AOS1929}
}

@article{li2017accumulation,
  title={Accumulation tests for FDR control in ordered hypothesis testing},
  author={Li, Ang and Barber, Rina Foygel},
  journal={Journal of the American Statistical Association},
  volume={112},
  number={518},
  pages={837--849},
  year={2017},
  publisher={Taylor \& Francis}
}

@misc{miles2021optimal,
      title={Optimal tests of the composite null hypothesis arising in mediation analysis}, 
      author={Caleb H. Miles and Antoine Chambaz},
      year={2021},
      eprint={2107.07575},
      archivePrefix={arXiv},
      primaryClass={math.ST}
}

@article{Karmakar,
author = {Bikram Karmakar and Dylan S. Small and Paul R. Rosenbaum},
title = {{Reinforced Designs: Multiple Instruments Plus Control Groups as Evidence Factors in an Observational Study of the Effectiveness of Catholic Schools}},
journal = {Journal of the American Statistical Association},
volume = {116},
number = {533},
pages = {82-92},
year  = {2021},
publisher = {Taylor & Francis},
doi = {10.1080/01621459.2020.1745811},

URL = { 
        https://doi.org/10.1080/01621459.2020.1745811
    
},
eprint = { 
        https://doi.org/10.1080/01621459.2020.1745811
    
}

}

@article{voxels,
  title={{Conjunction Group Analysis: An Alternative to Mixed/Random Effect Analysis}},
  author={Ruth Heller and Yulia Golland and Rafael Malach and Yoav Benjamini},
  journal={NeuroImage},
  year={2007},
  volume={37},
  pages={1178-1185}
}

@article{Rietveld,
author = {Cornelius A. Rietveld and Dalton Conley and Nicholas Eriksson and Tõnu Esko and Sarah E. Medland and Anna A. E. Vinkhuyzen and Jian Yang and Jason D. Boardman and Christopher F. Chabris and Christopher T. Dawes and Benjamin W. Domingue and David A. Hinds and Magnus Johannesson and Amy K. Kiefer and David Laibson and Patrik K. E. Magnusson and Joanna L. Mountain and Sven Oskarsson and Olga Rostapshova and Alexander Teumer and Joyce Y. Tung and Peter M. Visscher and Daniel J. Benjamin and David Cesarini and Philipp D. Koellinger and the Social Science Genetics Association Consortium},
title ={{Replicability and Robustness of Genome-Wide-Association Studies for Behavioral Traits}},
journal = {Psychological Science},
volume = {25},
number = {11},
pages = {1975-1986},
year = {2014},
doi = {10.1177/0956797614545132},
    note ={PMID: 25287667},

URL = { 
        https://doi.org/10.1177/0956797614545132
    
}
}

@article{odonovan,
  title={{Identification of Loci Associated with Schizophrenia by Genome-wide Association and Follow-up}},
  author={Michael C O'Donovan and Nicholas Craddock and Nadine Norton and Hywel Williams and Timothy Peirce and Valentina Moskvina and Ivan Nikolov and Marian Hamshere and Liam Carroll and Lyudmila Georgieva and Sarah Dwyer and Peter Holmans and Jonathan L Marchini},
  journal={Nature Genetics},
  year={2008},
  volume={40},
  pages={1053–1055},
  URL = {https://doi.org/10.1038/ng.201}
}

@misc{dickhaus,
  title = {{Multiple Testing of Partial Conjunction Null Hypotheses, With Application to Replicability Analysis of High Dimensional Studies}},
      author={Thorsten Dickhaus and Ruth Heller and Anh-Tuan Hoang},
      year={2023},
      eprint={2110.06692},
      archivePrefix={arXiv},
      primaryClass={stat.ME}
}

@article{zuo,
    doi = {DOI: 10.1371/journal.pone.0026726},
    author = {Zuo, Lingjun AND Zhang, Clarence K. AND Wang, Fei AND Li, Chiang-Shan R. AND Zhao, Hongyu AND Lu, Lingeng AND Zhang, Xiang-Yang AND Lu, Lin AND Zhang, Heping AND Zhang, Fengyu AND Krystal, John H. AND Luo, Xingguang},
    journal = {PLOS ONE},
    publisher = {Public Library of Science},
    title = {{A Novel, Functional and Replicable Risk Gene Region for Alcohol Dependence Identified by Genome-Wide Association Study}},
    year = {2011},
    month = {11},
    volume = {6},
    url = {https://doi.org/10.1371/journal.pone.0026726},
    pages = {1-8},
    number = {11},

}

@article{storey,
 title = {{Strong Control, Conservative Point Estimation and Simultaneous Conservative Consistency of False Discovery Rates: A Unified Approach}},
 ISSN = {13697412, 14679868},
 URL = {http://www.jstor.org/stable/3647634},
 author = {John D. Storey and Jonathan E. Taylor and David Siegmund},
 journal = {Journal of the Royal Statistical Society. Series B (Statistical Methodology)},
 number = {1},
 pages = {187-205},
 publisher = {Royal Statistical Society, Wiley},
 volume = {66},
 year = {2004}
}

@article{storey2002,
 ISSN = {13697412, 14679868},
 URL = {http://www.jstor.org/stable/3088784},
 author = {John D. Storey},
 journal = {Journal of the Royal Statistical Society. Series B (Statistical Methodology)},
 number = {3},
 pages = {479--498},
 publisher = {[Royal Statistical Society, Wiley]},
 title = {{A Direct Approach to False Discovery Rates}},
 urldate = {2022-07-25},
 volume = {64},
 year = {2002}
}

@misc{adaptgmm, 
  doi = {arXiv: 2106.15812},
  
  url = {https://arxiv.org/abs/2106.15812},
  
  author = {Chao, Patrick and Fithian, William},
  
  title = {{AdaPT-GMM: Powerful and Robust Covariate-Assisted Multiple Testing}},
  
  publisher = {arXiv},
  
  year = {2021},
  
  copyright = {arXiv.org perpetual, non-exclusive license}
}

@article{hdmt,
author = {James Y. Dai and Janet L. Stanford and Michael LeBlanc},
title = {{A Multiple-Testing Procedure for High-Dimensional Mediation Hypotheses}},
journal = {Journal of the American Statistical Association},
volume = {0},
number = {0},
pages = {1-16},
year  = {2020},
publisher = {Taylor & Francis},
doi = {DOI: 10.1080/01621459.2020.1765785},

URL = { 
        https://doi.org/10.1080/01621459.2020.1765785
    
},
eprint = { 
        https://doi.org/10.1080/01621459.2020.1765785
    
}

}

@article{Barfield,
author = {Barfield, Richard and Shen, Jincheng and Just, Allan C and Vokonas, Pantel S and Schwartz, Joel and Baccarelli, Andrea A and VanderWeele, Tyler J and Lin, Xihong},
address = {United States},
copyright = {2017 WILEY PERIODICALS, INC.},
issn = {0741-0395},
journal = {Genetic Epidemiology},
language = {eng},
number = {8},
pages = {824-833},
publisher = {Wiley Subscription Services, Inc},
title = {{Testing for the Indirect Effect Under the Null for Genome‐Wide Mediation Analyses}},
volume = {41},
year = {2017},
}

@article{zhang,
    author = {Zhang, Haixiang and Zheng, Yinan and Zhang, Zhou and Gao, Tao and Joyce, Brian and Yoon, Grace and Zhang, Wei and Schwartz, Joel and Just, Allan and Colicino, Elena and Vokonas, Pantel and Zhao, Lihui and Lv, Jinchi and Baccarelli, Andrea and Hou, Lifang and Liu, Lei},
    title = {{Estimating and Testing High-Dimensional Mediation Effects in Epigenetic Studies}},
    journal = {Bioinformatics},
    volume = {32},
    number = {20},
    pages = {3150-3154},
    year = {2016},
    month = {06},
    issn = {1367-4803},
    doi = {DOI: 10.1093/bioinformatics/btw351},
    url = {https://doi.org/10.1093/bioinformatics/btw351},
    eprint = {https://academic.oup.com/bioinformatics/article-pdf/32/20/3150/25040558/btw351.pdf},
}

@article{sesia,
author = {Matteo Sesia  and Stephen Bates  and Emmanuel Candès  and Jonathan Marchini  and Chiara Sabatti },
title = {{False Discovery Rate Control in Genome-Wide Association Studies With Population Structure}},
journal = {Proceedings of the National Academy of Sciences},
volume = {118},
number = {40},
pages = {e2105841118},
year = {2021},
doi = {10.1073/pnas.2105841118},

URL = {https://www.pnas.org/doi/abs/10.1073/pnas.2105841118},
eprint = {https://www.pnas.org/doi/pdf/10.1073/pnas.2105841118}
}

@article{kraft,
author = {Peter Kraft and Eleftheria Zeggini and John P. A. Ioannidis},
title = {{Replication in Genome-Wide Association Studies}},
volume = {24},
journal = {Statistical Science},
number = {4},
publisher = {Institute of Mathematical Statistics},
pages = {561 -- 573},
year = {2009},
doi = {10.1214/09-STS290},
URL = {https://doi.org/10.1214/09-STS290}
}

@article{hirsch,
    author = {Joel N. Hirschhorn and David Altshuler},
    title = {{Once and Again—Issues Surrounding Replication in Genetic Association Studies}},
    journal = {The Journal of Clinical Endocrinology and Metabolism},
    volume = {87},
    number = {10},
    pages = {4438-4441},
    year = {2002},
    month = {10},
    issn = {0021-972X},
    doi = {DOI: 10.1210/jc.2002-021329},
    url = {https://doi.org/10.1210/jc.2002-021329},
    eprint = {https://academic.oup.com/jcem/article-pdf/87/10/4438/9154944/jcem4438.pdf},
}

@article{ioannidis,
    author = {John P. A. Ioannidis},
    title = {{Why Most Published Research Findings Are False}},
    journal = {PLoS Medicine},
    volume = {2},
    number = {8},
    pages = {e124},
    year = {2005},
    month = {8},
    doi = {DOI: 10.1371/journal.pmed.0020124}
}

@article{nci,
    author = {{NCI-NHGRI Working Group on Replication in Association Studies}},
    title = {{Replicating Genotype–Phenotype Associations}},
    journal = {Nature},
    volume = {447},
    pages = {655-660},
    year = {2007},
    month = {6},
    doi = {DOI: 10.1038/447655a}
}

@article{sanford,
author = {Sanford L. Braver and Felix J. Thoemmes and Robert Rosenthal},
title ={{Continuously Cumulating Meta-Analysis and Replicability}},
journal = {Perspectives on Psychological Science},
volume = {9},
number = {3},
pages = {333-342},
year = {2014},
doi = {10.1177/1745691614529796},
    note ={PMID: 26173268},

URL = { 
        https://doi.org/10.1177/1745691614529796
    
},
eprint = { 
        https://doi.org/10.1177/1745691614529796
    
}
}

@article{
camerer,
author = {Colin F. Camerer  and Anna Dreber  and Eskil Forsell  and Teck-Hua Ho  and Jürgen Huber  and Magnus Johannesson  and Michael Kirchler  and Johan Almenberg  and Adam Altmejd  and Taizan Chan  and Emma Heikensten  and Felix Holzmeister  and Taisuke Imai  and Siri Isaksson  and Gideon Nave  and Thomas Pfeiffer  and Michael Razen  and Hang Wu },
title = {{Evaluating Replicability of Laboratory Experiments in Economics}},
journal = {Science},
volume = {351},
number = {6280},
pages = {1433-1436},
year = {2016},
doi = {10.1126/science.aaf0918},
URL = {https://www.science.org/doi/abs/10.1126/science.aaf0918},
eprint = {https://www.science.org/doi/pdf/10.1126/science.aaf0918}
}

@article{Benjamini1995,
  author = {Benjamini, Yoav and Hochberg, Yosef},
  doi = {DOI: http://dx.doi.org/10.2307/2346101},
  journal = {Journal of the Royal Statistical Society Series B (Methodological)},
  number = 1,
  pages = {289-300},
  title = {{Controlling the False Discovery Rate: A Practical and Powerful Approach
	to Multiple Testing}},
  url = {http://dx.doi.org/10.2307/2346101},
  volume = 57,
  year = 1995
}

@article{jinandcai,
 ISSN = {01621459},
 URL = {http://www.jstor.org/stable/27639880},
 author = {Jiashun Jin and T. Tony Cai},
 journal = {Journal of the American Statistical Association},
 number = {478},
 pages = {495--506},
 publisher = {[American Statistical Association, Taylor & Francis, Ltd.]},
 title = {{Estimating the Null and the Proportion of Nonnull Effects in Large-Scale Multiple Comparisons}},
 urldate = {2022-07-25},
 volume = {102},
 year = {2007}
}

@article{dreyfuss,
  author = {Jonathan M. Dreyfuss and Yixing Yuchi and Xuehong Dong and Vissarion Efthymiou and Hui Pan andDonald C. Simonson and Ashley Vernon and Florencia Halperin and Pratik Aryal and Anish Konkar and Yinong Sebastian and Brandon W. Higgs and Joseph Grimsby and Cristina M. Rondinone and Simon Kasif and Barbara B. Kahn and Kathleen Foster and Randy Seeley and Allison Goldfine and Vera Djordjilović and Mary Elizabeth Patti},
  title = {{High-Throughput Mediation Analysis of Human Proteome and Metabolome Identifies Mediators of Post-Variatric Surgical Diabetes Control}},
  year = {2021},
  journal = {Nature Communications},
volume ={21},
pages={6951},
doi = {10.1038/s41467-021-27289-2},
url = {https://doi.org/10.1038/s41467-021-27289-2}
}

@article{baronkenny,
  author = {R M. Baron and D A. Kenny},
  title = {{The Moderator–Mediator Variable Distinction in Social Psychological Research: Conceptual, Strategic, and Statistical Considerations}},
  year = {1986},
  journal = {Journal of Personality
and Social Psychology},
volume ={51},
pages={1173}
}

@article{gd_conv,
author = {Kiwiel, Krzysztof and Murty, Katta},
year = {1996},
month = {04},
pages = {},
title = {Convergence of the Steepest Descent Method for Minimizing Quasiconvex Functions},
volume = {89},
journal = {Journal of Optimization Theory and Applications},
doi = {10.1007/BF02192649}
}

@misc{sgd_nonconvex,
  doi = {DOI: 10.48550/ARXIV.2104.00423},
  
  url = {https://arxiv.org/abs/2104.00423},
  
  author = {Patel, Vivak and Zhang, Shushu},
  
  title = {{Stochastic Gradient Descent on Nonconvex Functions with General Noise Models}},
  
  publisher = {arXiv},
  
  year = {2021},
  
  copyright = {arXiv.org perpetual, non-exclusive license}
}

@article{beg,
 ISSN = {0581572X},
 URL = {http://www.jstor.org/stable/25050725},
 author = {R. B. Bapat and M. I. Beg},
 journal = {Sankhyā: The Indian Journal of Statistics, Series A (1961-2002)},
 number = {1},
 pages = {79--93},
 publisher = {Springer},
 title = {{Order Statistics for Nonidentically Distributed Variables and Permanents}},
 urldate = {2022-12-18},
 volume = {51},
 year = {1989}
}

@article{ernst2004permutation,
  title={{Permutation Methods: A Basis for Exact Inference}},
  author={Ernst, Michael D},
  journal={Statistical Science},
  pages={676--685},
  year={2004},
  publisher={JSTOR}
}

@misc{isotonic_reg,
  doi = {DOI: 10.48550/ARXIV.1708.09468},
  
  url = {https://arxiv.org/abs/1708.09468},
  
  author = {Han, Qiyang and Wang, Tengyao and Chatterjee, Sabyasachi and Samworth, Richard J.},
  
  title = {{Isotonic Regression in General Dimensions}},
  
  publisher = {arXiv},
  
  year = {2017},
  
  copyright = {arXiv.org perpetual, non-exclusive license}
}

@inbook{vaart_1998,
place={Cambridge},
series={Cambridge Series in Statistical and Probabilistic Mathematics},
title={{Asymptotic Statistics}}, 
DOI={10.1017/CBO9780511802256}, 
publisher={Cambridge University Press}, 
author={{van der Vaart}, Aad W.}, 
year={1998}, 
Chapter = {21},
collection={Cambridge Series in Statistical and Probabilistic Mathematics}
}

@book{folland,
  title={Advanced Calculus},
  author={Folland, G.B.},
  isbn={9780130652652},
  lccn={2001055359},
  series={Featured Titles for Advanced Calculus Series},
  url={https://books.google.com/books?id=iatzQgAACAAJ},
  year={2002},
  publisher={Prentice Hall}
}

@book{Lee,
  title={Introduction to Smooth Manifolds},
  author={Lee, J.M.},
  isbn={9780387954486},
  lccn={2002070454},
  series={Graduate Texts in Mathematics},
  url={https://books.google.com/books?id=eqfgZtjQceYC},
  year={2003},
  publisher={Springer}
}

@book{wedhorn,
  title={Manifolds, Sheaves, and Cohomology},
  author={Wedhorn, T.},
  isbn={9783658106331},
  lccn={2016947452},
  series={Springer Studium Mathematik - Master},
  url={https://books.google.com/books?id=oEG8DAAAQBAJ},
  year={2016},
  publisher={Springer Fachmedien Wiesbaden}
}

@book{kelley,
  title={General Topology},
  author={Kelley, J.L.},
  series={Graduate texts in mathematics},
  url={https://books.google.com/books?id=nTIMzgEACAAJ},
  year={1955},
  publisher={D. Van Nostrand}
}
\clearpage
\appendix

\section{cPCH Adjustment Look-Up Table}\label{sec:lookup}
   \begin{table}[h]
       \centering
     \begin{tabular}{llcccc}
   $m$ & $r$ & $\alpha$ & Combination Method & $a(\alpha)$ & Maximum Type I Error at $a(\alpha)$\\
2 & 2 & 0.01 & Fisher & 0.0075 & 0.0096 \\
 2 & 2 & 0.01 & Simes & 0.0075 & 0.0096 \\
2 & 2 & 0.05 & Fisher &  0.0425 & 0.049 \\
2 & 2 & 0.05 & Simes &  0.0425 & 0.050 \\
2 & 2 & 0.1 & Fisher & 0.085 & 0.098 \\
2 & 2 & 0.1 & Simes &  0.089 & 0.10 \\
3 & 2 & 0.01 & Fisher & 0.0075 & 0.0096 \\
3 & 2 & 0.01 & Simes & 0.0075 & 0.0096 \\
3 & 2 & 0.05 & Fisher &  0.0425 & 0.049 \\
3 & 2 & 0.05 & Simes &   0.0425 & 0.050 \\
3 & 2 & 0.1 & Fisher &  0.089 & 0.10 \\
3 & 2 & 0.1 & Simes &  0.089 & 0.10 \\
3 & 3 & 0.01 & Fisher  & 0.0075 & 0.0096 \\
3 & 3 & 0.01 & Simes  & 0.0075 & 0.0096 \\
3 & 3 & 0.05 & Fisher &  0.0425 & 0.050 \\
3 & 3 & 0.05 & Simes &  0.0425 & 0.050 \\
3 & 3 & 0.1 & Fisher &  0.085 & 0.10 \\
3 & 3 & 0.1 & Simes &  0.085 & 0.10 \\
     \end{tabular}
       \caption{Adjustment values $a(\alpha)$ for the cPCH test computed using the SGD algorithm described in Supplementary Material~\ref{sec:sgd_comp}, for various common values of $\alpha$ for $m=2, 3$.}
       \label{tab:lookup}
   \end{table}


\clearpage
\section*{Further Details on Computing $a(\alpha)$}\label{sm:a_alpha}
One way to compute $a(\alpha)$ is via a grid search, i.e., for a given $\alpha$, estimate the maximum Type I error of the unadjusted cPCH test by computing the corresponding p-value across $n$ replicates over a grid of $\theta \in \Theta^{r/m}_0$, using binary search to efficiently find $a(\alpha)$. To further speed up computations, we can estimate the maximum Type I error via stochastic gradient descent (SGD) on the Type I error curve over $\bm{\theta} \in \Theta_0^{r/m}$. In fact, we generate the adjustment table used in our implementation of the cPCH test (as provided in our Github and of which Table~\ref{tab:lookup} is a subset) using this SGD approach. As exhibited by the corresponding maximum Type I error estimates at each $a(\alpha)$ in Table~\ref{tab:lookup} and the full adjustment table provided in our Github, we find that this approach tends to finds the $a(\alpha)$ such that the estimated maximum Type I error is close to $\alpha$. We provide further the computational details for this approach and justification of the use of SGD in this setting in the following subsection.

 Hence, by setting $n$ sufficiently large, we can estimate $a(\alpha)$ with arbitrarily high precision. Therefore, the cPCH test using the resulting  $a(\alpha)$ would offer essentially exact Type I error control (within nearly negligible sampling error). This approach is made computationally tractable by the generally small dimension of the null space for most common PCH testing scenarios, such as causal mediation analysis, where $m = 2$, and replicability analysis, where $m$ is often $\leq 5$ \cite[]{followup, HY2014, adafilt}, and the efficiency of computing unadjusted cPCH p-value (see Section~\ref{section:computing}). 


\begin{figure}[htp]
\centering

\begin{subfigure}{\textwidth}
  \centering
  \includegraphics[width=0.95\linewidth]{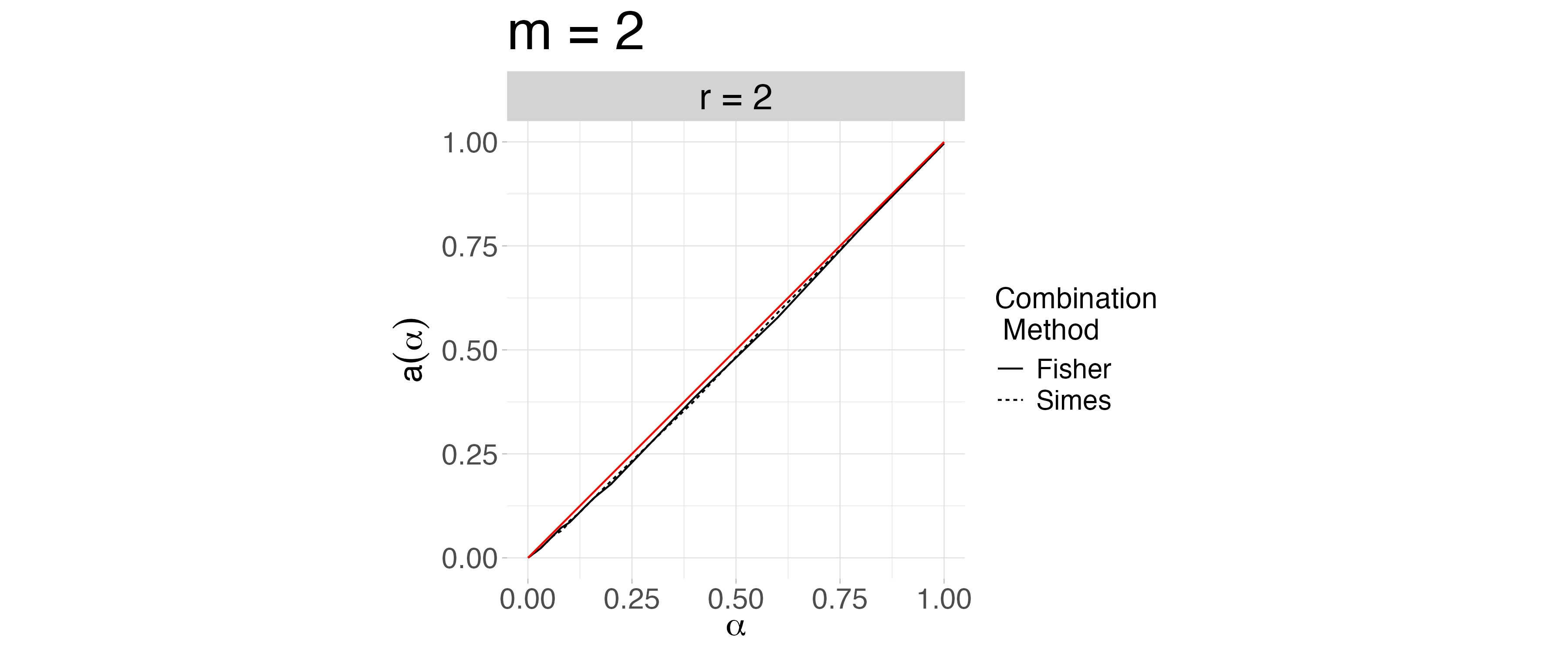}
  \caption{}
  \label{subfig:fig1}
\end{subfigure}

\begin{subfigure}{\textwidth}
  \centering
  \includegraphics[width=0.9\linewidth]{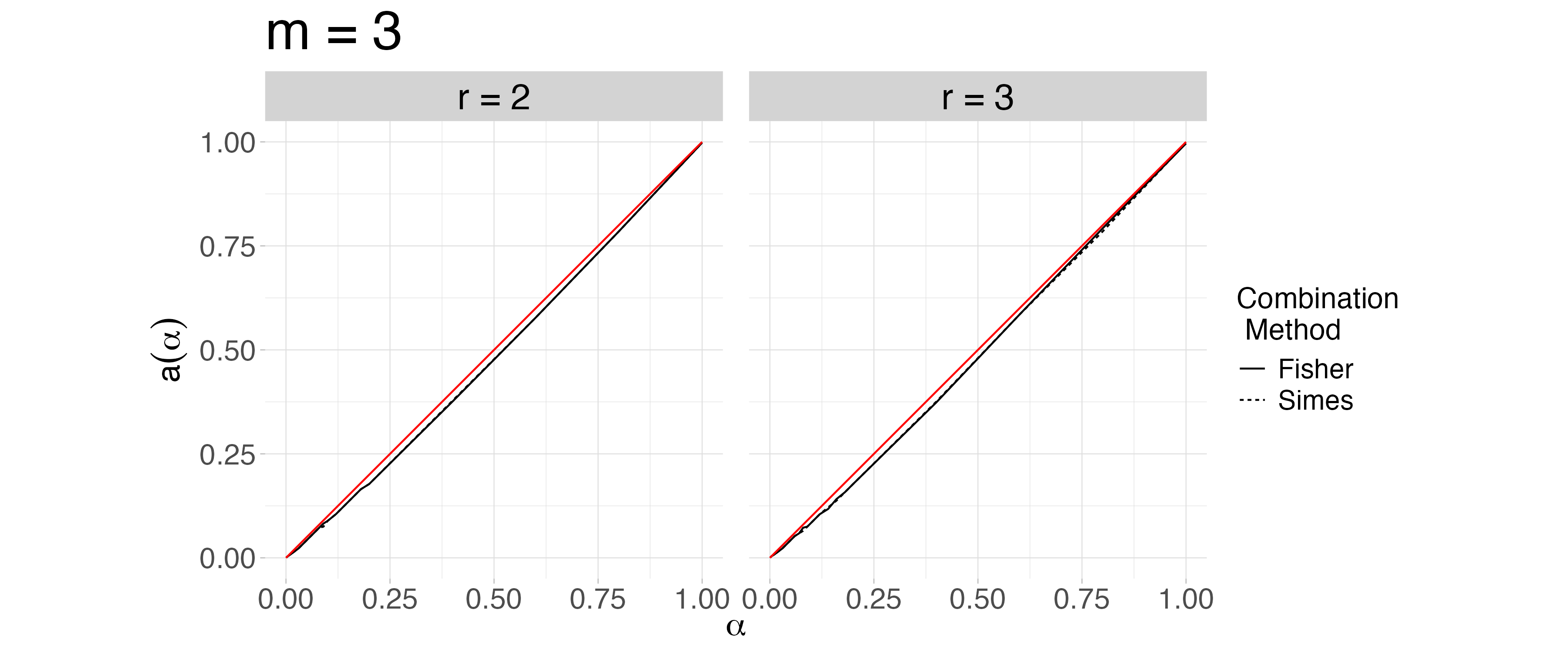}
  \caption{}
  \label{subfig:fig2}
\end{subfigure}

\begin{subfigure}{\textwidth}
  \centering
  \includegraphics[width=\linewidth]{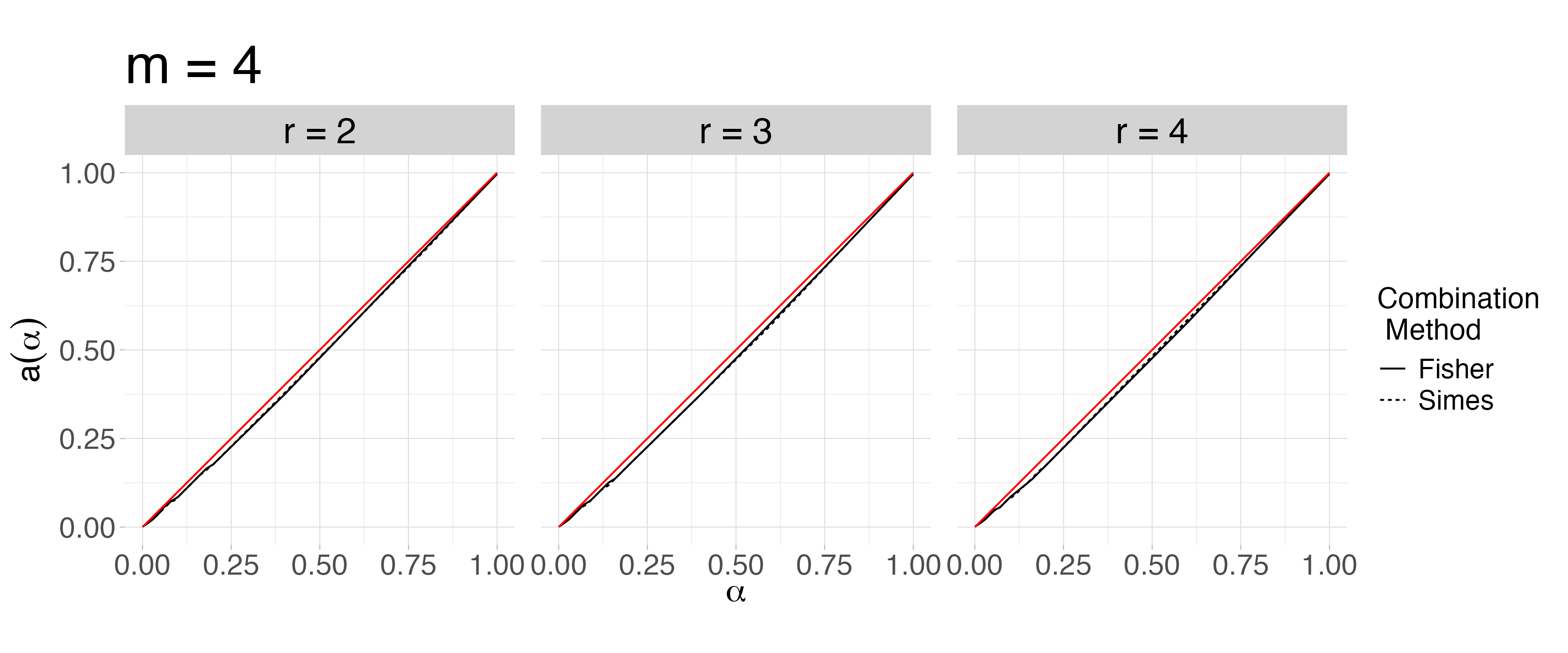}
  \caption{}
  \label{subfig:fig3}
\end{subfigure}
\caption{Level adjustment $a(\alpha)$ for cPCH with Fisher's and Sime's combining functions across various values of $m$ and $r$. $\alpha$ represents the nominal level and $a(\alpha)$ represents the level to run the unadjusted cPCH test to have Type I error control at level $\alpha$. The red line depicts the $\alpha = a(\alpha)$ line. Hence, the adjustment is rather small across all values of $\alpha$.}
\label{fig:alpha_prim_plot}

\end{figure}

Additionally, as shown in Figure~\ref{fig:alpha_prim_plot}, $a(\alpha)$ varies smoothly in $\alpha$, and hence, for $\alpha$ values not present in the adjustment table provided in our Github, we can well approximate the corresponding $a(\alpha)$ value by linearly interpolating between the nearest $\alpha$ values considered in the adjustment table provided in our Github. This linear interpolation is incorporated in our implementation of the cPCH test so that the test can be computed for any $\alpha \in [0, 1]$. 
Hence, there is no computational cost on users to compute $a(\alpha)$ themselves for any $\alpha \in [0, 1]$, as long as they are considering an $m \leq 5$, which is common in PCH testing. Though our adjustment table is computed under the setting of Section~\ref{section:preliminaries}, where each of the $T_i$'s are single, independent unit-variance Gaussians, our simulations in Section~\ref{section:robustness} suggest that the following results would still hold under other distributional assumptions for the base test statistics. 

Though the adjustment table provided in our Github (of which Table~\ref{tab:lookup} is a subset) goes up to $m=5$, we provide all scripts necessary to find $a(\alpha)$ for larger $m$. 

\subsection{The SGD Algorithm for Quantifying Maximum Type I Error}
\label{sec:sgd_comp}
In this section, we present the justification and computation details of the stochastic gradient descent (SGD) analysis for use in computing the level adjustments for the cPCH test. 
To apply SGD to the Type I error of the cPCH test, we write the Type I error as a function of $\bm\theta$:
\begin{equation}\label{eq:type1error} 
	E(\btheta) = \PP_{\btheta}\left( \mytest \right), ~ \bm\theta \in \nspace.
\end{equation} 
 In the above notation, we suppress the Type I error's dependence on $\alpha$ and the combining function $f$ for ease of presentation. It is of interest to quantify $\max_{\btheta \in \nspace} E(\btheta)$ relative to $\alpha$. Though $E(\btheta)$ does not admit an analytical form, we can estimate the gradient of $E(\btheta)$ empirically via the Monte Carlo sampling described above, then use SGD on $-E(\btheta)$ to estimate the maximum Type I error. 
 By the symmetry of the testing problem, we can conduct the SGD algorithm to search over the subset of $\Theta_0^{r/m}$ where the non-zero means are all positive.

As suggested by Figure~\ref{fig:t1error}, $-E(\btheta)$ is not a convex function of $\btheta$. To justify our use of SGD in this non-convex setting, we first note that when $r=m=2$, $-E(\btheta)$ appears to be smooth and quasi--convex.
For continuously differentiable, quasi-convex functions, gradient descent will converge to a stationary point \cite[]{gd_conv}.  
Since the Type I error curve appears to have a single stationary point occurring at the global minimum (where $\theta_{(2)} \approx 2$), if the SGD algorithm converges to a finite solution, that solution is likely close to the global minimum \cite[]{sgd_nonconvex}. For $r=m=2$, we find that across multiple, independent initializations, the SGD algorithm consistently converges to solutions where $\theta_{(2)} \approx 2$, which is consistent with the actual location of the maximum Type I error of the cPCH test for $r=m=2$.

As discussed in Supplementary Materials~\ref{appendix:approx_valid}, we see evidence that the behavior of the cPCH test for $r=m=2$ generalizes to different $m$ and $r$. Therefore, we expect that, for any $m$ and $r$, the Type I error will also be quasi-convex and the maximum Type I error will occur when $\theta_{(j)} \approx 2$ for all $j = m-r+2, ..., m$. In fact, for $m = 3, 4$, we found that the SGD algorithm consistently converges to solutions where $\theta_{(j)} \approx 2$ for all $j = m-r+2, ..., m$, thus supporting our claim.

We now provide the computational details of the SGD algorithm used. First, we simplify the notation of the cPCH test in Definition~\ref{cpchdef} and write it as a function of the data $\bT$ i.e., $\varphi^{\text{cPCH}}_\alpha\left(\bT \right):=  \indic{\mytest}$. Expressing $E(\btheta)$ in terms of the integral
$$
E(\btheta)  
= \int_{\RR^m} \varphi^{\text{cPCH}}_\alpha\left(\bm{t} \right) \prod_{i=1}^m \phi(t_i - \theta_i )d t_i
$$
and noticing the fact that $ \frac{d}{d \theta_i} \phi(t_i - \theta_i) = -( t_i-\theta_i) \phi(t_i - \theta_i) $, we have the gradient of $-E(\btheta)$ equals
\begin{equation}\nonumber
	-\nabla E(\btheta) = \Ep{\btheta}{(\bT - \btheta ) \varphi^{\text{cPCH}}_\alpha\left(\bm{T} \right)} =  \EE{\bm{Z}  \varphi^{\text{cPCH}}_\alpha(\bZ + \btheta)},
\end{equation}
where $\bZ = (\bT - \btheta ) \stackrel{d}{\sim} \cN(\bm{0}, \Ib_m)$. This gradient can be unbiasedly estimated by
\begin{equation}\nonumber
	\frac{1}{n} \sum_{i=1}^{n} \bZ_i \ \varphi^{\text{cPCH}}_\alpha(\bZ_i + \btheta) 
\end{equation}
where the $Z_i, i=1, ..., \repl$ are $\repl$ i.i.d. samples from $\cN(\bm{0}, \Ib_m)$ and we can choose the number of repeated samples $\repl$ to be sufficiently large. 

Due to the symmetry of the problem, we note
$\min_{\btheta \in \nspace} -E(\btheta) = \min_{\btheta \in \ccone} -E(\btheta)$
  where $\ccone = \{\btheta \in \nspace: \theta_1\ge \cdots \theta_{r-1} \ge 0 = \theta_{r} = \cdots = \theta_m \}$ is a convex cone and its volume is only $\frac{1}{\binom{m}{r-1}(r-1)! 2^{r-1}}$ of $\nspace$. Therefore, restricting $\btheta$ in the much smaller searching space $\ccone$  will speed up the convergence to a local minimum. With such a constrained parameter space of $\btheta$, we project via isotonic regression \cite[]{isotonic_reg} to ensure the move at each step is still within $\ccone$. We also make use of random initializations to help adequately search the parameter space for estimating the global minimum $\inf_{\btheta \in \nspace} -E(\btheta)$. The whole procedure is spelled out in Algorithm \ref{algo:sgd}. For the adjustment table provided in our Github, we use an exponentially decaying learning rate and terminate the algorithm after $200$ batches.


 \begin{algorithm}[!htb]
 \caption{SGD estimation of the cPCH test's maximum Type I error}
\label{algo:sgd}
\begin{algorithmic}[1]
\REQUIRE $m, r, f, N$, a level $\alpha$, the number of repeated samples $\repl$, the maximum number of batches $T$, the learning rate $\gamma$.
\vspace{0.05cm}
\STATE Initialize $\btheta$: Draw $\theta_1$ from $\cN(0, \sigma^2)$ for some $\sigma$, then for $k = 2, \cdots, r-1$, draw $\theta_k$ from $\text{Trunc--Norm}(0, \sigma^2, |\theta_{k-1}|)$. For $\theta_i$, $i=1, .., m$, set $\theta_i$ to its absolute value when $i \leq r-1$ and $0$ otherwise. Set $\bm{\theta} = (\theta_1, ..., \theta_m)$.
\REPEAT
\STATE Sample i.i.d. samples $\{\bZ_i\}_{i=1}^n$ from $\cN(\bm{0}, \Ib_m)$.
\STATE Run conditional PCH testing on data $\{\bZ_i + \btheta\}_{i=1}^n$ and obtain $\{\varphi^{\text{cPCH}}_\alpha(\bZ_i + \btheta)\}_{i=1}^n$.
\STATE Compute the gradient estimate: $\bm{g} \leftarrow n^{-1}\sum_{i=1}^{n} \bZ_i \varphi^{\text{cPCH}}_\alpha(\bZ_i + \btheta)$.
\STATE Update $\btheta \leftarrow \btheta  - \gamma \bm{g}$
\STATE Reset $\btheta$ to be the projection of $\btheta$ on $\ccone$. 
\UNTIL{t = T or reaching the stopping criterion.}
\ENSURE Estimate of the maximum Type I error: $ n^{-1}\sum_{i=1}^{n}\varphi^{\text{cPCH}}_\alpha(\bZ_i + \btheta)$.
\end{algorithmic}
\end{algorithm}

\clearpage

\section{Further Details on Computing Unadjusted cPCH p-values}\label{section:computing_further_details}

Recall from Section \ref{section:computing} that the unadjusted cPCH p-value can be represented as
$$\mathbb{P}_{\hat{\btheta}}\left(f(\bm{T}_{(1:m-r+1)}) \geq f_{\text{obs}} \,\mid\, \bT_{(m-r+2:m)}\right), $$
where $\hat{\btheta} = \left(0, ..., 0, T_{(m-r+2)}, ..., T_{(m)}\right)$. 
Our sampling approach for computing unadjusted cPCH p-values involves conditioning on extra events about which of the $T_1, ..., T_m$ correspond to the order statistics $\bm{T}_{(1:m-r+1)}$ and $\bT_{(m-r+2:m)}$. To formally describe how we condition on these extra events, we denote by $S$ the set of all possible ways to observe some unordered set of the $T_i$'s corresponding to the entries of $\bT_{(1:m-r+1)}$ and some ordered set of the remaining $T_j$'s corresponding to the entries of $\bT_{(m-r+2:m)}$. For example, when $m=3, r=2$, we have 
$$
S = \left\{(\{T_1, T_2\}, T_3), (\{T_2, T_3\}, T_1), (\{T_1, T_3\}, T_2)\right\}
$$ 
where the inner set corresponds to $\{T_{(1)}, T_{(2)}\}$ and the remaining term corresponds to $T_{(3)}$. Here we pause to highlight a fact: the cardinality of $S$ is $\frac{m!}{(m-r+1)!}$ which also equals the number of mixture components. Although in general the cardinality of $S$ grows exponentially in $m$, when $r$ is small, such as when $r=2$ for replicability analysis, the complexity of $\frac{m!}{(m-r+1)!}$ is a linear or a low-order polynomial in $m$, e.g., $\frac{m!}{(m-2+1)!} = O(m)$. Additionally, $m$ is often $\leq5$ in many common PCH testing scenarios \cite[]{ followup, HY2014, adafilt}, so the number of mixture components to compute tends to be reasonably small. 
For each term $S_\ell$ from the set $S$, for $\ell \in \left\{1,\dots, \frac{m!}{(m-r+1)!}\right\}$, we define the event
$$
B_\ell \coloneqq \left\{S_{\ell} = (\{T_{(1)}, ...,T_{(m-r+1)}\}, \bT_{(m-r+2:m)}) \right\}.
$$ 
For example, in the $m=3, r=2$ case above, $B_1 = \left\{(\{T_1, T_2\}, T_3) = (\{T_{(1)}, T_{(2)}\}, T_{(3)}) \right\}$. Now, we express the conditional distribution of $f(\bT_{(1:m-r+1)}) \mid \bT_{(m-r+2:m)}$ in the following mixture form:
\begin{align*}
    &\mathbb{P}_{\hat{\btheta}}\left( f(\bm{T}_{(1:m-r+1)}) \geq f_{\text{obs}} \given \bT_{(m-r+2:m)} \right) \\
   &= \sum_{\ell = 1}^{\frac{m!}{(m-r+1)!}}\mathbb{P}_{\hat{\btheta}}\left( B_\ell \given \bT_{(m-r+2:m)}\right) \mathbb{P}_{\hat{\btheta}}\left(f(\bm{T}_{(1:m-r+1)}) \geq f_{\text{obs}}  \given B_\ell, \bT_{(m-r+2:m)} \right).
\end{align*}
From here, the computational details of our strategy can be summarized in two steps: 
\begin{enumerate}
	\item We derive the analytic form of the mixture weights $\mathbb{P}_{\hat{\btheta}}\left( B_\ell \;\middle|\; \bT_{(m-r+2:m)}\right)$, which can be expressed using only evaluations of the standard normal cumulative distribution $\Phi$ and density $\phi$;
    \item We develop a method for sampling from the distribution of $f\left(\bm{T}_{(1:m-r+1)}\right) \mid B_\ell, \bT_{(m-r+2:m)}$ which relies solely on sampling from truncated-Normal distributions.
\end{enumerate}

We now provide the computational details for the above parts. 
Without loss of generality, we illustrate using $S_1 = (\{T_1, ..., T_{m-r+1}\}, \bT_{m-r+2:m})$ and 
$$
B_1 = \left\{(S_1 = (\{T_{(1)}, ..., T_{(m-r+1)}\}, \bT_{(m-r+2:m)})\right\}.
$$ 

\subsection{Analytical expressions of the mixture weights.\label{section:comp_first_term}}

As mentioned in Section~\ref{section:computing}, we can express the mixture weights $\mathbb{P}_{\hat{\btheta}}\left( B_\ell \;\middle|\;\bT_{(m-r+2:m)}\right)$ using only evaluations of the standard normal cumulative distribution $\Phi$ and density $\phi$. 
For example, given the observed $\bT_{(m-r+2:m)} =  \bm{t}_{m-r+2:m}$, we have \footnote{We employ a slight abuse of notation by writing $\mathbb{P}_{\hat{\btheta}}(\bT_{(m-r+2:m)} = t_{m-r+2:m})$ to denote the PDF of $\bT_{(m-r+2:m)}$ evaluated at $t_{m-r+2:m}$ and $\mathbb{P}_{\hat{\btheta}}(T_{j} = t_j)$ to denote the PDF of $T_j$ evaluated at $t_j$, both under the model $\bT \sim \mathcal{N}(\hat{\btheta}, I_m)$.}
\begin{align}
    &\mathbb{P}_{\hat{\btheta}}\left(B_1 \given \bT_{(m-r+2:m)} = t_{m-r+2:m}\right) \nonumber\\ 
    & =   \frac{\mathbb{P}_{\hat{\btheta}}\left( B_1, \bT_{(m-r+2:m)} =t_{m-r+2:m} \right)}{\mathbb{P}_{\hat{\btheta}}\left( \bT_{(m-r+2:m)} = t_{m-r+2:m}\right)} \nonumber\\ 
        &=   \frac{\mathbb{P}_{\hat{\btheta}}\left( |T_1|< |t_{m-r+2}|, ..., |T_{m-r+1}| < |t_{m-r+2}|, T_{m-r+2} = t_{m-r+2}, ..., T_{m} = t_{m} \right)}{\mathbb{P}_{\hat{\btheta}}\left( \bT_{(m-r+2:m)} =t_{m-r+2:m} \right)} \nonumber\\ 
        &=   \frac{\prod_{h=1}^{m-r+1}\mathbb{P}_{\hat{\btheta}}\left( |T_h|< |t_{m-r+2}|\right) \prod_{j = m-r+2}^{m}\mathbb{P}_{\hat{\btheta}}\left(T_{j} = t_{j} \right)}{\mathbb{P}_{\hat{\btheta}}\left( \bT_{(m-r+2:m)} = t_{m-r+2:m} \right)} \label{eq:mixweights_end}
\end{align}
where
\begin{align*}
\mathbb{P}_{\hat{\btheta}}\left( |T_h| < |t_{m-r+2}|\right) & = \Phi(|t_{m-r+2}|) - \Phi(-|t_{m-r+2}|), h = 1, ..., m-r+1\\
\mathbb{P}_{\hat{\btheta}}\left(T_{j} = t_j\right)  & = \phi(t_j-\hat\theta_{j}) , j = m-r+2, ..., m\\
\mathbb{P}_{\hat{\btheta}}\left( \bT_{(m-r+2:m)} = t_{m-r+2:m} \right) &= \sum_{\ell=1}^{\binom{m!}{(m-r+1)!}} \mathbb{P}_{\hat{\btheta}}\left( B_\ell,  \bT_{(m-r+2:m)} = t_{m-r+2:m} \right).
\end{align*}
The summands in the last line have the same form as the numerator in Equation~\eqref{eq:mixweights_end}, and hence can be computed in the same way, although the means of $T_{1:m-r+1}$ above are all $0$ since we calculate the above probability for $B_1$. The means will be different for different $B_\ell$ since, conditional on $B_\ell$, we assume a particular subset of $T_{1:m}$ correspond to $T_{(1:m-r+1)}$.

\subsection{Distributions of the mixture components.}
By the conditioning on $B_1$,

$\mathbb{P}_{\hat{\btheta}}\left(f(\bm{T}_{(1:m-r+1)}) \geq f_{\text{obs}} \;\middle|\; B_1, \bT_{(m-r+2:m)} \right) = \mathbb{P}_{\hat{\btheta}}\left(f(\bm{T}_{1:m-r+1}) \geq f_{\text{obs}} \;\middle|\; B_1, \bT_{(m-r+2:m)} \right)$ for any $f$ that is permutation invariant (e.g., Fisher's, Simes'). Note, we can exactly specify the distribution of $ T_{1}, .., T_{m-r+1}$ conditional on $B_1$ and  $\bT_{(m-r+2:m)}$:
 $$
 T_{1},..., T_{m-r+1} \mid B_1, \bT_{(m-r+2:m)}  \iid \text{Trunc--Norm}(0, 1, T_{(m-r+2)}),
 $$
where $\text{Trunc--Norm}(\mu, 1, t)$ is the truncated normal distribution with location $\mu$ and scale $1$ truncated at $|t|$ and $-|t|$. 
 Therefore, we can utilize standard sampling procedures for truncated normal distributions to generate $N$ independent copies $\left\{\tilde{\bT}_{1}^{(k)}\right\}_{k=1}^N$ from the distribution of $T_{1}, ..., T_{m-r+1} \mid B_1, \bT_{(m-r+2:m)}$, where the ``$1$'' subscript on $\tilde{\bT}_{1}^{(k)}$ denotes the conditioning on $B_1$. Let $X^{(k)}_{1} = f\left(\tilde{\bT}_{1}^{(k)}\right)$. Then, we estimate $\mathbb{P}_{\hat{\btheta}}\left(f(\bm{T}_{(1:m-r+1)}) \geq f_{\text{obs}} \;\middle|\; B_1, \bT_{(m-r+2:m)} \right)$ using
\begin{equation*}
   g\left(f_{\text{obs}}, \left\{X_{1}^{(k)}\right\}_{k=1}^N\right) \coloneqq \frac{1}{N+1}\left(1 + \sum_{k=1}^{N}\mathbbm{1}\left\{X_{1}^{(k)} \geq f_{\text{obs}} \right\}\right),
   \end{equation*}
  where the ``$+1$" in the numerator and denominator above is standard for Monte Carlo p-values; see, e.g., \cite{ernst2004permutation}
for details.

In the above notation, we suppress $g\left(f_{\text{obs}}, \left\{\copies\right\}_{k=1}^N\right)$'s dependence on $\hat{\btheta}$ for ease of presentation. Taking $N$ large, we can estimate $\mathbb{P}_{\hat{\btheta}}\left(f(\bm{T}_{(1:m-r+1)}) \geq f_{\text{obs}} \given B_\ell, \bT_{(m-r+2:m)} \right)$ with high accuracy. We then apply this logic to each $B_\ell$ to get estimates for each mixture component. 

Therefore, we compute the unadjusted cPCH p-value as:
$$
p_{\text{ucPCH}}^{r/m}(\bT)\coloneqq \sum_{\ell = 1}^{\frac{m!}{(m-r+1)!}}\mathbb{P}_{\hat{\btheta}}\left( B_\ell \;\middle|\; \bT_{(m-r+2:m)}\right)  g\left(f_{\text{obs}}, \left\{\copies\right\}_{k=1}^N\right).
$$
Note that the smallest $p_{\text{ucPCH}}^{r/m}(\bT)$ can be is $\frac{1}{N+1}$. In multiple testing settings where we have $M$ unadjusted cPCH p-values, many FDR controlling procedures compare p-values to thresholds on the order of $\alpha/M$.
Therefore, in multiple testing settings, $N$ must be set large enough to ensure that Unadjusted cPCH p-values can attain values below these thresholds to make any discoveries. Thus, computing $g\left(f_{\text{obs}}, \left\{\copies\right\}_{k=1}^N\right)$ is the main computational component of calculating unadjusted cPCH p-values; however, since it is possible to sample efficiently from a truncated Normal distribution, it is computationally feasible to set $N$ very large. See Figure~\ref{fig:comptimes} of Appendix~\ref{section:singlepch_additionalresults} for further details on computation time.

\subsection{Calculating Unadjusted cPCH p-values in the $r=m=2$ setting}\label{section:pvalscomp2}
In the $r=m=2$ setting, unadjusted cPCH p-values can be calculated without sampling using the following derivation. Since $r=m=2$, $S = \left\{ \left(\{T_1\}, T_2\right), \left(\{T_2\}, T_1\right)\right\}$. As in Section~\ref{section:computing}, let $S_{\ell}$ be the $\ell^{th}$ set in $S$ and
$$B_\ell = \{S_{\ell} = (\{T_{(1)}\}, T_{(2)}) \}.$$ Recall that the test statistic in this setting is $|T_{(1)}|$. Then, given the observed $|T_{(1)}| = f^{\text{obs}}$,
\begin{align}
   \mathbb{P}_{\hat{\btheta}}\left(|T_{(1)}| > f^{\text{obs}} \;\middle|\; T_{(2)} \right)  & = 
   \mathbb{P}_{\hat{\btheta}}\left(B_1 \;\middle|\;T_{(2)} \right)\mathbb{P}_{\hat{\btheta}}\left(|T_{(1)}| \geq  f^{\text{obs}}\;\middle|\; T_{(2)}, B_1\right) \nonumber \\ 
    & \quad\quad +  \mathbb{P}_{\hat{\btheta}}\left(B_2 \;\middle|\;T_{(2)} \right)\mathbb{P}_{\hat{\btheta}}\left(|T_{(1)}| \geq f^{\text{obs}} \;\middle|\; T_{(2)}, B_2 \right) \label{eq:cpch_pval_2_2}
\end{align}
where $\hat{\btheta}= (0, T_{(2)})$.
The mixture weights $\mathbb{P}_{\hat{\btheta}}\left(B_i \mid T_{(2)} \right)$ can be calculated as in Equation~\eqref{eq:mixweights_end}. For the $r=m=2$ setting, we can derive the analytic form of the mixture components $\mathbb{P}_{\hat{\btheta}}\left(|T_{(1)}| \geq  f^{\text{obs}} \mid T_{(2)}, B_1\right)$ as well. 
For example, conditional on $B_1$,
\begin{align}
\mathbb{P}_{\hat{\btheta}}\left(|T_{(1)}| \geq f^{\text{obs}} \given T_{(2)}, B_1 \right) = \frac{\mathbb{P}_{\hat{\btheta}}\left(f^{\text{obs}} \leq |T_1| \leq |T_{(2)}| \given T_{(2)}\right)}{\mathbb{P}_{\hat{\btheta}}\left(|T_1| \leq |T_{(2)}| \given T_{(2)}\right)}
=  \frac{2\left(\Phi\left(|T_{(2)}|\right) - \Phi\left(|f^{\text{obs}}|\right)\right)}{\Phi(|T_{(2)}|) - \Phi(-|T_{(2)}|)} \label{eq:mix1}
\end{align}
 Analogously, conditional on $B_2$, we get
\begin{align}
&\mathbb{P}_{\hat{\btheta}}\left(|T_{(1)}| \geq f^{\text{obs}} \given T_{(2)}, B_2 \right) \nonumber \\
&= \frac{\mathbb{P}_{\hat{\btheta}}\left(f^{\text{obs}} \leq |T_2| \leq |T_{(2)}| \given T_{(2)}\right)}{\mathbb{P}_{\hat{\btheta}}\left(|T_2| \leq |T_{(2)}| \given T_{(2)}\right)}\nonumber \\
&=  \frac{\left(\Phi\left(|T_{(2)}|-\hat{\theta}_{2}\right) - \Phi\left(f^{\text{obs}}-\hat{\theta}_{2}\right)\right) + \left(\Phi\left(-f^{\text{obs}}-\hat{\theta}_{2}\right) - \Phi\left(-|T_{(2)}|-\hat{\theta}_{2}\right)\right)}{\Phi(|T_{(2)}|-\hat{\theta}_{2}) - \Phi(-|T_{(2)}|-\hat{\theta}_{2})} \label{eq:mix2}
\end{align}

Plugging in Equation~\eqref{eq:mix1} and \eqref{eq:mix2} to Equation~\eqref{eq:cpch_pval_2_2} gives the final form for $  \mathbb{P}_{\hat{\btheta}}\left(|T_{(1)}| \geq f^{\text{obs}} \;\middle|\; T_{(2)} \right)$. Note, the same strategy could be applied for $r=m$, but we do not pursue this idea here.

\clearpage
\section{Robustness to Model Misspecification}\label{section:robustness}
To assess the cPCH test's robustness to model misspecification, we assess the performance of the cPCH test (which assumes $\bm{T}$ is normally distributed) when $\bm{T}$ is generated from a the t-distribution with a fixed scale and degrees of freedom.

Specifically, let $\text{t}(\theta, 1, \nu)$ be the generalized t-distribution with centrality parameter $\theta$, scale $1$, and degrees-of-freedom (DOF) $\nu$. Let $\Psi$ be the CDF of the $\text{t}(0, 1, \nu)$ distribution. We generate data by sampling $\bm{T}$ independently from the model:
$$  T_h \stackrel{i.i.d.}{\sim}\text{t}(\theta, 1, \nu),\; h = 1, ..., r^\star, \text{\hspace{0.5cm}}
    T_l  \stackrel{i.i.d.}{\sim}\text{t}(0, 1, \nu),\; l = r^\star+1, ..., m.$$
The (correctly specified) cPCH p-value ($N=10,000)$ for testing $H_0^{r/m}$ is computed from $\bm{T}$ where we use a t-distribution in the computation procedure described in Section~\ref{section:computing}. We then convert each base test statistic into its corresponding base p-value $p_i = 2(1-\Psi(|T_i|), i = 1, ..., m$ and produce the test statistics $W_i  = \Phi^{-1}(1-\frac{p_i}{2})$, from which we compute the (misspecified) cPCH p-value ($N = 10,000$) using a normal distribution in the computation procedure. This simulation emulates the situation in which the analyst only views the base p-values $p_1, ..., p_m$ and incorrectly assumes that the underlying base test statistics follow a normal distribution.

    \begin{figure}[!htb]
    \centering
    \includegraphics[width = \textwidth]{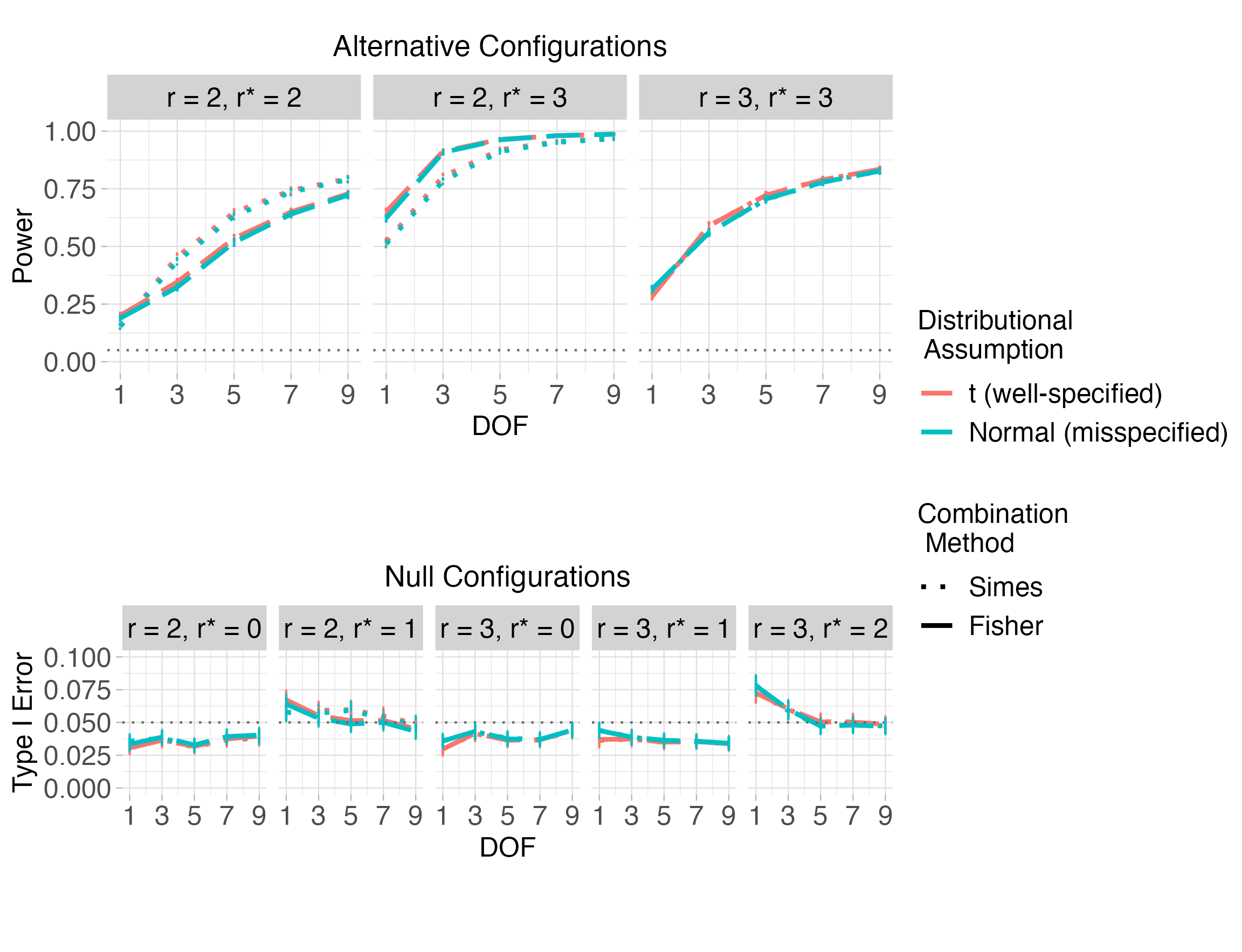}
   \caption{Power and Type I error of the properly specified (t-distributional assumption on $\bm{T}$) and the misspecified (normal distributional assumption on $\bm{T}$) cPCH tests. The nominal level is $\alpha = 0.05$ (dotted grey line). Each point represents the proportion of cPCH p-values below $\alpha$ over $5000$ independent replicates of the data generating procedure described in Section~\ref{section:robustness} for a given $(r^\star, r, \nu)$ with $m=3$ and $\theta=4$. Error bars depict $\pm 2$ standard errors.} 
    \label{fig:robust}
\end{figure}

Figure~\ref{fig:robust} shows that the cPCH test remains powerful and maintains Type I error control for most null configurations under model misspecification, even in some settings where the base test statistics have Cauchy distribution ($\nu=1$). Notably, both the misspecified and properly specified cPCH tests maintain Type I error control for all null configurations when $\nu \geq 5$. In most applied settings, the degrees of freedom of a t-distributed base test statistic reflects the underlying sample size used to compute the base test statistic. Thus, excluding settings where the sample size is extremely small such that $\nu < 5$, we expect the cPCH test to be robust to model misspecification. 

\clearpage
\section{Additional Single PCH Testing Results}\label{section:singlepch_additionalresults} 

        \begin{figure}[!htb]
        \centering
        \includegraphics[width = \textwidth]{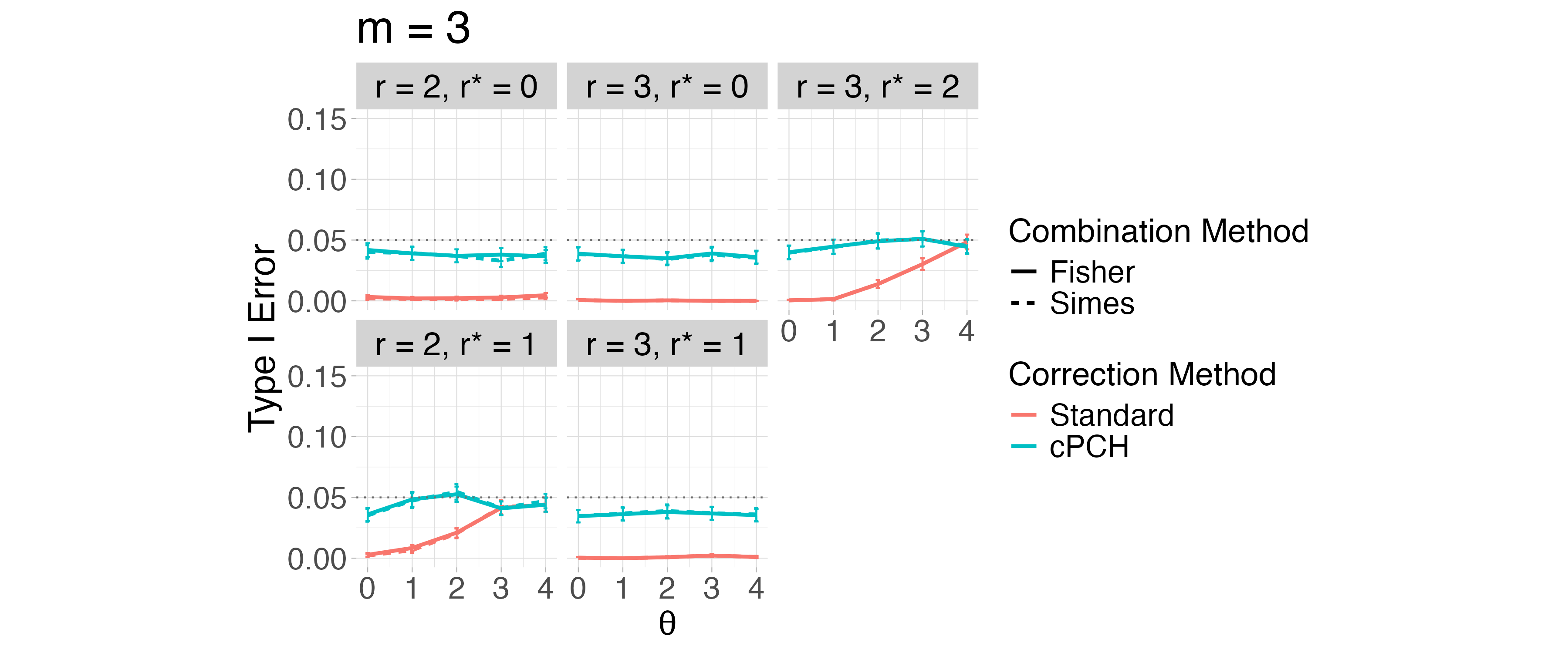}
 \caption{Type I error of the cPCH test across all null cases $(r^\star < r)$ at nominal level $\alpha = 0.05$ (dotted black line) for $m=3$. Recall that the Type I error of the cPCH Oracle test is exactly equal to the nominal level. Each point represents the proportion of cPCH p-values below $\alpha$ over $5000$ replicates of the data generating procedure outlined in Section~\ref{section:singlepower} for each $(r^\star, r, \theta)$ triplet. Error bars depict $\pm 2$ standard errors.
} 
        \label{fig:adjusted_single_pc_null_plot_m_3}
    \end{figure}
    
        \begin{figure}[!htb]
        \centering
        \includegraphics[width = \textwidth]{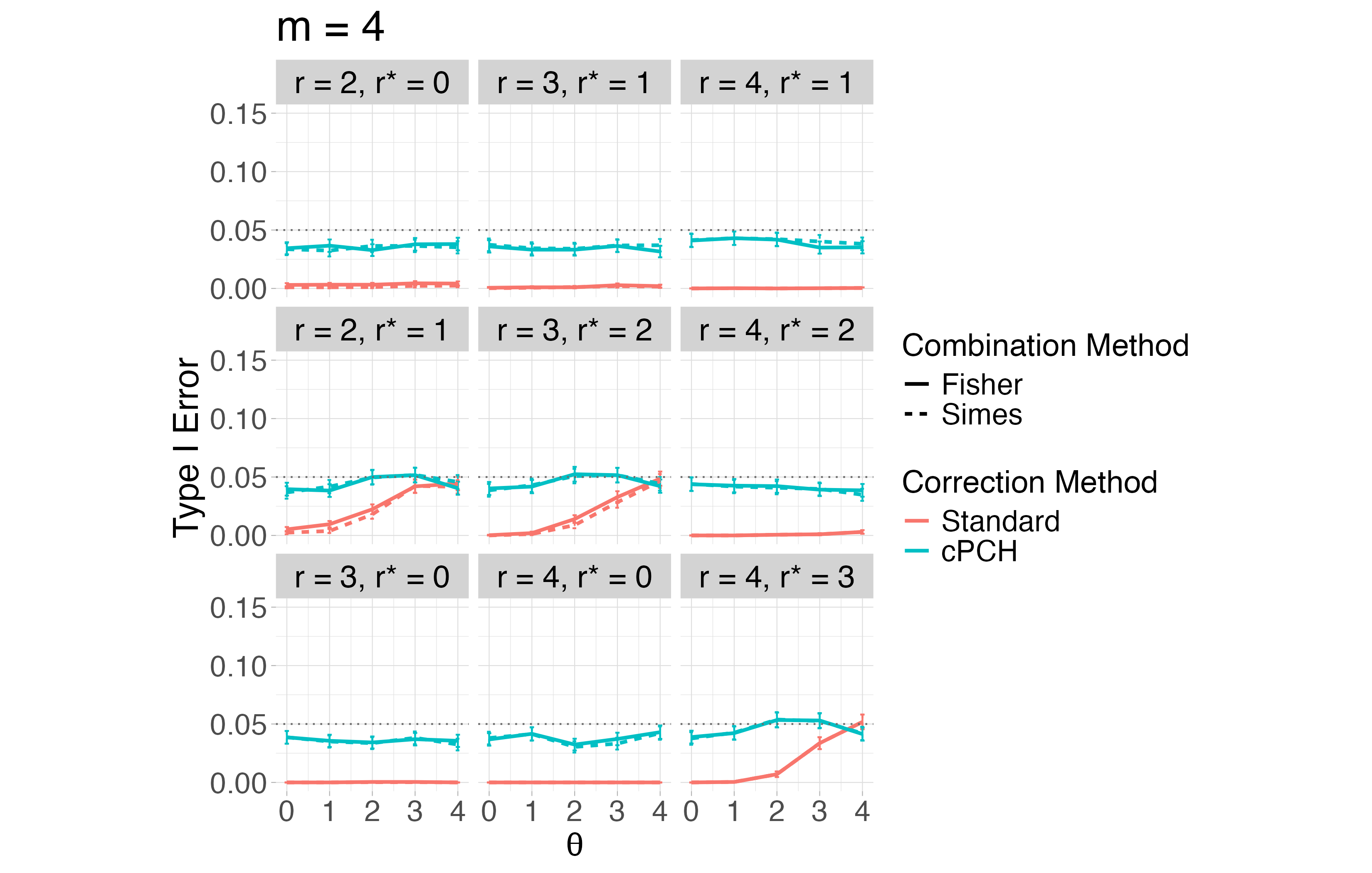}
 \caption{Type I error of the cPCH test across all null cases $(r^\star < r)$ at nominal level $\alpha = 0.05$ (dotted black line) for $m=4$. Recall that the Type I error of the cPCH Oracle test is exactly equal to the nominal level. Each point represents the proportion of cPCH p-values below $\alpha$ over $5000$ replicates of the data generating procedure outlined in Section~\ref{section:singlepower} for each $(r^\star, r, \theta)$ triplet. Error bars depict 2 standard errors.
} 
    \end{figure}

        \begin{figure}[!htb]
        \centering
        \includegraphics[width = \textwidth]{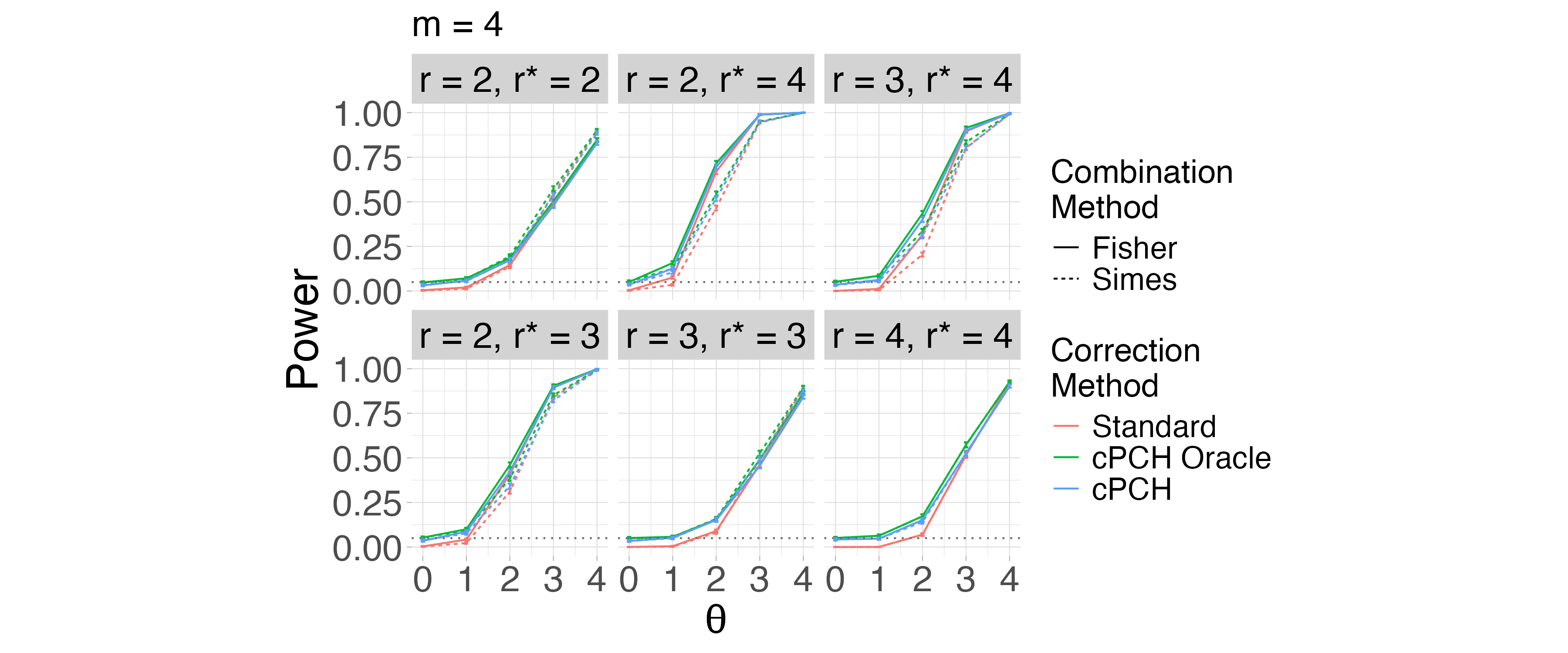}
 \caption{Power of the cPCH test across all alternative cases $(r^\star \geq r)$ at nominal level $\alpha = 0.05$ (dotted black line) for $m=4$. Each point represents the proportion of cPCH p-values below $\alpha$ over $5000$ replicates of the data generating procedure outlined in Section \ref{section:singlepower} for each $(r^\star, r, \theta)$ triplet. Error bars depict 2 standard errors.
} 
    \end{figure}

        \begin{figure}[!htb]
        \centering
        \includegraphics[width = \textwidth]{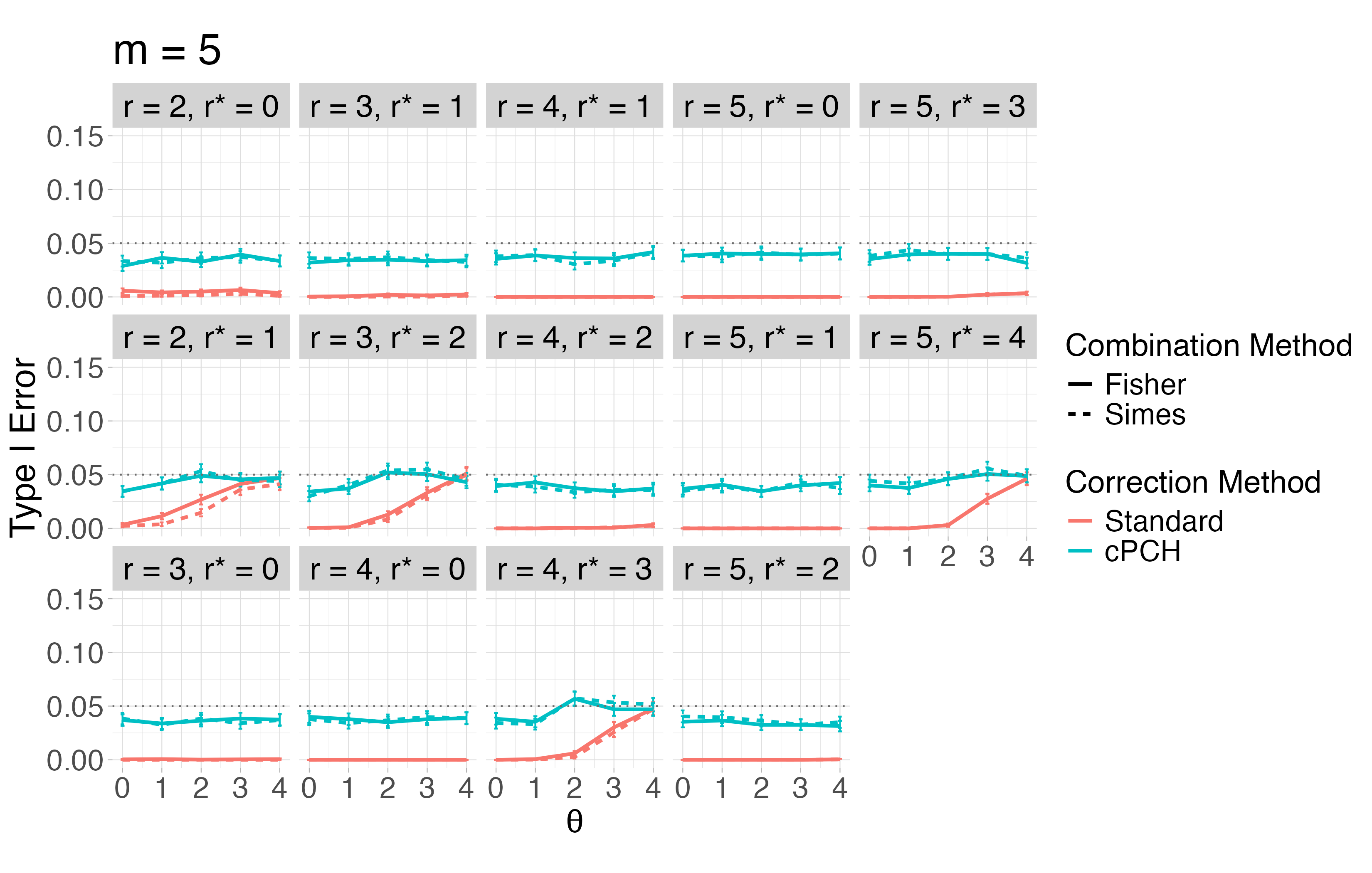}
 \caption{Type I error of the cPCH test across all null cases $(r^\star < r)$ at nominal level $\alpha = 0.05$ (dotted black line) for $m=5$. Recall that the Type I error of the cPCH Oracle test is exactly equal to the nominal level. Each point represents the proportion of cPCH p-values below $\alpha$ over $5000$ replicates of the data generating procedure outlined in Section~\ref{section:singlepower} for each $(r^\star, r, \theta)$ triplet. Error bars depict $\pm 2$ standard errors.
} 
    \end{figure}
    
        \begin{figure}[!htb]
        \centering
        \includegraphics[width =0.95\textwidth]{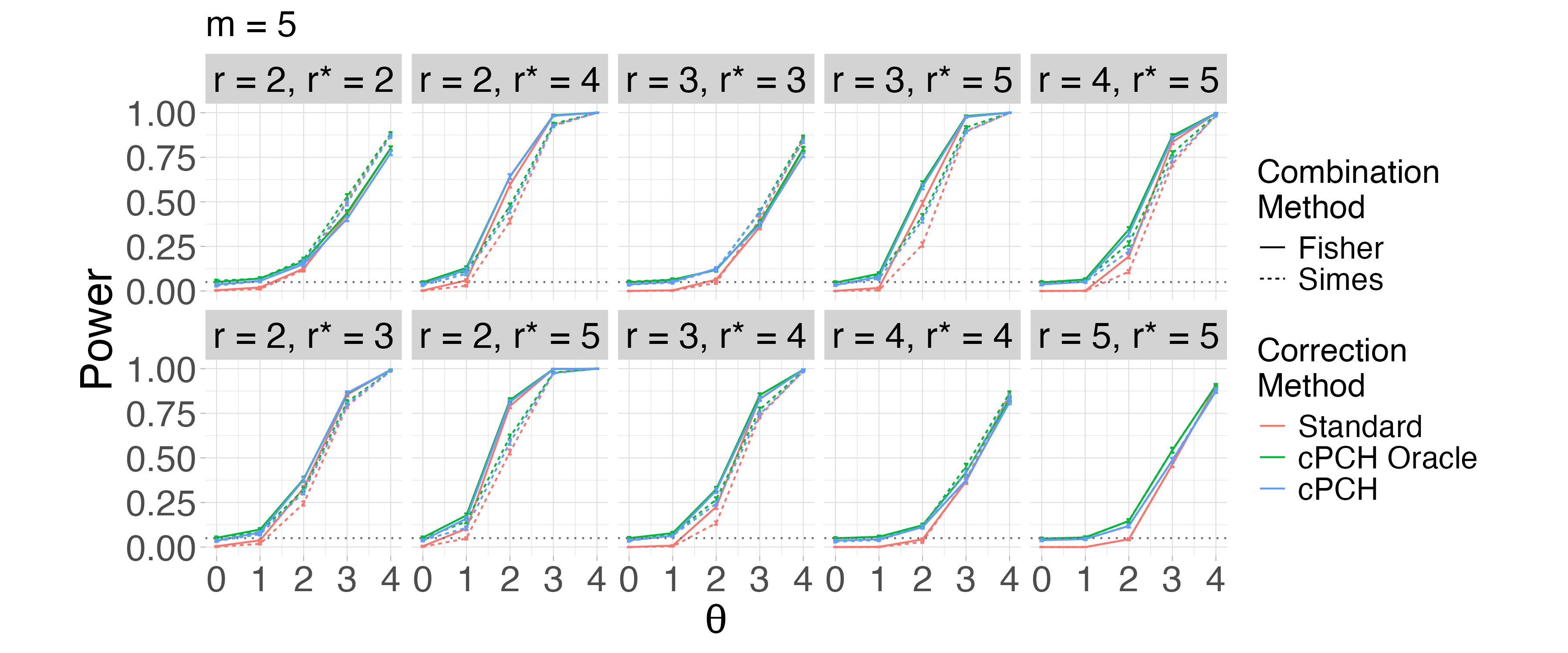}
 \caption{Power of the cPCH test across all alternative cases $(r^\star \geq r)$ at nominal level $\alpha = 0.05$ (dotted black line) for $m=5$. Each point represents the proportion of cPCH p-values below $\alpha$ over $5000$ replicates of the data generating procedure outlined in Section~\ref{section:singlepower} for each $(r^\star, r, \theta)$ triplet. Error bars depict $\pm 2$ standard errors.
} 
    \end{figure}

        \begin{figure}[!htb]
        \centering
        \includegraphics[width = 0.6\textwidth]{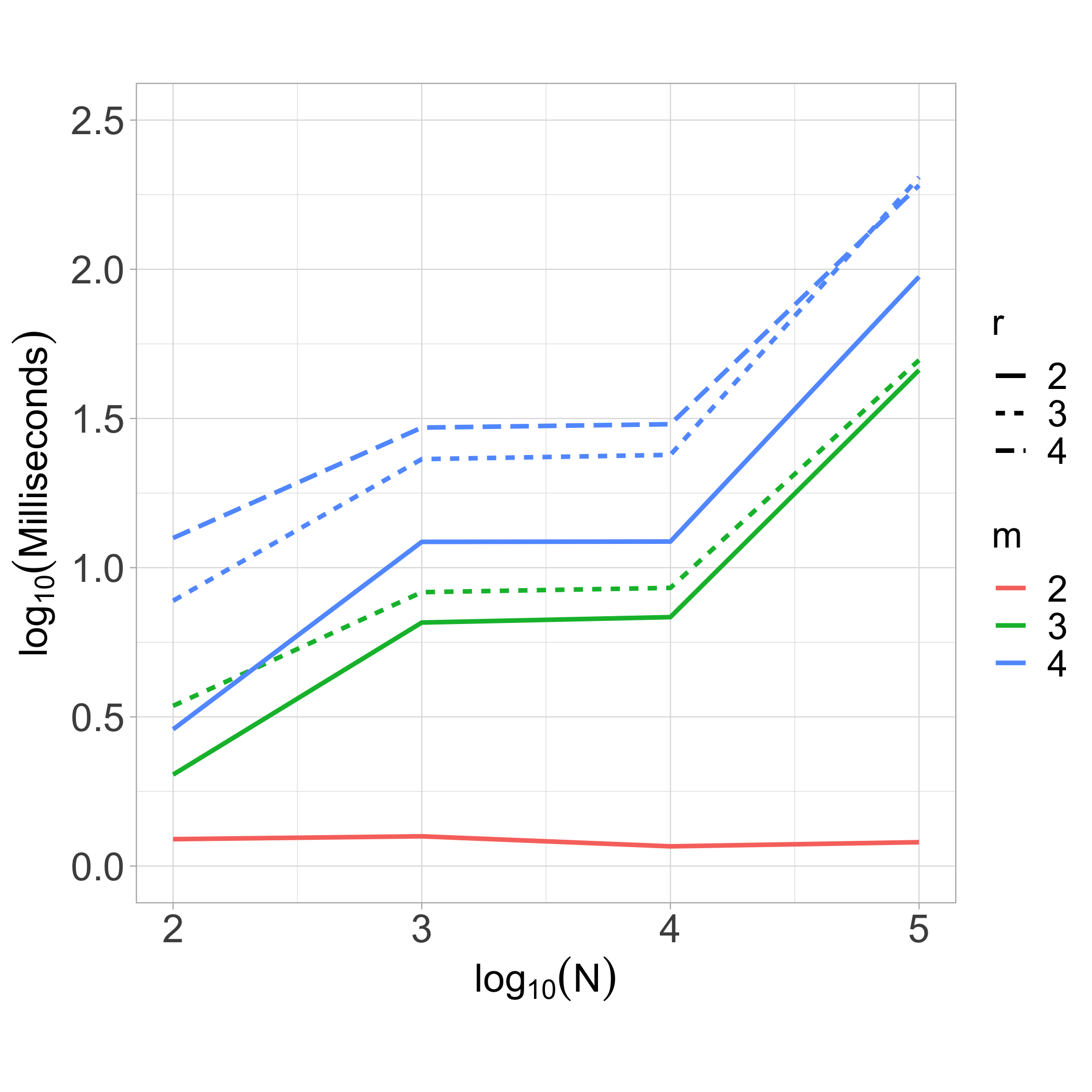}
 \caption{Computation time for a single cPCH p-value. Each point represents the log$_{10}$ of the average time in milliseconds to compute a single cPCH p-value (as described in Section~\ref{section:computing}) with Fisher's combining function over $100$ replicates for each combination of $m$, $r$, and $N$. We use Fisher's combining function since we found that the computation times were similar across Fisher's and Simes' combining functions. The computation times for $r=m=2$ are especially small ($\approx 1$ millisecond) because we are able to compute cPCH p-values in this case analytically using only evaluations of the standard normal CDF and PDF, which can be computed very efficiently; see the Supplementary Materials~\ref{section:computing_further_details} for further details.
} 
\label{fig:comptimes}
    \end{figure} 

\begin{figure}[htb!]
  \includegraphics[width=\linewidth]{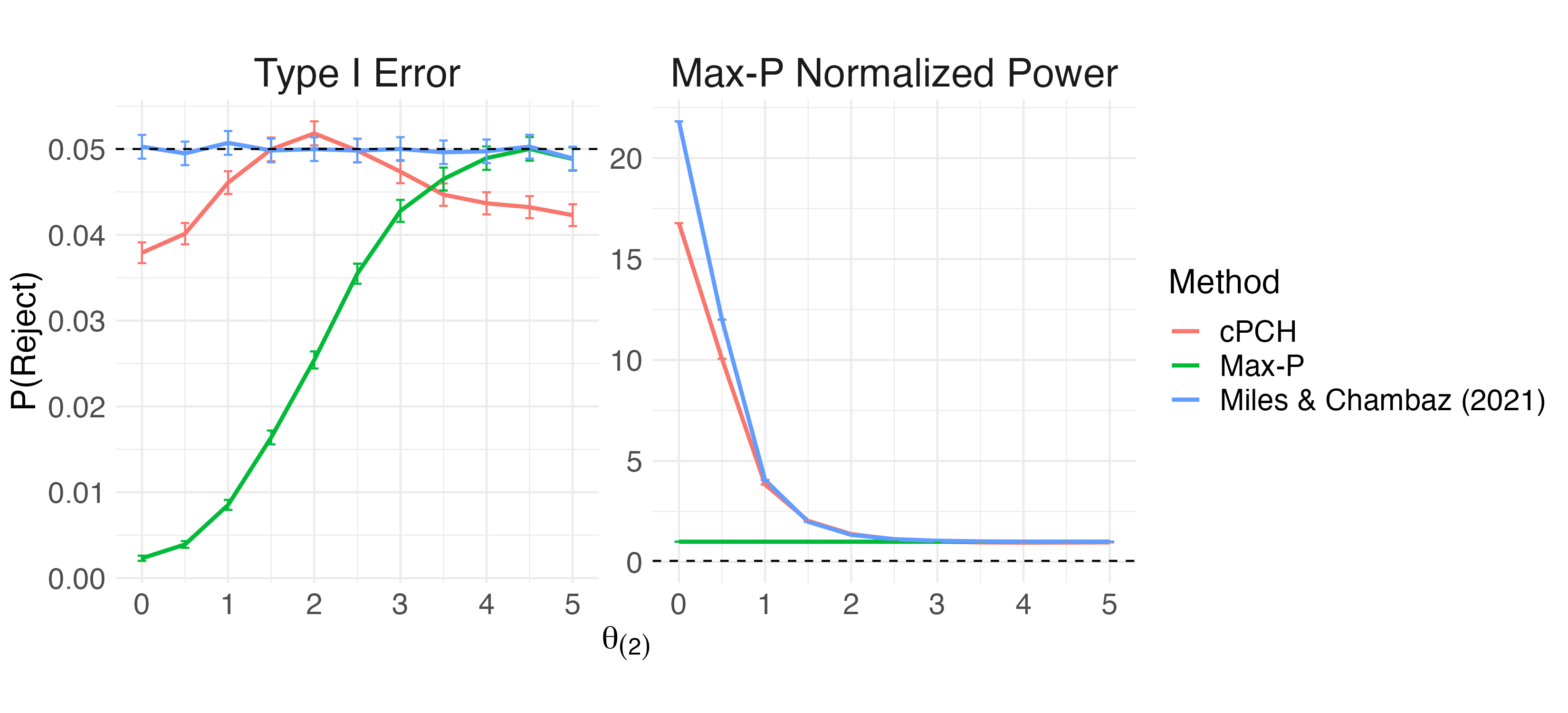}
\caption{ Type I error and \re{Max-P Normalized Power (i.e., the power of each test divided by the power of the Max-P test)} of the Max-P test, cPCH test, and the test of \cite{miles2021optimal} for $H_0^{2/2}$ at level $\alpha=0.05$ (dotted black line). For the Type I error plot, each point represents $100{,}000$ independent simulations with $T_i \sim \mathcal{N}(\theta_i, 1)$ where $T_1$ and $T_2$ are independent and $\left(\theta_1, \theta_2\right) = \left(0, \theta_{(2)}\right)$. For the power plot, $T_1$ and $T_2$ are independent and $\left(\theta_1, \theta_2\right) = \left(\frac{\theta_{(2)}}{2}, \theta_{(2)}\right)$. Error bars depict $\pm 2$ standard errors.}
\label{fig:mmcomp}
\end{figure}

\clearpage

 \section{Exactness of the Unadjusted cPCH Test under the Least Favorable Null Case}\label{appendix:lfn_valid}
 We first discuss the behavior of the cPCH test under the least favorable null case, which we introduced in Section~\ref{section:classical_intro} as the setting where, informally, there are exactly $r-1$ $\theta_i$'s with $\left|\theta_i\right|$ equaling infinity. In this section, we formalize this initial description and prove a validity result for the unadjusted cPCH test under the least favorable null case.

First, we provide some intuition for the least favorable null case. As discussed in Section~\ref{section:preliminaries}, the data for each base hypothesis is commonly summarized as an asymptotically normal parameter estimator such as a sample mean or ordinary least squares coefficient. Under our assumed setting in Section~\ref{section:preliminaries}, we expect the non-null $T_i$'s under fixed alternatives to approach infinity in magnitude as their underlying sample size approaches infinity. For example, if each base hypothesis test reports a sample mean of $n$ i.i.d. observations each approximated by a $\mathcal{N}(\mu_i, 1)$ distribution, then the standard $z$-test statistic is $T_i \sim \mathcal{N}(\theta_i, 1)$ where $\theta_i = \sqrt{n}\mu_i$. Therefore, as $n$ approaches infinity, the non-null $\left|\theta_i\right|$'s will also approach infinity, and thus, so will their corresponding $\left|T_i\right|$'s.  

When the base test statistics are independent and the null base p-values are uniformly distributed, the standard Fisher and Simes single PCH test are exact tests under the least favorable null case. We show that the  \rev{unadjusted} cPCH test exhibits a similar property under the least favorable null case.
First, we formalize our initial description of the least favorable null case by defining an \emph{least favorable null sequence}:

\begin{definition}[Least Favorable Null Sequence]
An least favorable null sequence $(\LFNseq)$ is a sequence in $\Theta_{0}^{r/m}$ such that $\left|\LFNseq_{(j)}\right| \to \infty$ for $j = m-r+2, ..., m$ as $n \to \infty$.
\end{definition}

We will prove that, under the least favorable null case, the limiting Type I 
error of the following more general version of the unadjusted cPCH test (for any m) is exactly $\alpha$ as $n \to \infty$.  

  \begin{definition}[Generalized cPCH test] The generalized cPCH test rejects $H_0^{r/m}$ when 
  
  $f\left(\bT_{(1:m-r+1)}\right) > c_\alpha\left(\hat{\btheta}_{(m-r+2:m)}, \bT_{ (m-r+2:m)}\right)$.
    \end{definition}
    
   In particular, Assumption~\ref{assump3} allows for more general sequences of estimators for $\btheta^{(n)}_{(m-r+2:m)}$ when setting the rejection threshold of the unadjusted cPCH test. Setting $\hat{\btheta}_{(m-r+2:m)} = \bT_{(m-r+2:m)}$ recovers the original cPCH test as in Definition~\ref{cpchdef}. 

Let $\Tseq$ denote a test statistic vector, with the superscript $(n)$ now allowing us to vary $\Tseq$'s distribution. Let $\varphi^{\text{ucPCH}}_\alpha\left(\Tseq\right) \coloneqq \mathbbm{1}\left\{f\left(\Tseq_{(1:m-r+1)}\right) > c_\alpha\left(\Tseq_{ (m-r+2:m)}, \Tseq_{ (m-r+2:m)}\right)\right\}$
and
$\varphi^{\text{PCHOrac}}_\alpha\left(\Tseq\right) \coloneqq \mathbbm{1}\left\{f\left(\Tseq_{(1:m-r+1)}\right) > c_\alpha\left(\LFNseq_{(m-r+2:m)}\right)\right\}$ denote the decisions made by the \rev{unadjusted} cPCH test and PCH Oracle test, respectively.  Throughout all proofs, we will use ``$\inp$'' to denote convergence in probability and the superscript ``$^c$'' to denote the complement of a set. 



\begin{customassumption}{1}\label{assump1}
 Assume $\left(\LFNseq\right)$ is a LFN sequence and that $\Tseq \sim \mathcal{N}(\btheta^{(n)}, I_m)$.
\end{customassumption}

\begin{customassumption}{2}\label{assump2}
Assume $f:\mathbb{R}^{m-r+1} \to \mathbb{R}$ is permutation invariant, continuously differentiable, and has $\grad f \neq 0$ except on a set whose closure has measure zero.
\end{customassumption}

\begin{customassumption}{3}\label{assump3}
Assume $\left(\hattheta\right)$ is a sequence of estimators for the LFN sequence $\left(\LFNseq_{(m-r+2:m)}\right)$ with
the following property: For any $K > 0$, $\lim_{n \to \infty} \mathbb{P}_{\LFNseq}\left(\left|\hatthetaelem_{(i)} \right| > K  \right) = 1$ for $i = m-r+2,..., m$.
\end{customassumption}
\begin{theorem}[Exactness of the \rev{unadjusted} cPCH test under the least favorable null case]\label{thrm1} Under Assumptions~\ref{assump1}--\ref{assump3}, for any $\alpha \in (0, 1)$,
$$\lim_{n \to \infty}\mathbb{P}_{\LFNseq}\left( \varphi^{\text{cPCH}}_\alpha\left(\Tseq\right) = \varphi^{\text{PCHOrac}}_\alpha\left(\Tseq\right) \right)  = 1.$$
In particular, the above implies that the generalized cPCH test's limiting Type I error under any LFN sequence is exactly $\alpha$.
\end{theorem}
Theorem~\ref{thrm1} shows not only that the  \rev{unadjusted} cPCH test achieves exactly the nominal Type I error rate in the least favorable null case, but that it in fact behaves \emph{identically} to the PCH Oracle test in such a case. Note as well that the Fisher and Simes combining functions satisfy the conditions on $f$ specified in the theorem statement.
The proof is provided below.

\begin{proof}
Without loss of generality, assume $\LFNseq= \left(0, ...0, \theta^{(n)}_{m-r+2}, ..., \theta^{(n)}_{m}\right)$ where $\theta^{(n)}_{i} \to \infty$ for $i = m-r+2, ..., m$.
Let $R_f \coloneqq \text{Range}(f)$. For $x \in R_f$, let $G\left(x, \LFNseq_{(m-r+2:m)}\right) \coloneqq \mathbb{P}_{\LFNseq}\left(f\left(\Tseq_{(1:m-r+1)}\right) \leq x \right)$.\footnote{By Assumption~\ref{assump2}, $G$ is permutation invariant with respect to $\LFNseq$, so $\mathbb{P}_{\LFNseq}\left(f\left(\Tseq_{(1:m-r+1)}\right) \leq x \right) = \mathbb{P}_{\sigma(\LFNseq)}\left(f\left(\Tseq_{(1:m-r+1)}\right) \leq x \right)$ for any permutation of the elements of $\LFNseq$, $\sigma(\LFNseq)$. Therefore, we can use $\LFNseq_{(m-r+2:m)}$ to represent $\LFNseq$ in the input to $G$.} 
Let $H(x) \coloneqq \mathbb{P}_{\LFNseq}\left(f\left(\Tseq_{1:m-r+1}\right) \leq x \right)$; note that $\Tseq_{1:m-r+1} \sim \mathcal{N}(\bm{0}, I_{m-r+1})$ for all $n$, so $H$ does not vary with $n$. For $a \in (0, 1)$, let

$c_a\left(\LFNseq_{(m-r+2:m)}\right) = \inf\left\{x: G\left(x, \LFNseq_{(m-r+2:m)}\right) \geq 1-a\right\}$ and $c_a^* =\inf\left\{x: H(x) \geq 1-a\right\}$.

Lemma~\ref{lemma1} (stated below and proved in Supplementary Material~\ref{section:additional_lemmas}) establishes that $c_\alpha\left(\LFNseq_{(m-r+2:m)}\right)$, the rejection threshold of the level $\alpha$ PCH Oracle test, converges to $c_\alpha^*$:

\begin{lemma}\label{lemma1}
Under Assumptions~\ref{assump1}-\ref{assump2}, for $\alpha \in (0, 1)$,
\begin{equation}
c_\alpha\left(\LFNseq_{(m-r+2:m)}\right) \to c_\alpha^*.
\end{equation}
\end{lemma}

Lemma~\ref{lemma2} (stated below and proved in Supplementary Material~\ref{section:additional_lemmas}) establishes that 

$c_\alpha\left(\hattheta, \Tseq_{(m-r+2:m)}\right)$, the rejection threshold of the level $\alpha$ generalized cPCH test, converges to $c_\alpha^*$ in probability:\footnote{Unlike $c_\alpha\left(\LFNseq_{(m-r+2:m)}\right)$, which is a fixed real number for each $n$, $c_\alpha\left(\hattheta, \Tseq_{(m-r+2:m)}\right)$ is a random variable that is a function of $\Tseq$ both through the conditioning event and $\hattheta$. }

\begin{lemma}\label{lemma2}
Under Assumptions~\ref{assump1}-\ref{assump3}, for $\alpha \in (0, 1)$,
\begin{equation*}
    c_\alpha\left(\hattheta, \Tseq_{(m-r+2:m)}\right) \inp c_\alpha^*.
\end{equation*}
\end{lemma}

Fix $\alpha \in (0, 1)$ and $\epsilon > 0$.  Let $Q_n = \mathbb{P}_{\LFNseq}\left(|c_\alpha\left(\hattheta, \Tseq_{(m-r+2:m)}\right) - c_\alpha^*| \leq \epsilon \right)$. By Lemma~\ref{lemma2}, $  \lim_{n \to \infty}Q_n = 1$. So,
\begin{align}\label{eq:thrm1_eq0}
& \lim_{n \to \infty}\mathbb{P}_{\LFNseq}\left( \varphi^{\text{cPCH}}_\alpha\left(\Tseq\right) \neq \varphi^{\text{PCHOrac}}_\alpha\left(\Tseq\right)  \right)\nonumber \\
& = \lim_{n \to \infty} \mathbb{P}_{\LFNseq}\left( \varphi^{\text{cPCH}}_\alpha\left(\Tseq\right) \neq \varphi^{\text{PCHOrac}}_\alpha\left(\Tseq\right) \given |c_\alpha\left(\hattheta, \Tseq_{(m-r+2:m)}\right) - c_\alpha^*| \leq \epsilon \right)Q_n + \nonumber \\
& \qquad \qquad  \mathbb{P}_{\LFNseq}\left( \varphi^{\text{cPCH}}_\alpha\left(\Tseq\right) \neq \varphi^{\text{PCHOrac}}_\alpha\left(\Tseq\right)  \given |c_\alpha\left(\hattheta, \Tseq_{(m-r+2:m)}\right) - c_\alpha^* | > \epsilon \right)(1-Q_n) \nonumber \\
& = \lim_{n \to \infty} \mathbb{P}_{\LFNseq}\left( \varphi^{\text{cPCH}}_\alpha\left(\Tseq\right) \neq \varphi^{\text{PCHOrac}}_\alpha\left(\Tseq\right) \given |c_\alpha\left(\hattheta, \Tseq_{(m-r+2:m)}\right) - c_\alpha^*| \leq \epsilon \right),
\end{align}
 where the last line follows from Lemma~\ref{lemma2}. By Lemma~\ref{lemma1}, there exists $N_\epsilon \in \mathbb{N}$ such that for all $n \geq N_\epsilon$,
$|c_\alpha\left(\LFNseq_{(m-r+2:m)}\right) - c_\alpha^*| < \epsilon$. So, for $n \geq N_\epsilon$,
\begin{align*}
   & \mathbb{P}_{\LFNseq}\left( \varphi^{\text{cPCH}}_\alpha\left(\Tseq\right) \neq \varphi^{\text{PCHOrac}}_\alpha\left(\Tseq\right) \given |c_\alpha\left(\hattheta, \Tseq_{(m-r+2:m)}\right) - c_\alpha^*| \leq \epsilon \right)  \nonumber \\
  & \leq \mathbb{P}_{\LFNseq}\left(  f\left(\bT^{(n)}_{(1:m-r+1)}\right) \in  \left( c_\alpha^* - \epsilon,c_\alpha^* + \epsilon \right) \given |c_\alpha\left(\hattheta, \Tseq_{(m-r+2:m)}\right) - c_\alpha^*| \leq \epsilon \right). 
\end{align*}
Since  $  \lim_{n \to \infty}Q_n = 1$,
\begin{align}\label{eq:thrm1_eq1}
       & \lim_{n \to \infty}\mathbb{P}_{\LFNseq}\left(  f\left(\bT^{(n)}_{(1:m-r+1)}\right) \in  \left( c_\alpha^* - \epsilon,c_\alpha^* + \epsilon \right) \given |c_\alpha\left(\hattheta, \Tseq_{(m-r+2:m)}\right) - c_\alpha^*| \leq \epsilon \right) \nonumber \\
    & =  \lim_{n \to \infty} \mathbb{P}_{\LFNseq}\left(  f\left(\bT^{(n)}_{(1:m-r+1)}\right) \in  \left( c_\alpha^* - \epsilon,c_\alpha^* + \epsilon \right) \right)
\end{align}
where the last line follows by Lemma~\ref{lemma3}, (stated below and proved in Supplementary Materials~\ref{section:additional_lemmas}), assuming the above limit exists, which we show in Equation~\eqref{eq:thrm1_eq2}:
\begin{lemma}\label{lemma3}
Let $\left(\Omega, \mathcal{F}, \mathbb{P}\right)$ be a probability space. Let $A_n$ and $B_n$ be a sequence of events in $\Omega$ where $\lim_{n \to \infty} \mathbb{P}\left(B_n\right) = 1$ and $\lim_{n \to \infty} \mathbb{P}\left(A_n\right) = a$ for some $a \in [0, 1]$. Then, $\mathbb{P}\left(A_n \cap B_n\right) \to a$ and
$\mathbb{P}\left(A_n \given B_n\right) \to a$.
\end{lemma}

Let $B_n\left(\LFNseq, \Tseq \right)$ be the event that, for $\Tseq \sim \mathcal{N}(\LFNseq, I_m)$, $\left\{T^{(n)}_{(m-r+2)}, ..., T^{(n)}_{(m)}\right\} = \left\{T^{(n)}_{m-r+2}, ..., T^{(n)}_{m}\right\}$.
The set $\left\{T^{(n)}_{(m-r+2)}, ..., T^{(n)}_{(m)}\right\}$ will almost surely be uniquely defined because we will almost surely have no ties among the $T^{(n)}_i$'s since $\Tseq \sim \mathcal{N}(\LFNseq, I_m)$. We show in the proof of Lemma~\ref{lemma1} that $\lim_{n \to \infty }\mathbb{P}_{\LFNseq}\left(B_n\left(\LFNseq, \Tseq \right)\right) = 1$. So,
\begin{align}\label{eq:thrm1_eq2}
& \lim_{n \to \infty} \mathbb{P}_{\LFNseq}\left(  f\left(\bT^{(n)}_{(1:m-r+1)}\right) \in  \left( c_\alpha^* - \epsilon,c_\alpha^* + \epsilon \right) \right) \nonumber \\
    & = \lim_{n \to \infty} \mathbb{P}_{\LFNseq}\left(  f\left(\bT^{(n)}_{(1:m-r+1)}\right) \in  \left( c_\alpha^* - \epsilon,c_\alpha^* + \epsilon \right) \given B_n\left(\LFNseq, \Tseq \right) \right)\mathbb{P}_{\LFNseq}\left(B_n\left(\LFNseq, \Tseq \right)\right) + \nonumber \\
    & \qquad \qquad    \mathbb{P}_{\LFNseq}\left(  f\left(\bT^{(n)}_{(1:m-r+1)}\right) \in  \left( c_\alpha^* - \epsilon,c_\alpha^* + \epsilon \right) \given B_n\left(\LFNseq, \Tseq \right)^c \right)\mathbb{P}_{\LFNseq}\left(B_n\left(\LFNseq, \Tseq \right)^c\right) \nonumber \\
    & = \lim_{n \to \infty} \mathbb{P}_{\LFNseq}\left(  f\left(\bT^{(n)}_{(1:m-r+1)}\right) \in  \left( c_\alpha^* - \epsilon,c_\alpha^* + \epsilon \right) \given B_n\left(\LFNseq, \Tseq \right) \right) \nonumber \\
    & = \lim_{n \to \infty} \mathbb{P}_{\LFNseq}\left(  f\left(\bT^{(n)}_{1:m-r+1}\right) \in  \left( c_\alpha^* - \epsilon,c_\alpha^* + \epsilon \right) \given B_n\left(\LFNseq, \Tseq \right) \right) \nonumber \\
    & =  H(c_\alpha^* + \epsilon) - H(c_\alpha^* - \epsilon),
\end{align}
where the fourth line follows from the fact that $\lim_{n \to \infty }\mathbb{P}_{\LFNseq}\left(B_n\left(\LFNseq, \Tseq \right)\right) = 1$ and the fifth line follows from the conditioning on $B_n\left(\LFNseq, \Tseq \right)$ and the permutation invariance of $f$ in Assumption~\ref{assump2}.
The last line follows from applying Lemma~\ref{lemma3} using the facts that
$\lim_{n \to \infty }\mathbb{P}_{\LFNseq}\left(B_n\left(\LFNseq, \Tseq \right)\right) = 1$ and
$\lim_{n \to \infty} \mathbb{P}_{\LFNseq}\left(  f\left(\bT^{(n)}_{1:m-r+1}\right) \in  \left( c_\alpha^* - \epsilon, c_\alpha^* + \epsilon \right) \right) =  H(c_\alpha^* + \epsilon) - H(c_\alpha^* - \epsilon)$.
Combining Equations~\eqref{eq:thrm1_eq0}-\eqref{eq:thrm1_eq2}, we have that
$$\lim_{n \to \infty}\mathbb{P}_{\LFNseq}\left( \varphi^{\text{cPCH}}_\alpha\left(\Tseq\right) \neq \varphi^{\text{PCHOrac}}_\alpha\left(\Tseq\right)  \right) \leq   H(c_\alpha^* + \epsilon) - H(c_\alpha^* - \epsilon).$$
Since $\epsilon$ was arbitrary, the above is true for any $\epsilon > 0$. Additionally, $H$ is continuous on $R_f$ as a result of Lemma~\ref{lemma4} (stated below and proved in Appendix~\ref{section:additional_lemmas}):

\begin{lemma}\label{lemma4}
If $\bm{X} = (X_1, ..., X_l)$ is a vector of continuous random variables supported on $\mathbb{R}^l$ with density $p_{\bm{X}}(\bm{x}) > 0$ for all $\bm{x} \in \mathbb{R}^l$ and $f:\mathbb{R}^l \to \mathbb{R}$ is continuously differentiable function with $\grad f \neq 0$ except on a set whose closure has measure zero, then the CDF of $f(\bm{X})$ is continuous and strictly increasing on the range of $f$.
\end{lemma}

So, by the continuity of $H$, as $\epsilon \to 0$, $ H(c_\alpha^* + \epsilon) - H(c_\alpha^* - \epsilon) \to 0$.

Therefore, we conclude that $\lim_{n \to \infty}\mathbb{P}_{\LFNseq}\left( \varphi^{\text{cPCH}}_\alpha\left(\Tseq\right) \neq \varphi^{\text{PCHOrac}}_\alpha\left(\Tseq\right)  \right) = 0$ and hence,
$$\lim_{n \to \infty}\mathbb{P}_{\LFNseq}\left( \varphi^{\text{cPCH}}_\alpha\left(\Tseq\right) = \varphi^{\text{PCHOrac}}_\alpha\left(\Tseq\right)  \right) = 1.$$

\end{proof}

\section{Lemmas for Proving Theorem S1}\label{section:additional_lemmas}

Lemmas~\ref{lemma1}-\ref{lemma2} are the main results that allow us to prove Theorem~\ref{thrm1}, with Lemmas~\ref{lemma3}-\ref{lemma4} providing useful results that help to complete the proof of Theorem~\ref{thrm1} and are used in the proofs of Lemmas~\ref{lemma1}-\ref{lemma2}. The proof of Lemma~\ref{lemma2} additionally relies on Lemmas~\ref{lemma5}-\ref{lemma6}. The proofs of Lemmas~\ref{lemma1}-\ref{lemma6} are below.

\subsection{Proof of Lemma~\ref{lemma1}}
\begin{proof}
As in Theorem~\ref{thrm1}, let  $\LFNseq = \left(0, ..., 0, \theta_{m-r+2}^{(n)}, ..., \theta_{m}^{(n)}\right)$ where $\theta^{(n)}_{i} \to \infty$ and define $B_n\left(\LFNseq, \Tseq \right)$ as the event that, for $\Tseq \sim \mathcal{N}(\LFNseq, I_m)$, $\left\{T^{(n)}_{(m-r+2)}, ..., T^{(n)}_{(m)}\right\} = \left\{T^{(n)}_{m-r+2}, ..., T^{(n)}_{m}\right\}$. We will first show that $\lim_{n \to \infty}\mathbb{P}_{\LFNseq}\left(B_n\left(\LFNseq, \Tseq \right) \right)  = 1$.
Let $K > 0$. 
Then, for $m-r+2 \leq i \leq m$,
\begin{align}
  \lim_{n \to \infty} \mathbb{P}_{\LFNseq}\left( \left|T^{(n)}_{i}\right| > K\right) 
=  \lim_{n \to \infty} 1-\Phi\left( K - \theta^{(n)}_{i}\right) +  \Phi\left( -K - \theta^{(n)}_{i}\right) = 1, \label{eq:lem1_eq0} 
\end{align}
where the last line follows from Assumption~\ref{assump1}. Let $F^{(n)}_{K} \coloneqq \left\{\min_{i = m-r+2, ..., m}\left|T^{(n)}_{i}\right| > K\right\}$. Then,
\begin{equation}\label{eq:lem1_eqfk}
\mathbb{P}_{\LFNseq} \left(F^{(n)}_{K}\right) = \mathbb{P}_{\LFNseq} \left(\bigcap_{i=m-r+2}^m \left\{\left|T^{(n)}_{i}\right| > K\right\} \right) = \prod_{i=m-r+2}^m\mathbb{P}_{\LFNseq} \left(\left|T^{(n)}_{i}\right| > K \right)\to 1,
\end{equation}
 where the second equality follows from the fact that $T^{(n)}_{m-r+2}, ..., T^{(n)}_{m}$ are independent and the last result follows from Equation~\eqref{eq:lem1_eq0}. So,
\begin{align*}
\mathbb{P}_{\LFNseq}\left(B_n\left(\LFNseq, \Tseq \right) \right) & = \mathbb{P}_{\LFNseq}\left( \left\{T^{(n)}_{(m-r+2)}, ..., T^{(n)}_{(m)}\right\} = \left\{T^{(n)}_{m-r+2}, ..., T^{(n)}_{m}\right\}  \right)  \\ 
& = \mathbb{P}_{\LFNseq}\left( \max_{j = 1, ..., m-r+1}\left|T^{(n)}_{j}\right| <  \min_{i = m-r+2, ..., m}\left|T^{(n)}_{i}\right| \right)  \\
    & = \mathbb{P}_{\LFNseq}\left( \max_{j = 1, ..., m-r+1}\left|T^{(n)}_{j}\right| <  \min_{i = m-r+2, ..., m}\left|T^{(n)}_{i}\right| \given  F^{(n)}_{K} \right) \mathbb{P}_{\LFNseq}\left( F^{(n)}_{K} \right) \\ 
    & \qquad \qquad + \mathbb{P}_{\LFNseq}\left( \max_{j = 1, ..., m-r+1}\left|T^{(n)}_{j}\right| <  \min_{i = m-r+2, ..., m}\left|T^{(n)}_{i}\right| \given F^{(n)c}_{K} \right) \mathbb{P}_{\LFNseq}\left( F^{(n)c}_{K} \right).
\end{align*}
By Equation~\eqref{eq:lem1_eqfk}, 
$$\lim_{n \to \infty} \mathbb{P}_{\LFNseq}\left(B_n\left(\LFNseq, \Tseq \right) \right) = \lim_{n \to \infty} \mathbb{P}_{\LFNseq}\left( \max_{j = 1, ..., m-r+1}\left|T^{(n)}_{j}\right| <  \min_{i = m-r+2, ..., m}\left|T^{(n)}_{i}\right| \given  F^{(n)}_{K}  \right),$$ where, for any $n$,
\begin{align*}
        & \mathbb{P}_{\LFNseq}\left( \max_{j = 1, ..., m-r+1}\left|T^{(n)}_{j}\right| <  \min_{i = m-r+2, ..., m}\left|T^{(n)}_{i}\right| \given  F^{(n)}_{K}  \right) \\
        & \geq \mathbb{P}_{\LFNseq}\left( \max_{j = 1, ..., m-r+1}\left|T^{(n)}_{j}\right| <  K \given  F^{(n)}_{K}  \right) \\
    & = \mathbb{P}_{\LFNseq}\left( \max_{j = 1, ..., m-r+1}\left|T^{(n)}_{j}\right| <  K  \right)  \\ 
     & = (\Phi(K)- \Phi(-K))^{m-r+1}. 
\end{align*}

Therefore, 
$$\lim_{n \to \infty} \mathbb{P}_{\LFNseq}\left(B_n\left(\LFNseq, \Tseq \right) \right) \geq  (\Phi(K)- \Phi(-K))^{m-r+1}.$$
Since $K$ was arbitrary, the above line holds for any $K>0$. 
Note that as $K$ gets arbitrarily large, $(\Phi(K)- \Phi(-K))^{m-r+1} \to 1$. Therefore, 
\begin{equation}\label{eq:lem1_eq2}
\lim_{n \to \infty}\mathbb{P}_{\LFNseq}\left(B_n\left(\LFNseq, \Tseq \right) \right) = 1.
\end{equation}

Let $x \in R_f$. Then,
\begin{align*}
    \lim_{n \to \infty} G\left(x, \LFNseq_{(m-r+2:m)}\right) & =\lim_{n \to \infty} \mathbb{P}_{\LFNseq}\left(f\left(\Tseq_{(1:m-r+1)}\right) \leq x\right)  \nonumber\\
    &=   \lim_{n \to \infty} \mathbb{P}_{\LFNseq}\left( f\left(\Tseq_{(1:m-r+1)}\right) \leq x \given B_n\left(\LFNseq, \Tseq \right)\right) \mathbb{P}_{\LFNseq}\left( B_n\left(\LFNseq, \Tseq \right)\right)+  \nonumber\\
    & \qquad \qquad \mathbb{P}_{\LFNseq}\left( f\left(\Tseq_{(1:m-r+1)}\right) \leq x \given B_n\left(\LFNseq, \Tseq \right)^c \right) \mathbb{P}_{\LFNseq}\left( B_n\left(\LFNseq, \Tseq \right)^c \right)\\
    & = \lim_{n \to \infty}\mathbb{P}_{\LFNseq}\left( f\left(\Tseq_{(1:m-r+1)}\right) \leq x \given B_n\left(\LFNseq, \Tseq \right)\right)\\
    &= \lim_{n \to \infty}\mathbb{P}_{\LFNseq}\left( f\left(\Tseq_{1:m-r+1}\right) \leq x \given B_n\left(\LFNseq, \Tseq \right)\right) \nonumber\\
 & = H(x),
 \end{align*}

where the fourth line follows from Equation~\eqref{eq:lem1_eq2}, and the fifth line follows from the conditioning on $B_n\left(\LFNseq, \Tseq \right)$ and the permutation invariance of $f$ in Assumption~\ref{assump2}. 
 The last line follows from Lemma~\ref{lemma3} using Equation~\eqref{eq:lem1_eq2} and the fact that $\lim_{n \to \infty} \mathbb{P}_{\LFNseq}\left( f\left(\Tseq_{1:m-r+1}\right) \leq x \right) = H(x)$. 

So,
\begin{equation*}
   G\left(x, \LFNseq_{(m-r+2:m)}\right)  \to H(x). 
\end{equation*}
Since $x$ was arbitrary, the above is true for any $x \in R_f$.
 Additionally, $H$ is continuous on $R_f$ as a result of Lemma~\ref{lemma4}, 
where $f$ satisfies the conditions of Lemma~\ref{lemma4} by Assumption~\ref{assump2}. Therefore, by Lemma 21.2 of \cite{vaart_1998}, for any $\alpha \in (0, 1)$,
  \begin{equation*}
c_\alpha\left(\LFNseq_{(m-r+2:m)}\right) \to c_\alpha^*.
\end{equation*} 
\end{proof}

\subsection{Proof of Lemma~\ref{lemma2}}

\begin{proof}
Without loss of generality, let $\LFNseq =\left(0, ..., 0, \theta^{(n)}_{(m-r+2)}, ..., \theta^{(n)}_{(m)}\right)$ where $\theta^{(n)}_{i} \to \infty$
and let $\hat{\btheta}^{(n)} = \left(0, ..., 0, \hat{\theta}^{(n)}_{(m-r+2)}, ..., \hat{\theta}^{(n)}_{(m)}\right)$. Let

$\tilde{F}\left(x, \Tseq\right) \coloneqq \mathbb{P}_{\tilde{\bT}^{(n)} \sim \hat{\btheta}^{(n)}}\left(  f\left(\tilde{\bT}^{(n)}_{(1:m-r+1)}\right) \leq x \given \tilde{\bT}^{(n)}_{(m-r+2:m)}  = \Tseq_{(m-r+2:m)}, \Tseq \right)$ represent the probability that $f\left(\tilde{\bT}^{(n)}_{(1:m-r+1)}\right) \leq x$ conditional on $\tilde{\bT}^{(n)}_{(m-r+2:m)}  = \Tseq_{(m-r+2:m)}$ and $\Tseq$ where the subscript on the probability denotes that $\tilde{\bT}^{(n)} \mid \hat{\btheta}^{(n)}_{(m-r+2:m)}  \sim \mathcal{N}(\hat{\btheta}^{(n)}, I_m)$. Note, this probability is only over $\tilde{\bT}^{(n)}$, which is a function of $\hat{\btheta}^{(n)}_{(m-r+2:m)}$. Therefore, $\tilde{F}\left(x, \Tseq\right)$ is a random variable since it is a function of $\Tseq$ both through the conditioning event and through $\hat{\btheta}^{(n)}_{(m-r+2:m)}$. We consider $\tilde{F}\left(x, \Tseq\right)$ because $c_\alpha\left(\hattheta, \Tseq_{(m-r+2:m)}\right) = \inf\left\{x: \tilde{F}\left(x, \Tseq\right) \geq 1-\alpha \right\}$. Thus, to show our final result, we will first show that, for any $x \in R_f$, $ \tilde{F}\left(x, \Tseq\right) \inp H(x)$.

Showing $\tilde{F}\left(x, \Tseq\right) \inp H(x)$ relies on Lemma~\ref{lemma5} (stated below and proved in Supplementary Materials~\ref{sec:lem5_proof}):
\begin{lemma}\label{lemma5}
Under Assumptions~\ref{assump1} and \ref{assump3},
\begin{equation*}
   \mathbb{P}_{\tilde{\bT}^{(n)} \sim \hat{\btheta}^{(n)}}\left(B_n\left(\hat{\btheta}^{(n)},\tilde{\bT}^{(n)}\right) \given \tilde{\bT}^{(n)}_{(m-r+2:m)}  = \Tseq_{(m-r+2:m)}, \Tseq \right) \inp 1.
\end{equation*}
\end{lemma}

Here, $B_n\left(\hat{\btheta}^{(n)},\tilde{\bT}^{(n)}\right)$ represents the event that, for  $\tilde{\bT}^{(n)} \mid \hat{\btheta}^{(n)}_{(m-r+2:m)}  \sim \mathcal{N}(\hat{\btheta}^{(n)}, I_m)$,  

$\left\{\tilde{T}^{(n)}_{(m-r+2)}, ..., \tilde{T}^{(n)}_{(m)}\right\} = \left\{\tilde{T}^{(n)}_{m-r+2}, ..., \tilde{T}^{(n)}_{m}\right\} $. The set $\left\{\tilde{T}^{(n)}_{(m-r+2)}, ..., \tilde{T}^{(n)}_{(m)}\right\}$ will almost surely be uniquely defined since the $\tilde{T}^{(n)}_i$'s will almost surely have no ties. 

Note, $\mathbb{P}_{\tilde{\bT}^{(n)} \sim \hat{\btheta}^{(n)}}\left(B_n\left(\hat{\btheta}^{(n)},\tilde{\bT}^{(n)}\right) \given \tilde{\bT}^{(n)}_{(m-r+2:m)}  = \Tseq_{(m-r+2:m)}, \Tseq \right)$ is also a random variable as it is a function of $\Tseq$ both through the conditioning event and through $\hat{\btheta}^{(n)}_{(m-r+2:m)}$. For ease of notation, let 

$\tildeg \coloneqq \mathbb{P}_{\tilde{\bT}^{(n)} \sim \hat{\btheta}^{(n)}}\left(  f\left(\tilde{\bT}^{(n)}_{(1:m-r+1)}\right) \leq x \given     B_n\left(\hat{\btheta}^{(n)},\tilde{\bT}^{(n)}\right), \tilde{\bT}^{(n)}_{(m-r+2:m)}  = \Tseq_{(m-r+2:m)}, \Tseq \right)$,
and let
$P_B\left(\Tseq\right) = \mathbb{P}_{\tilde{\bT}^{(n)} \sim \hat{\btheta}^{(n)}}\left(     B_n\left(\hat{\btheta}^{(n)},\tilde{\bT}^{(n)}\right) \given \tilde{\bT}^{(n)}_{(m-r+2:m)}  = \Tseq_{(m-r+2:m)}, \Tseq \right)$
and analogously,
$P_B\left(\Tseq\right)^c = \mathbb{P}_{\tilde{\bT}^{(n)} \sim \hat{\btheta}^{(n)}}\left(     B_n\left(\hat{\btheta}^{(n)},\tilde{\bT}^{(n)}\right)^c \given \tilde{\bT}^{(n)}_{(m-r+2:m)}  = \Tseq_{(m-r+2:m)}, \Tseq \right)$.
Fix $x \in R_f$. Then,
\begin{align}
 \tilde{F}\left(x, \Tseq\right)  & = \tildeg P_B\left(\Tseq\right) + \nonumber \\
    & \mathbb{P}_{\tilde{\bT}^{(n)} \sim \hat{\btheta}^{(n)}}\left(  f\left(\tilde{\bT}^{(n)}_{(1:m-r+1)}\right) \leq x \given     B_n\left(\hat{\btheta}^{(n)},\tilde{\bT}^{(n)}\right)^c, \tilde{\bT}^{(n)}_{(m-r+2:m)}  = \Tseq_{(m-r+2:m)}, \Tseq \right) P_B\left(\Tseq\right)^c. 
\label{eq:lemma2_eq0}
\end{align}

Note the second term in the summation above converges in probability to $0$ by Lemma~\ref{lemma5} and the fact $P_B(\Tseq)^c$ is a bounded random variable. 

We will now show that $\tildeg \inp H(x)$. By the conditioning on $B_n\left(\hat{\btheta}^{(n)},\tilde{\bT}^{(n)}\right)$ and the permutation invariance of $f$ in Assumption~\ref{assump2},
$$\tildeg = \mathbb{P}_{\tilde{\bT}^{(n)} \sim \hat{\btheta}^{(n)}}\left(  f\left(\tilde{\bT}^{(n)}_{1:m-r+1}\right) \leq x \given     B_n\left(\hat{\btheta}^{(n)},\tilde{\bT}^{(n)}\right), \tilde{\bT}^{(n)}_{(m-r+2:m)}  = \Tseq_{(m-r+2:m)}, \Tseq \right).$$ $\tilde{T}^{(n)}_{1},...,\tilde{T}^{(n)}_{m-r+1}$ are i.i.d. standard normally distributed by assumption, so, 
$$\tilde{T}^{(n)}_{1},...,\tilde{T}^{(n)}_{m-r+1}\mid B_n\left(\hat{\btheta}^{(n)},\tilde{\bT}^{(n)}\right),  \tilde{\bT}^{(n)}_{(m-r+2:m)}  = \Tseq_{(m-r+2:m)}, \Tseq  \iid \text{Trunc--Norm}\left(0, 1, T^{(n)}_{(m-r+2)}\right),$$
where $\text{Trunc--Norm}(0, 1, c)$ is the standard normal distribution truncated at $-|c|$ and $|c|$ for $c \in \mathbb{R}$.
Let $ p_{\tilde{\bT}^{(n)} \sim \hat{\btheta}^{(n)}}\left(\bm{t} \given B_n\left(\hat{\btheta}^{(n)},\tilde{\bT}^{(n)}\right),  \tilde{\bT}^{(n)}_{(m-r+2:m)}  = \Tseq_{(m-r+2:m)}, \Tseq \right)$ be the PDF of $ \tilde{T}^{(n)}_{1}, ...,\tilde{T}^{(n)}_{m-r+1}$ conditional on $B_n\left(\hat{\btheta}^{(n)},\tilde{\bT}^{(n)}\right),  \tilde{\bT}^{(n)}_{(m-r+2:m)}  = \Tseq_{(m-r+2:m)}, \Tseq$ evaluated at $\bm{t} \in \mathbb{R}^{m-r+1}$. 
By definition of a truncated normal random variable, we have for any $\bm{t} \in \mathbb{R}^{m-r+1}$,
\begin{align*}
  &p_{\tilde{\bT}^{(n)} \sim \hat{\btheta}^{(n)}}\left(\bm{t} \given B_n\left(\hat{\btheta}^{(n)},\tilde{\bT}^{(n)}\right),  \tilde{\bT}^{(n)}_{(m-r+2:m)}  = \Tseq_{(m-r+2:m)}, \Tseq \right) =\\
  & \qquad \prod_{i=1}^{m-r+1} \frac{\phi(t_i) \mathbbm{1}_{T^{(n)}_{(m-r+2)}}\left(t_i\right)}{\Phi\left(\left|T^{(n)}_{(m-r+2)}\right|\right) - \Phi\left(-\left|T^{(n)}_{(m-r+2)}\right|\right)},
\end{align*}
where $\mathbbm{1}_{T^{(n)}_{(m-r+2)}}\left(t\right) \coloneqq \indic{t \in \left(-\left|T^{(n)}_{(m-r+2)}\right|, \left|T^{(n)}_{(m-r+2)}\right|\right)}$ for $t \in \mathbb{R}$.

Set $\epsilon > 0$ and define $S_x = \{\bm{t} \in \mathbb{R}^{m-r+1}: f(\bm{t}) < x\}$. By Assumption~\ref{assump2}, where the $\nabla f \neq 0$ condition guarantees the montonicity of $f$, there exists $K_{x} > 0$ such that for any $\bm{t}\in S_x$ , $|t_i| < K_x$ for all $i = 1, ..., m-r+1$. So, for any $K_{x} > 0$ satisfying this condition,
\begin{align}
   & \mathbb{P}_{\LFNseq}\left(|\tildeg - H(x)| > \epsilon \right) \nonumber \\
    & = \mathbb{P}_{\LFNseq}\left(|\tildeg- H(x)| > \epsilon \given \left|T^{(n)}_{ (m-r+2)}\right| > K_x \right) \mathbb{P}_{\LFNseq}\left(\left|T^{(n)}_{(m-r+2)}\right| > K_x \right) + \nonumber\\
    & \qquad \qquad \mathbb{P}_{\LFNseq}\left(|\tildeg - H(x)| > \epsilon \given \left|T^{(n)}_{(m-r+2)}\right| \leq K_x \right) \mathbb{P}_{\LFNseq}\left(\left|T^{(n)}_{(m-r+2)}\right| \leq K_x \right)\label{eq:lemma2_eq1} 
\end{align}
where
\begin{align*}
&\mathbb{P}_{\LFNseq}\left(|\tildeg - H(x)| > \epsilon \given \left|T^{(n)}_{(m-r+2)}\right| > K_x \right) \\
    & = \mathbb{P}_{\LFNseq}\left( \left|\int_{S_x} \left(\prod_{i=1}^{m-r+1} \frac{\phi(t_i) \mathbbm{1}_{T^{(n)}_{(m-r+2)}}\left(t_i\right)}{\Phi\left(\left|T^{(n)}_{(m-r+2)}\right|\right) - \Phi\left(-\left|T^{(n)}_{(m-r+2)}\right|\right)} - \prod_{i=1}^{m-r+1} \phi(t_i)\right) d\bm{t}  \right|> \epsilon \given \left|T^{(n)}_{(m-r+2)}\right|  > K_x \right) \\
    & = \mathbb{P}_{\LFNseq}\left( \left|\int_{S_x} \left(\prod_{i=1}^{m-r+1} \frac{\phi(t_i)}{\Phi\left(\left|T^{(n)}_{(m-r+2)}\right|\right) - \Phi\left(-\left|T^{(n)}_{(m-r+2)}\right|\right)} - \prod_{i=1}^{m-r+1} \phi(t_i)\right) d\bm{t}  \right|> \epsilon \given \left|T^{(n)}_{(m-r+2)}\right| > K_x \right) \\
    & = \mathbb{P}_{\LFNseq}\left( \int_{S_x} \left(\prod_{i=1}^{m-r+1} \frac{\phi(t_i)}{\Phi\left(\left|T^{(n)}_{(m-r+2)}\right|\right) - \Phi\left(-\left|T^{(n)}_{(m-r+2)}\right|\right)} - \prod_{i=1}^{m-r+1} \phi(t_i)  \right) d\bm{t} > \epsilon \given \left|T^{(n)}_{(m-r+2)}\right|  > K_x \right) \\
    & \leq \mathbb{P}_{\LFNseq}\left( \int_{S_x} \left(\prod_{i=1}^{m-r+1} \frac{\phi(t_i)}{\Phi(K_x) - \Phi(-K_x)} - \prod_{i=1}^{m-r+1} \phi(t_i)  \right) d\bm{t} > \epsilon \given \left|T^{(n)}_{(m-r+2)}\right|  > K_x \right) \\
    & = \mathbb{P}_{\LFNseq}\left( \int_{S_x} \left(\prod_{i=1}^{m-r+1} \frac{\phi(t_i)}{\Phi(K_x) - \Phi(-K_x)} - \prod_{i=1}^{m-r+1} \phi(t_i)  \right)d\bm{t} > \epsilon \right)\\
 &= \mathbbm{1}\left\{ \left(\frac{1}{\Phi(K_x) - \Phi(-K_x)} - 1\right)^{m-r+1}\int_{S_x} \prod_{i=1}^{m-r+1}\phi(t_i)d\bm{t} > \epsilon \right\}\\
  & = \mathbbm{1}\left\{ \left(\frac{1}{\Phi(K_x) - \Phi(-K_x)} - 1\right)^{m-r+1}H(x) > \epsilon \right\}.
\end{align*}
In the third line, we drop the indicator in the numerator because of the conditioning on $ \left|T^{(n)}_{(m-r+2)}\right| > K_x$. The fourth line follows because the expression within the absolute value is non-negative almost surely. 
Note, there exists $K_0> 0$ such that for all $K > K_0$,
\begin{align}\label{eq:kx_cond}
    \Phi(K) - \Phi(-K) > \frac{1}{1+\left(\frac{\epsilon}{H(x)}\right)^{\frac{1}{m-r+1}}}
\end{align}
and therefore,
\begin{align*}
   \left(\frac{1}{\Phi(K) - \Phi(-K)} - 1\right)^{m-r+1}H(x) < \left(\frac{\epsilon}{H(x)}\right)H(x) = \epsilon.
\end{align*}
For any $K_x$ satisfying $|t_i| < K_x$ for any $\bm{t} \in S_x$, any $K > K_x$ also satisfies this condition. Therefore, we can set $K_x$ sufficiently large such that it satisfies this condition and Equation~\eqref{eq:kx_cond}. Therefore, for such a $K_x$,
\begin{equation}\label{eq:lemma2_eq2}
\mathbb{P}_{\LFNseq}\left(|\tildeg - H(x)| > \epsilon \given \left|T^{(n)}_{(m-r+2)}\right| > K_x \right) = 0    
\end{equation}

As a result of Lemma~\ref{lemma6} (stated below and proved in Supplementary Materials~\ref{sec:lem6_proof}), for any $K > 0 $, $\lim_{n \to \infty}\mathbb{P}_{\LFNseq}\left(\left|T^{(n)}_{(m-r+2)}\right| > K \right) = 1$:

\begin{lemma}\label{lemma6}
Under Assumption~\ref{assump1}, $\hattheta = \Tseq_{(m-r+2:m)}$ satisfies Assumption~\ref{assump3}. 
\end{lemma}

Therefore, setting $K_x$ in Equation~\eqref{eq:lemma2_eq1} sufficiently large to satisfy Equation~\eqref{eq:kx_cond} and $|t_i| < K_x$ for any $\bm{t} \in S_x$, and applying Equation~\eqref{eq:lemma2_eq2} and Lemma~\ref{lemma6}, we have that
$$\lim_{n \to \infty}\mathbb{P}_{\LFNseq}\left(\left|\tildeg - H(x)\right| > \epsilon \right) = 0.$$
Since $\epsilon$ was arbitrary, we conclude that
$$\tildeg \inp H(x).$$
Since $x$ was arbitrary, the above holds for any $x \in R_f$. Combining the above result with Equation~\eqref{eq:lemma2_eq0} and Lemma~\ref{lemma5}, we have that, for any $x \in R_f$,
\begin{equation}\label{eq:lemma2_eq3}
\tilde{F}\left(x, \Tseq\right)  \inp  H(x).
\end{equation}

To show the final result, we first note that $H$ is strictly increasing and continuous on $R_f$ as a result of Lemma~\ref{lemma4}, where $f$ satisfies the conditions of Lemma~\ref{lemma4} by Assumption~\ref{assump2}. Therefore, by Lemma~\ref{lemma4} and the fact that Equation~\eqref{eq:lemma2_eq3} holds for all $x \in R_f$, $H(c^*_\alpha) = 1-\alpha$  for any $\alpha \in (0, 1)$. Set $ 0 < \epsilon < \min(\alpha, 1-\alpha)$. Then, $H(c^*_{\alpha+\epsilon}) = 1-\alpha - \epsilon$ and $H(c^*_{\alpha-\epsilon})  = 1-\alpha + \epsilon$. Set $\delta = \epsilon/2$ and let $\gamma > 0$. Then, by Equation~\eqref{eq:lemma2_eq3}, there exists $N_1 \in \mathbb{N}$ such that for all $n \geq N_1$, 
$$\mathbb{P}_{\LFNseq}\left(\left|\tilde{F}\left(c^*_{\alpha+\epsilon}, \Tseq\right) - (1-\alpha - \epsilon)  \right| \leq \delta \right) \geq 1-\gamma $$
and there exists $N_2  \in \mathbb{N}$ such that for all $n \geq N_2$, 
$$\mathbb{P}_{\LFNseq}\left(\left|\tilde{F}\left(c^*_{\alpha-\epsilon}, \Tseq\right) - (1-\alpha +\epsilon)  \right| \leq \delta \right) \geq 1-\gamma. $$

Let $N = \max(N_1, N_2)$. Then, for $n \geq N$,
$\tilde{F}\left(c^*_{\alpha+\epsilon}, \Tseq\right)$ is contained in the interval $[1-\alpha - \frac{3\epsilon}{2}, 1-\alpha-\frac{\epsilon}{2} ]$ and $\tilde{F}\left(c^*_{\alpha-\epsilon}, \Tseq\right)$ is contained in the interval $[1-\alpha + \frac{\epsilon}{2}, 1-\alpha+\frac{3\epsilon}{2} ]$ with probability $\geq 1-\gamma$. Therefore, for any $n \geq N$,
\begin{align}\label{eq:lem2_conv}
   & \mathbb{P}_{\LFNseq}\left(c_\alpha\left(\hattheta, \Tseq_{(m-r+2:m)}\right) \in (c^*_{\alpha + \epsilon},c^*_{\alpha - \epsilon} ]\right) \nonumber \\
    & \geq  \mathbb{P}_{\LFNseq}\left( \left\{\tilde{F}\left(c^*_{\alpha+\epsilon}, \Tseq\right) < 1-\alpha\right\} \cap \left\{\tilde{F}\left(c^*_{\alpha-\epsilon}, \Tseq\right)\geq 1-\alpha\right\}\right)  \nonumber \\
        & \geq  \mathbb{P}_{\LFNseq}\Big( \left\{\left|\tilde{F}\left(c^*_{\alpha+\epsilon}, \Tseq\right) - (1-\alpha -\epsilon)  \right| \leq \delta\right\} \cap \left\{ \left|\tilde{F}\left(c^*_{\alpha-\epsilon}, \Tseq\right) - (1-\alpha +\epsilon)  \right| \leq \delta \right\} \Big)  \nonumber \\
    & \geq 1-\gamma
\end{align}
Since $\gamma$ was arbitrary, for $ 0 < \epsilon < \min(\alpha, 1-\alpha)$,
$$\lim_{n \to \infty} \mathbb{P}_{\LFNseq}\left(c_\alpha\left(\hattheta, \Tseq_{(m-r+2:m)}\right) \in (c^*_{\alpha + \epsilon},c^*_{\alpha - \epsilon} ]\right) = 1.$$ 


Since $H$ is strictly increasing on $R_f$, $c^*_\alpha$ is continuous for $\alpha \in (0, 1)$. Then, by the continuity of $c^*_\alpha$ and the fact that the above holds for any arbitrary $0 < \epsilon < \min(\alpha, 1-\alpha)$, we conclude that
\begin{equation*}
    c_\alpha\left(\hattheta, \Tseq_{(m-r+2:m)}\right) \inp c_\alpha^*.
\end{equation*}
\end{proof}


\subsection{Proof of Lemma~\ref{lemma3}}\label{sec:lem3_proof}

\begin{proof}
\begin{equation*}
    \mathbb{P}\left(A_n \given B_n \right) = \frac{\mathbb{P}\left(A_n \cap B_n \right)}{\mathbb{P}\left( B_n \right)} =  \frac{\mathbb{P}\left(A_n \right) - \mathbb{P}\left(A_n \cap B_n^c\right)}{\mathbb{P}\left( B_n \right)},
\end{equation*}
where $B_n^c$ represents the complement of $B_n$. Since $ \mathbb{P}\left(B_n^c \right) \to 0$ by assumption, 
\begin{equation*}
   \lim_{n \to \infty}\mathbb{P}\left(A_n \cap B_n^c\right) = \lim_{n \to \infty}\mathbb{P}\left(A_n \given  B_n^c\right)\mathbb{P}\left( B_n^c\right) \to 0.
\end{equation*}
So, we conclude that 
$$ \mathbb{P}\left(A_n \right) - \mathbb{P}\left(A_n \cap B_n^c\right) \to a,$$ and hence,
$$\mathbb{P}\left(A_n \given B_n \right) \to a.$$
\end{proof}

\subsection{Proof of Lemma~\ref{lemma4}}\label{sec:lem4_proof}

\begin{proof}
Let $\mu$ represent the Lebesgue measure on $\mathbb{R}^l$.
 Let $R_f = \text{Range}(f)$. Let $x \in R_f$ and $P_x \coloneqq \{\bm{y}: \bm{y} \in f^{-1}(x)\}$. Let $H$ represent the CDF of $f(\bm{X})$.
By the continuity of $f$ and the fact that $p_{\bm{X}}(\bm{x}) > 0$ for all $\bm{x} \in \mathbb{R}^l$, the CDF of $f(\bm{X})$ must be strictly increasing. 
Assume for the sake of contradiction that the CDF of $f(\bm{X})$ is not strictly increasing. Then, since $f$ is continuous, $R_f$ must be a connected set in $\mathbb{R}$, i.e., $R_f$ must be an interval. Therefore, if the CDF of  $f(\bm{X})$ is not strictly increasing, there exists some open interval $(a, b) \in R_f$ such that 
 \begin{align*}
   0 & = \mathbb{P}\left(f(\bm{X}) \in (a, b) \right)\\
   & = \mathbb{P}\left(\bm{X} \in f^{-1}\left((a, b)\right) \right) \\
   & = \int_{\bm{x} \in  f^{-1}\left((a, b)\right) }p_{\bm{X}}(\bm{x})d\mu 
 \end{align*}
 However, $\int_{\bm{x} \in  f^{-1}\left((a, b)\right) }p_{\bm{X}}(\bm{x})d\mu  > 0$ since $p_{\bm{X}}(\bm{x}) > 0$ for all $\bm{x} \in \mathbb{R}^l$ and $f^{-1}\left((a, b)\right)$ is some open set in $\mathbb{R}^l$ (since for continuous functions, the pre-image of open sets are open sets). Therefore, the CDF of $f(\bm{X})$ must be strictly increasing.
 
To show continuity, we will show that, for any $x \in R_f$, $\mu(P_x) = 0$. If this is true, then, for any fixed $x \in R_f$,
\begin{align*}
    0 = \mathbb{P}(\bm{X} \in P_x)  & = \lim_{n \to \infty} \mathbb{P}\left(\bm{X} \in f^{-1}\left((x-{1/n}, x +{1/n}]\right)\right)\\
     & = \lim_{n \to \infty} \mathbb{P}\left(f(\bm{X}) \in (x-{1/n}, x +{1/n}] \right)\\
      & = \lim_{n \to \infty} \mathbb{P}\left(f(\bm{X}) \leq x +{1/n} \right) - \mathbb{P}\left(f(\bm{X}) \leq x-{1/n} \right)\\
& = \lim_{n \to \infty} H(x +{1/n}) - H(x -{1/n})
\end{align*}
Let $\epsilon > 0$. The last line above implies that there exists $N$ sufficiently large such that for all $n \geq N$, $H(x +{1/n}) - H(x -{1/n}) < \epsilon$. So for any $0 < \delta <1/n$, if $x-\delta < x' < x+\delta$, then
\begin{align*}
    |H(x) - H(x')| &\leq   H(x + \delta) - H(x - \delta) < H(x +{1/n}) - H(x -{1/n}) < \epsilon
\end{align*}
where the first inequality follows from the fact that $H$ is a CDF and hence is nondecreasing. Since $\epsilon$ was arbitrary, $H$ is continuous at $x$. Since the above holds for any $x \in R_f$, $H$ is continuous at every $x \in R_f$.


We will now show that, for any $x \in R_f$, $\mu(P_x) = 0$. Fix an arbitrary $x \in R_f$. Let $Q = \{\bm{t} \in \mathbb{R}^{l}: f \text{ is not continuous at }\bm{t} \} \cup \{\bm{t} \in \mathbb{R}^{l}: f \text{ is not differentiable at }\bm{t} \}\cup  \{\bm{t} \in \mathbb{R}^{l}:\grad{f} \text{ is not continuous at } \bm{t} \}\cup  \{\bm{t} \in \mathbb{R}^{l}:\grad{f}(\bm{t}) = 0 \}$. By assumption, the closure of $Q$, which we denote as $\overline{Q}$, has Lebesgue measure $0$. 
Let $\bm{y} \in P_x \setminus \overline{Q}$, 
Since $ \overline{Q}$ is closed, and $\bm{y}$ is in the complement of $\overline{Q}$, there must exist an open neighborhood of $\bm{y}$ that is in $\mathbb{R}^{l}\setminus \overline{Q}$. Also there must be at least one element of $\grad f(\bm{y})$ that is not zero, again, because $\bm{y}$ is in the complement of $\overline{Q}$. Without loss of generality, assume that the last element of $\grad f(\bm{y})$ is nonzero, i.e., $\grad f(\bm{y})_l \neq 0$.
Then, by the Implicit Function Theorem (Theorem 3.1, \cite{folland}), there exists open sets $U_{\bm{y}} \subset \mathbb{R}^{l-1}$ and $V_{\bm{y}} \subset \mathbb{R}$ such that $(y_1, ..., y_{l-1}) \in U_{\bm{y}}$ and $y_l \in V_{\bm{y}}$ and there exists a unique function $g: U_{\bm{y}} \to \mathbb{R}$ such that 
\begin{align*}
    \{y_1, ..., y_{l-1}, g(y_1, ..., y_{l-1}): (y_1, ..., y_{l-1}) \in  U_{\bm{y}} \} &=\{\bm{y} \in U_{\bm{y}} \times V_{\bm{y}} \mid f(\bm{y}) = x \}\\
    & = P_x\cap \left( U_{\bm{y}} \times V_{\bm{y}}\right) 
\end{align*}
and $g$ is continuously differentiable. In the above, $\times$ represents the Cartesian product. Let $W_{\bm{y}} =  U_{\bm{y}} \times V_{\bm{y}} $. Note, $\{y_1, ..., y_{l-1}, g(y_1, ..., y_{l-1}): (y_1, ..., y_{l-1}) \in  U_{\bm{y}} \}$ is the graph of the continuous function $g:U_{\bm{y}} \to \mathbb{R}$, and so has Lebesgue measure $0$ in $\mathbb{R}^l$  (Proposition 6.3, \cite{Lee}).
Therefore, $\forall \bm{y} \in P_x\setminus \overline{Q}$, there exists an open set $W_{\bm{y}} \subseteq \mathbb{R}^n$ such that 
\begin{equation}\label{eq:lem5_eq1}
    \mu\left(P_x \cap W_{\bm{y}} \right) = 0.
\end{equation}
We can write
\begin{align*}
P_x\setminus \overline{Q} = \bigcup_{\bm{y} \in P_x\setminus \overline{Q}} \left(P_x\setminus \overline{Q}\right) \cap W_{\bm{y}}  = \left(P_x\setminus \overline{Q}\right) \cap \left(\bigcup_{\bm{y} \in P_x\setminus \overline{Q}} W_{\bm{y}}\right).
\end{align*}
Note, $\bigcup_{y \in P_x\setminus \overline{Q}} W_{\bm{y}}$ is a open cover of $P_x\setminus \overline{Q} \subseteq \mathbb{R}^l$. By the fact that $\mathbb{R}^l$ is second-countable (Example 1.2, \cite{wedhorn}), and any subspace of $\mathbb{R}^l$ is second-countable (Remark 1.3, \cite{wedhorn}), by Lindel\"{o}f's Covering Theorem (Theorem 15, \cite{kelley}), there exists a countable subcover of $P_x\setminus \overline{Q}$, i.e., for some countable set $A \in P_x\setminus \overline{Q}$, $P_x\setminus \overline{Q} \subseteq \bigcup_{\bm{y} \in A} W_{\bm{y}}.$
Therefore,
\begin{align*}
     P_x\setminus \overline{Q} \cap \left(\bigcup_{\bm{y} \in P_x\setminus \overline{Q}} W_{\bm{y}}\right) & =  P_x\setminus \overline{Q} \cap \left(\bigcup_{\bm{y} \in A} W_{\bm{y}}\right) = \bigcup_{\bm{y} \in A} \left(P_x\setminus \overline{Q}\right) \cap W_{\bm{y}} 
\end{align*}
where 
\begin{align*}
\mu\left( \bigcup_{\bm{y} \in A} \left(P_x\setminus \overline{Q}\right) \cap W_{\bm{y}}\right) \leq \sum_{\bm{y} \in A}  \mu\left(\left(P_x\setminus \overline{Q}\right) \cap W_{\bm{y}}\right) \leq \sum_{\bm{y} \in A}  \mu\left(P_x \cap W_{\bm{y}}\right) = 0.
\end{align*}
The last equality follows from Equation~\eqref{eq:lem5_eq1} and the fact that $A$ is a countable set.

Therefore,  $\mu\left(  P_x\setminus \overline{Q} \right) = 0$ and
\begin{equation*}
    \mu(P_x) \le \mu\left(\left(P_x\setminus \overline{Q}\right) \cup \overline{Q}\right) = 0
\end{equation*}
Since we picked $x$ to be any arbitrary element of $R_f$, the above is true for all $x \in R_f$.
\end{proof}

\subsection{Proof of Lemma~\ref{lemma5}}\label{sec:lem5_proof}

\begin{proof}
Assume, without loss of generality, that $\hat{\btheta}^{(n)} = \left(0, ..., 0, \hat{\theta}^{(n)}_{(m-r+2)}, ..., \hat{\theta}^{(n)}_{(m)}\right)$. Let $B_n\left(\hat{\btheta}^{(n)},\tilde{\bT}^{(n)}\right)$ be the event that
$\left\{\tilde{T}^{(n)}_{(m-r+2)}, ..., \tilde{T}^{(n)}_{(m)}\right\} = \left\{\tilde{T}^{(n)}_{m-r+2}, ..., \tilde{T}^{(n)}_{m}\right\} $ for $\tilde{\bT}^{(n)} \mid \hat{\btheta}^{(n)}_{(m-r+2:m)}  \sim \mathcal{N}(\hat{\btheta}^{(n)}, I_m)$. 

Thus, 
$\mathbb{P}_{\tilde{\bT}^{(n)} \sim \hat{\btheta}^{(n)}}\left(B_n\left(\hat{\btheta}^{(n)},\tilde{\bT}^{(n)}\right) \given \tilde{\bT}^{(n)}_{(m-r+2:m)}  = \Tseq_{(m-r+2:m)}, \Tseq \right)$ is a random variable as it is a function of $\Tseq$ both through the conditioning event and through $\hat{\btheta}^{(n)}_{(m-r+2:m)}$.

Let $p_{\tilde{\bT}^{(n)} \sim \hat{\btheta}^{(n)}, \Tseq \sim \LFNseq}\left(\bm{t} \given \Tseq \right)$ be the PDF of the conditional distribution of $\tilde{\bT}^{(n)}_{(m-r+2:m)}$ given $\Tseq$ evaluated at $\tilde{\bT}^{(n)}_{(m-r+2:m)} = \bm{t}\in\mathbb{R}^{r-1}$ and let $p_{\tilde{\bT}^{(n)} \sim \hat{\btheta}^{(n)}, \Tseq \sim \LFNseq}\left(\bm{t} \given B_n\left(\hat{\btheta}^{(n)},\tilde{\bT}^{(n)}\right), \Tseq \right)$ be the analogous PDF of the conditional distribution of $\tilde{\bT}^{(n)}_{(m-r+2:m)}$ given $\Tseq$ and $B_n\left(\hat{\btheta}^{(n)},\tilde{\bT}^{(n)}\right)$,
where $\tilde{\bT}^{(n)} \mid \hat{\btheta}^{(n)}_{(m-r+2:m)}  \sim \mathcal{N}(\hat{\btheta}^{(n)}, I_m)$ and $\Tseq \sim \mathcal{N}(\LFNseq, I_m)$. 
Then,
\begin{align*}
& \mathbb{P}_{\tilde{\bT}^{(n)} \sim \hat{\btheta}^{(n)}}\left(B_n\left(\hat{\btheta}^{(n)},\tilde{\bT}^{(n)}\right) \given \tilde{\bT}^{(n)}_{(m-r+2:m)} = \Tseq_{(m-r+2:m)}, \Tseq \right) =\\
  & = \frac{p_{\tilde{\bT}^{(n)} \sim \hat{\btheta}^{(n)}, \Tseq \sim \LFNseq}\left(\Tseq_{(m-r+2:m)} \given B_n\left(\hat{\btheta}^{(n)},\tilde{\bT}^{(n)}\right), \Tseq \right) \mathbb{P}_{\tilde{\bT}^{(n)} \sim \hat{\btheta}^{(n)}}\left(B_n\left(\hat{\btheta}^{(n)},\tilde{\bT}^{(n)}\right) \given \Tseq \right)}{p_{\tilde{\bT}^{(n)} \sim \hat{\btheta}^{(n)}, \Tseq \sim \LFNseq}\left( \Tseq_{(m-r+2:m)} \given \Tseq \right)},
\end{align*}
where the numerator
\begin{align*}
   &p_{\tilde{\bT}^{(n)} \sim \hat{\btheta}^{(n)}, \Tseq \sim \LFNseq}\left(\Tseq_{(m-r+2:m)} \given B_n\left(\hat{\btheta}^{(n)},\tilde{\bT}^{(n)}\right), \Tseq \right) \mathbb{P}_{\tilde{\bT}^{(n)} \sim \hat{\btheta}^{(n)}}\left(B_n\left(\hat{\btheta}^{(n)},\tilde{\bT}^{(n)}\right) \given \Tseq \right)\\
   & = p_{\tilde{\bT}^{(n)} \sim \hat{\btheta}^{(n)}, \Tseq \sim \LFNseq}\left( \Tseq_{(m-r+2:m)} \given \Tseq \right) -\\
   & \qquad p_{\tilde{\bT}^{(n)} \sim \hat{\btheta}^{(n)}, \Tseq \sim \LFNseq}\left(\Tseq_{(m-r+2:m)} \given B_n\left(\hat{\btheta}^{(n)},\tilde{\bT}^{(n)}\right)^c, \Tseq \right) \mathbb{P}_{\tilde{\bT}^{(n)} \sim \hat{\btheta}^{(n)}}\left(B_n\left(\hat{\btheta}^{(n)},\tilde{\bT}^{(n)}\right)^c \given \Tseq \right).
\end{align*}

We are about to show that $\mathbb{P}_{\tilde{\bT}^{(n)} \sim \hat{\btheta}^{(n)}}\left( B_n\left(\hat{\btheta}^{(n)},\tilde{\bT}^{(n)}\right) \given \Tseq \right) \inp 1$, and therefore, the last line above (the term being subtracted) converges to $0$ in probability, thus giving our final result.

For $K > 0$, let $J^{(n)}_{K} = \left\{\bigcap_{i = m-r+2}^m \hat{\bm{\theta}}^{(n)}_{(i)} > K\right\}$. Set $\epsilon > 0$. For any $K > 0$,
\begin{align}
 & \mathbb{P}_{\LFNseq}\left(\left|1-\mathbb{P}_{\tilde{\bT}^{(n)} \sim \hat{\btheta}^{(n)}}\left(B_n\left(\hat{\btheta}^{(n)},\tilde{\bT}^{(n)}\right)\given \Tseq\right)\right| < \epsilon\right) \nonumber \\
&=\mathbb{P}_{\LFNseq}\left(\mathbb{P}_{\tilde{\bT}^{(n)} \sim \hat{\btheta}^{(n)}}\left(B_n\left(\hat{\btheta}^{(n)},\tilde{\bT}^{(n)}\right) \given \Tseq \right) > 1-\epsilon\right) \nonumber \\
&=\mathbb{P}_{\LFNseq}\left(\mathbb{P}_{\tilde{\bT}^{(n)} \sim \hat{\btheta}^{(n)}}\left(B_n\left(\hat{\btheta}^{(n)},\tilde{\bT}^{(n)}\right) \given \Tseq \right) > 1-\epsilon  \given J^{(n)}_{K}\right)\mathbb{P}_{\LFNseq}\left(J^{(n)}_{K}\right)\nonumber \\
   & \qquad \qquad + \mathbb{P}_{\LFNseq}\left(\mathbb{P}_{\tilde{\bT}^{(n)} \sim \hat{\btheta}^{(n)}}\left(B_n\left(\hat{\btheta}^{(n)},\tilde{\bT}^{(n)}\right) \given \Tseq\right) > 1-\epsilon \given J^{(n)c}_{K}\right)\mathbb{P}_{\LFNseq}\left(J^{(n)c}_{K}\right) \label{eq:l3_0}
\end{align}
where the second line follows because $\mathbb{P}_{\tilde{\bT}^{(n)} \sim \hat{\btheta}^{(n)}}\left(B_n\left(\hat{\btheta}^{(n)},\tilde{\bT}^{(n)}\right)\given \Tseq\right)$ is bounded between $0$ and $1$. 
 First, note that for any $K > 0$,
\begin{align}
   \lim_{n\to \infty}\mathbb{P}_{\LFNseq}\left(J^{(n)}_{K}\right) & =  \lim_{n\to \infty} 1- \mathbb{P}_{\LFNseq}\left(\bigcup_{i = m-r+2}^m \hat{\bm{\theta}}^{(n)}_{(i)} \leq K\right) \nonumber \\
   & \geq   1- \lim_{n\to \infty} \sum_{i=m-r+2}^{m}\mathbb{P}_{\LFNseq}\left(\hat{\bm{\theta}}^{(n)}_{(i)} \leq K\right) \nonumber \\
   & = 1 \label{eq:l3_1}
\end{align}
where the last line follows by Assumption~\ref{assump3}. We will now show that, for sufficiently large $K$, 

$\mathbb{P}_{\LFNseq}\left(\mathbb{P}_{\tilde{\bT}^{(n)} \sim \hat{\btheta}^{(n)}}\left(B_n\left(\hat{\btheta}^{(n)},\tilde{\bT}^{(n)}\right) \given \Tseq \right) > 1-\epsilon  \given J^{(n)}_{K}\right) = 1$ for any $n$. For $K > 0$, let $X_{i}^{(K)} \iid \mathcal{N}(K, 1)$, $i = 1, ..., r-1$ independent of $\Tseq$. 
Then,
\begin{align}
&\mathbb{P}_{\LFNseq}\left(\mathbb{P}_{\tilde{\bT}^{(n)} \sim \hat{\btheta}^{(n)}}\left(B_n\left(\hat{\btheta}^{(n)},\tilde{\bT}^{(n)}\right)  \given \Tseq \right) > 1-\epsilon  \given J^{(n)}_{K}\right) \nonumber \\
 & = \mathbb{P}_{\LFNseq}\left(\mathbb{P}_{\tilde{\bT}^{(n)} \sim \hat{\btheta}^{(n)}}\left( \max_{j = 1, ..., m-r+1}\left|\tilde{T}^{(n)}_{j}\right| <  \min_{i = m-r+2, ..., m}\left|\tilde{T}^{(n)}_{i}\right|  \given \Tseq \right) > 1-\epsilon  \given J^{(n)}_{K}\right) \nonumber\\
  & \geq \mathbb{P}_{\LFNseq}\left(\mathbb{P}_{\tilde{\bT}^{(n)} \sim \hat{\btheta}^{(n)}}\left( \max_{j = 1, ..., m-r+1}\left|\tilde{T}^{(n)}_{i} \right| <  \min_{i = 1, ..., r-1}\left|X_{i}^{(K)} \right|  \given \Tseq \right) > 1-\epsilon \given J_K^{(n)} \right) \nonumber\\
  & = \mathbb{P}\left(\mathbb{P}\left( \max_{j = 1, ..., m-r+1}\left|\tilde{T}^{(n)}_{i} \right| <  \min_{i = 1, ..., r-1}\left|X_{i}^{(K)} \right|\right) > 1-\epsilon \right) \nonumber\\
 & = \mathbbm{1}\left\{\mathbb{P}\left( \max_{j = 1, ..., m-r+1}\left|\tilde{T}^{(n)}_{j}\right| <  \min_{i = 1, ..., r-1}\left|X_{i}^{(K)} \right|\right) > 1-\epsilon \right\} \label{eq:lem5_convprob}
\end{align}
where the third line holds because for independent random variables $Z_j \iid \mathcal{N}(0, 1)$, $j = 1, ..., m-r+1$ and $Y_i \sim \mathcal{N}(\mu_i, 1)$, $i = 1, ..., r-1$ where all $\mu_i > K$, $$\mathbb{P}\left( \max_{j = 1, ..., m-r+1}\left|Z_j\right| <  \min_{i = 1, ..., r-1}\left|X_{i}^{(K)} \right|\right) < \mathbb{P}\left( \max_{j = 1, ..., m-r+1}\left|Z_j\right| <  \min_{i = 1, ..., r-1}\left|Y_i \right|\right).$$ 
In the second to last line and onward, we drop the conditioning events and the $\LFNseq$ and $\hat{\btheta}^{(n)}$ subscripts because the $\tilde{T}_{j}^{(n)}$'s, which are i.i.d. standard normally distributed regardless of $n$, and the $X_i^{(K)}$'s do not depend on $\Tseq$, and hence, do not depend on $\hat{\btheta}^{(n)}_{(m-r+2:m)}$ and $\LFNseq$.

Let $L_{K-\sqrt{K}} =\left\{ \min_{i = 1, ..., r-1}\left|X_{i}^{(K)}\right| > K - \sqrt{K} \right\}$. Then,
\begin{align*}
&\mathbb{P}\left(\max_{j = 1, ..., m-r+1}\left|\tilde{T}^{(n)}_{j}\right| <  \min_{i = 1, ..., r-1}\left|X_{i}^{(K)}\right| \right) \\
     & = \mathbb{P}\left(\max_{j = 1, ..., m-r+1}|\tilde{T}^{(n)}_{j}| <  \min_{i = 1, ..., r-1}\left|X_{i}^{(K)}\right| \given  L_{K-\sqrt{K}} \right)\mathbb{P}\left( L_{K-\sqrt{K}} \right)\\
     & \qquad \qquad + \mathbb{P}\left(\max_{j = 1, ..., m-r+1}\left|\tilde{T}^{(n)}_{j}\right| <  \min_{i = 1, ..., r-1}\left|X_{i}^{(K)}\right| \given L^{c}_{K-\sqrt{K}} \right)\mathbb{P}\left(  L^{c}_{K-\sqrt{K}} \right)
\end{align*}
where
\begin{align*}
    \mathbb{P}\left( L_{K-\sqrt{K}} \right) = \left(1-\Phi\left(-\sqrt{K}\right) + \Phi\left(-2K+\sqrt{K}\right)\right)^{r-1}.
\end{align*}
and
\begin{align*}
  \mathbb{P}\left(\max_{j = 1, ..., m-r+1}\left|\tilde{T}^{(n)}_{j}\right| <  \min_{i = 1, ..., r-1}\left|X_{i}^{(K)}\right| \given  L_{K-\sqrt{K}} \right) & \geq 
  \mathbb{P}\left(\max_{j = 1, ..., m-r+1}\left|\tilde{T}^{(n)}_{j}\right| <  K-\sqrt{K} \given  L_{K-\sqrt{K}} \right)\\
    & = \mathbb{P}\left(\max_{j = 1, ..., m-r+1}\left|\tilde{T}^{(n)}_{j}\right| <  K-\sqrt{K} \right) \\
   &  = \left(\Phi\left(K-\sqrt{K}\right) - \Phi\left(-(K-\sqrt{K})\right) \right)^{m-r+1}.
\end{align*}
As $K$ gets arbitrarily large,  $ \left(1-\Phi\left(-\sqrt{K}\right) + \Phi\left(-2K+\sqrt{K}\right)\right)^{r-1} \to 1$ and 

$\left(\Phi\left(K-\sqrt{K}\right) - \Phi\left(-(K-\sqrt{K})\right) \right)^{m-r+1} \to 1$ and therefore,
$$\mathbb{P}\left(\max_{j = 1, ..., m-r+1}\left|\tilde{T}^{(n)}_{j}\right| <  \min_{i = 1, ..., r-1}\left|X_{i}^{(K)}\right| \right) \to 1.$$

The above implies that there exists $K_\epsilon > 0$ such that for any $K > K_\epsilon$, 

$\mathbb{P}\left( \max_{j = 1, ..., m-r+1}\left|\tilde{T}^{(n)}_{j}\right| <  \min_{i = 1, ..., r-1}\left|X_{i}^{(K)}\right|\right) > 1-\epsilon$. So, for $K > K_\epsilon$,
$$\mathbbm{1}\left\{\mathbb{P}\left( \max_{j = 1, ..., m-r+1}\left|\tilde{T}^{(n)}_{j}\right| <  \min_{i = 1, ..., r-1}\left|X_{i}^{(K)} \right|\right) > 1-\epsilon \right\} = 1$$
and by Equation~\eqref{eq:lem5_convprob},
\begin{align}\label{eq:l3_3}
\mathbb{P}_{\LFNseq}\left(\mathbb{P}_{\tilde{\bT}^{(n)} \sim \hat{\btheta}^{(n)}}\left(B_n\left(\hat{\btheta}^{(n)},\tilde{\bT}^{(n)}\right) \given \Tseq \right) > 1-\epsilon  \given J^{(n)}_{K}\right) = 1   
\end{align}
Setting $K > K_\epsilon$ in Equation~\eqref{eq:l3_0}, by Equation~\eqref{eq:l3_3} and the fact that Equation~\eqref{eq:l3_1} holds for any $K>0$, we have that
$$\lim_{n \to \infty}\mathbb{P}_{\LFNseq}\left(\left|1-\mathbb{P}_{\tilde{\bT}^{(n)} \sim \hat{\btheta}^{(n)}}\left(B_n\left(\hat{\btheta}^{(n)},\tilde{\bT}^{(n)}\right)  \given \Tseq 
 \right)\right| > \epsilon\right) = 0. $$ 

Since $\epsilon$ was arbitrary, we conclude that
$$\mathbb{P}_{\tilde{\bT}^{(n)} \sim \hat{\btheta}^{(n)}}\left(B_n\left(\hat{\btheta}^{(n)},\tilde{\bT}^{(n)}\right) \given \Tseq  \right) \inp 1.$$
\end{proof}

\subsection{Proof of Lemma~\ref{lemma6}}\label{sec:lem6_proof}
\begin{proof}
Assume without loss of generality that $\LFNseq= \left(0, ...0, \theta^{(n)}_{m-r+2}, ..., \theta^{(n)}_{m}\right)$ where $\theta^{(n)}_{i} \to \infty$. Fix $K>0$ and $i \in \{m-r+2, ..., m\}$. We will show that $\lim_{n \to \infty} \mathbb{P}_{\LFNseq}\left(\left|T^{(n)}_{(i)}\right| > K \right) = 1$. 
\begin{align*}
  \lim_{n \to \infty}\mathbb{P}_{\LFNseq}\left(\left|T^{(n)}_{(i)}\right| > K \right) & =  \lim_{n \to \infty}\mathbb{P}_{\LFNseq}\left(\left|T^{(n)}_{(i)}\right|  > K \given B_n\left(\LFNseq, \Tseq \right) \right) \mathbb{P}_{\LFNseq}\left(B_n\left(\LFNseq, \Tseq \right) \right) +\\
  & \qquad \qquad \mathbb{P}_{\LFNseq}\left(\left|T^{(n)}_{(i)}\right| > K \given B_n\left(\LFNseq, \Tseq \right)^c \right) \mathbb{P}_{\LFNseq}\left(B_n\left(\LFNseq, \Tseq \right)^c \right)\\
   & = \lim_{n \to \infty} \mathbb{P}_{\LFNseq}\left( \left|T^{(n)}_{(i)}\right|  > K \given B_n\left(\LFNseq, \Tseq \right)\right)\\
  &\geq  \lim_{n \to \infty} \mathbb{P}_{\LFNseq}\left(  \left|T^{(n)}_{(m-r+2)}\right| > K \given B_n\left(\LFNseq, \Tseq \right) \right)\\
    &=  \lim_{n \to \infty} \mathbb{P}_{\LFNseq}\left( \bigcap_{i=m-r+2}^m \left\{\left|T^{(n)}_{i}\right| > K\right\} \given B_n\left(\LFNseq, \Tseq \right) \right)\\
    &= 1,
\end{align*}
where the third line follows from the fact that $ \lim_{n \to \infty}\mathbb{P}_{\LFNseq}\left(B_n\left(\LFNseq, \Tseq \right) \right) = 1$ as shown in Equation~\eqref{eq:lem1_eq2} of Lemma~\ref{lemma1}, the fifth line follows because of the conditioning on $B_n\left(\LFNseq, \Tseq \right)$, and the last line follows from applying Lemma~\ref{lemma3} with Equation~\eqref{eq:lem1_eq0} and 
Equation~\eqref{eq:lem1_eq2} of Lemma~\ref{lemma1}. Since we picked an arbitrary $i$ and $K$, we can conclude that for any $K > 0$ and $i = m-r+2, ..., m$, $\lim_{n \to \infty} \mathbb{P}_{\LFNseq}\left(\left|T^{(n)}_{(i)} \right| > K \right) = 1$.
\end{proof}
\clearpage

 \section{Approximate Validity of the Unadjusted cPCH Test}\label{appendix:approx_valid}
We find that, under the setting in Section~\ref{section:preliminaries}, for a fixed $m$, $r$, and $\alpha$, the unadjusted cPCH test has small and quantifiable Type I error inflation. Additionally, we find that for a fixed $m$ and $r$, the unadjusted cPCH test produces nearly uniform p-values under the null. Thus, we call the unadjusted cPCH test ``approximately valid and non-conservative''.

Our characterization of the unadjusted cPCH test's validity for finite $\theta_i$ relies fundamentally on the fact that, for a given choice of $m$ and $r$, we can estimate the unadjusted cPCH p-value distribution for a given $\bm\theta \in \Theta_0^{r/m}$ with high accuracy via Monte Carlo sampling. Specifically, given a $\bm\theta \in \Theta_0^{r/m}$, we generate samples $\tilde{\bT}^{(k)} \stackrel{i.i.d}{\sim} \mathcal{N}({\bm\theta}, I_m), k = 1, ..., N$ and compute the unadjusted cPCH p-value for each sample to empirically estimate the distribution of the unadjusted cPCH p-value under $\btheta$. Performing this sampling procedure across a sufficiently fine grid of $\Theta_0^{r/m}$ with many replicates $N$ allows us to obtain highly accurate estimates of the p-value distribution under the null for the desired $m$ and $r$. Given a fixed $\alpha$, we can also use the Monte Carlo samples to obtain highly accurate estimates of the Type I error for each $\bm\theta \in \Theta_0^{r/m}$ by computing the proportion of the sampled cPCH p-values at $\bm\theta \in \Theta_0^{r/m}$ that are below $\alpha$.

We pause to highlight that our approach in this section is somewhat different from a usual simulation study, which would typically explore a small (but hopefully at least somewhat representative) subset of all possible data-generating scenarios. In contrast, we provide an exhaustive characterization of the unadjusted cPCH p-value distribution at effectively \emph{every} null $\btheta$ for realistic values of $m$ and $r$. 
Such an exhaustive search is made possible by the small dimension of the null space for most common PCH testing scenarios, such as causal mediation analysis, where $m = 2$, and replicability analysis, where $m$ is often $\leq 5$ \cite[]{followup, HY2014, adafilt}. 
For all analyses in this section, we fix $\alpha = 0.05$. Thus, the following results can be interpreted as a computational proof that, for the configurations of $m$ and $r$ tested, the cPCH test produces approximately uniform p-values under the null and that, for $\alpha = 0.05$, the Type I error inflation of the cPCH test is small. We expect these results to generalize to larger $m$ and different choices of $\alpha$. Though our results assume the setting of Section~\ref{section:preliminaries} where each $T_i$ are single, independent unit-variance Gaussians, our simulations in Section~\ref{section:robustness} suggest that the following results would still hold under other distributional assumptions for the base test statistics.

\begin{figure}[!htb]
        \centering
        \includegraphics[width = \textwidth]{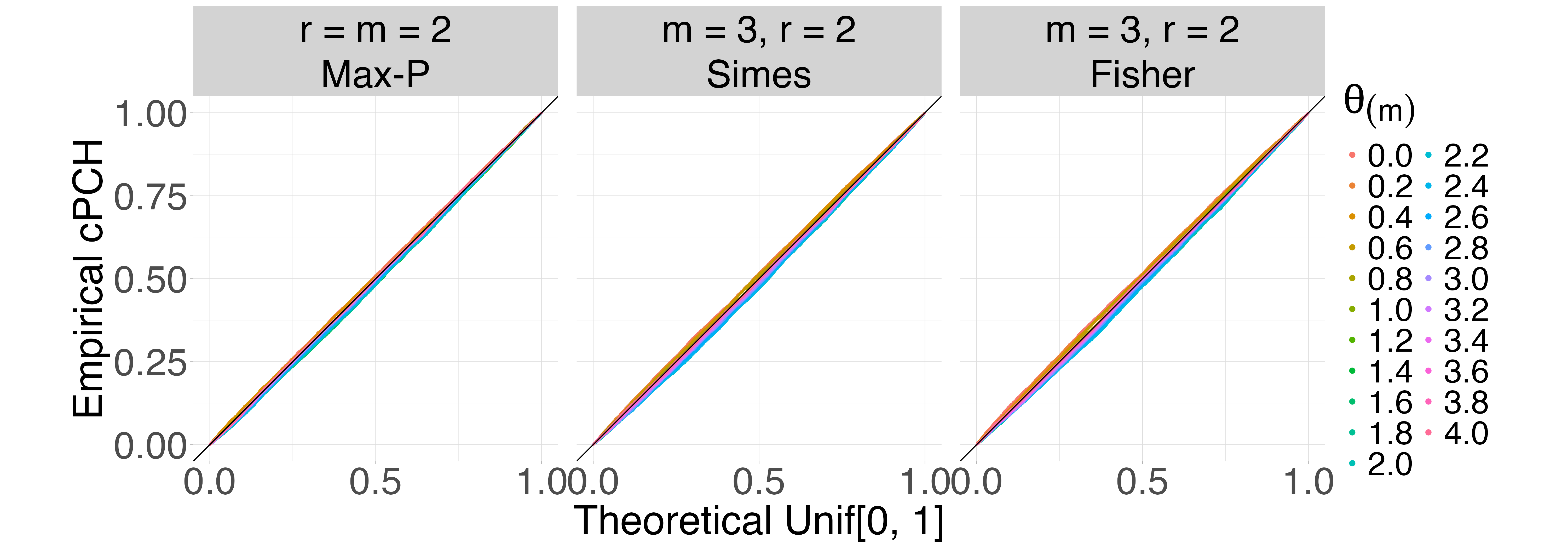}
 \caption{Quantile-quantile plot of the empirical cPCH p-value distribution vs. the theoretical Unif[0, 1] distribution. The cPCH p-value distributions were estimated using $n = 10,000$ independent replicates of the sampling scheme described in Section~\ref{appendix:approx_valid}.
 }  
 \label{fig:qqplot}
    \end{figure}

We first quantify the closeness of the null unadjusted cPCH p-value distribution to the Unif$[0, 1]$ distribution by computing the quantile-quantile plots and Kolmorgorov--Smirnov distances between the empirical cPCH p-value distribution (estimated using the Monte Carlo sampling scheme described above) and the theoretical Unif[0, 1] distribution for various $\bm\theta \in \Theta_0^{r/m}$. Our results show that the density of null cPCH p-values closely matches that of a Unif$[0, 1]$ distribution over a fine grid of $\bm\theta \in \Theta_0^{r/m}$ for any choice of combining function (Simes, or Fisher); see the qq--plots in Figure~\ref{fig:qqplot} as an example. To visually depict results, we focus on the $r=m=2$ and $m=3, r=2$ cases since any $\btheta \in \Theta_0^{2/2} \cup \Theta_0^{2/3}$ can be represented by a single scalar $\theta_{(m)}$.

           \begin{figure}[!htb]
        \centering
        \includegraphics[width = \textwidth]{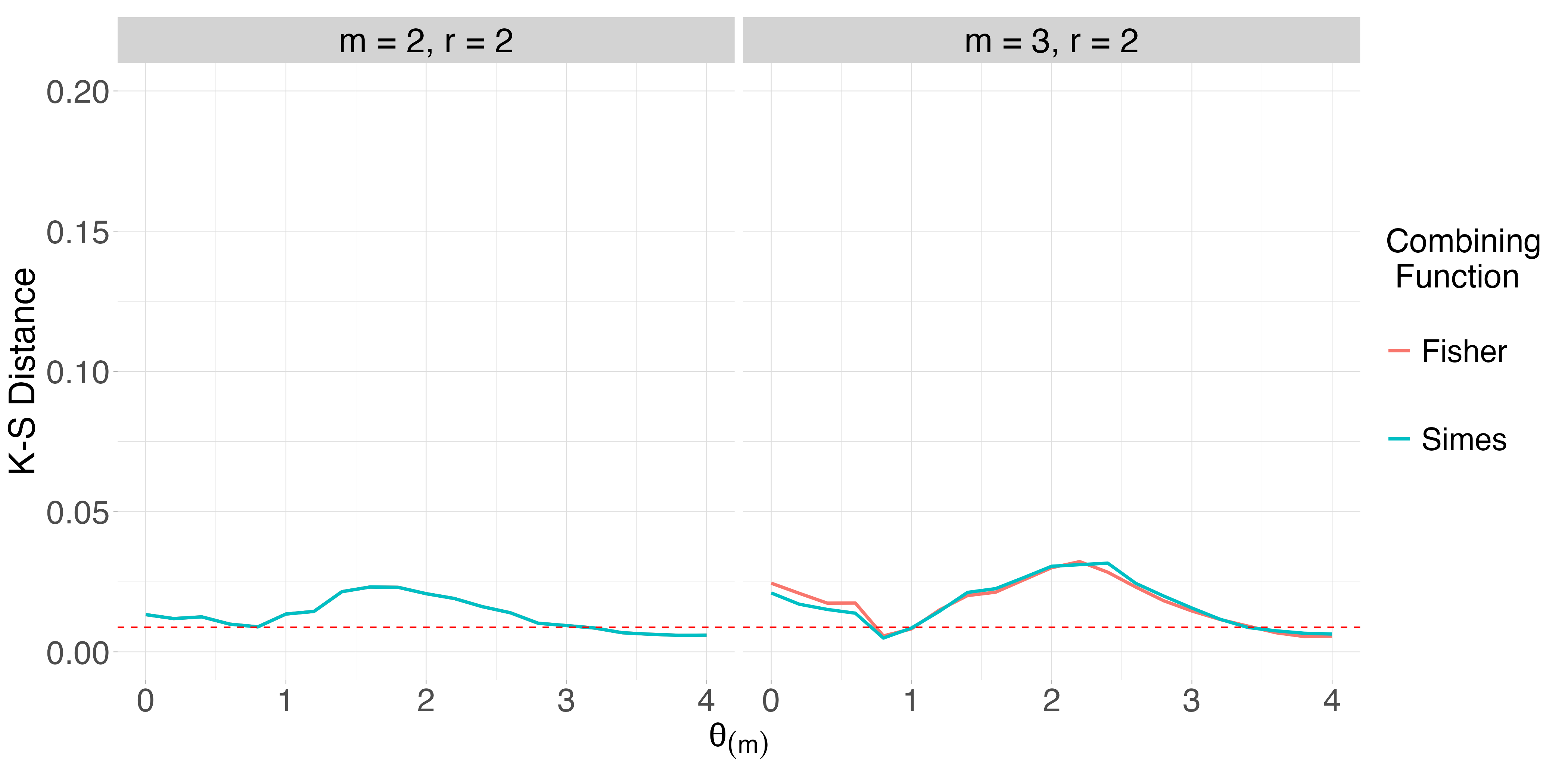}
 \caption{Kolmorgorov--Smirnov distances between the null cPCH p-value and Unif[0, 1] CDFs. Each point represents the Kolmorgorov--Smirnov distance between the Unif[0, 1] CDF and the empirically estimated cPCH p-value CDF for a given $\theta_{(m)}$ using $n = 10,000$ independent replicates for the Monte Carlo sampling scheme described in Section~\ref{appendix:approx_valid}. To provide a baseline for comparison, the red dotted line represents the average Kolmorgorov--Smirnov distance between the empirical CDF of the Unif[0, 1] distribution estimated over $10,000$ independent replicates and the theoretical Unif[0, 1] CDF. 
 } 
        \label{fig:ks_plot}
    \end{figure}

As shown in Figure~\ref{fig:ks_plot}, we see that the K--S distances between the cPCH null p-value distributions and the Unif$[0, 1]$ distribution are small, as expected based on the results in Figure~\ref{fig:qqplot}. The maximum Kolmorgorov--Smirnov distance for any choice of combining function, $m$, and $r$ occurs when $\theta_{(m)}$ is approximately between $1$ and $3$, with smaller deviations occurring under the global null, i.e., $\theta_{(m)} = 0$. Comparing Figure~\ref{fig:ks_plot} for $r=m=2$ with the corresponding Type I error plot for $r=m=2$ in Figure~\ref{fig:t1error}, we see that the deviation at the global null occurs because the cPCH test is slightly conservative under the global null, though far less so than its classical counterpart, the Max-P test. The deviations when $\theta_{(m)}$ is approximately between $1$ and $3$ occur due to the slight Type I error inflation of the cPCH test in this region. Note that the $m=3, r=2$ setting exhibits the same pattern of deviations as in the $r=m=2$ setting, and we expect a similar pattern (very slight conservativeness near the global null and slight anticonservativeness when the non-null $\theta_i$'s are between 1 and 3) to generalize to other $m$ and $r$.

We also use the SGD algorithm described in Supplementary Materials~\ref{sec:sgd_comp} to estimate the maximum Type I error inflation of the unadjusted cPCH test for various $m$, $r$, and $\alpha$. As shown in Table~\ref{tab:sgd}, the estimated maximum Type I error inflation of the cPCH test is generally small for various $2 \leq m \leq 4$ and $2 \leq r \leq m$ when $\alpha = 0.05$. 

 \begin{table}[ht]
      \centering
     \begin{tabular}{llll}
   $m$  & $r$ & cPCH-Fisher & cPCH-Simes\\
 \midrule
    2 & 2 &  {0.060}  &  \NA \\
    3 & 2 & 0.058 & 0.059 \\
         & 3 &  0.053 & 0.053 \\  
   4 &  2 & 0.057 & 0.059 \\
         &  3 &   0.058 & 0.060 \\
          &  4 &  0.062 &  0.062 \\ 
     \end{tabular}
     \caption{Maximum Type I error of the cPCH test computed via SGD. All standard errors are below 0.0015, where standard errors are calculated over many independent replicates of the cPCH p-value ($N = 10,000$) at the $\btheta^* \in \Theta_0^{r/m}$ for which the SGD algorithm terminated.     }
      \label{tab:sgd}
  \end{table} 
  
\clearpage

\section{PCH Multiple Testing Simulation Study}\label{section:pch_mt_sim_study}
In this section, we present the full details of the PCH multiple testing simulation study summarized in Section~\ref{section:mt}. Recall, we assume there are $M$ PCH's being simultaneously tested and denote $T_{ij}$ and $p_{ij}$ as the $i$th base test statistic and p-value, respectively, for the $j$th PCH being tested, $i = 1, ..., m$, $j = 1, ..., M$.

\subsection{An Overview of Methods Under Comparison}\label{section:methodsundercomparision_appendix}
Our study includes two state-of-the-art Empirical Bayes methods, the Divide-Aggregate Composite-null Test (DACT) \cite[]{dact} and High Dimensional Multiple Testing (HDMT) \cite[]{hdmt} and two state-of-the-art filtering methods, AdaFilter \cite[]{adafilt} and the method presented in \cite{dickhaus}.

Both DACT and HDMT adapt existing empirical Bayes frameworks for estimating the proportion of each null configuration \cite[]{efron2001, storey2002, storey}. They both guarantee FDR control as $M \to \infty$ under specific regularity conditions, which are primarily set to guarantee that the estimators for the proportions of each null configuration are consistent. Additionally, by the linear structural model commonly used for causal mediation analysis \cite[]{baronkenny}, both DACT and HDMT assume that the base test statistics are normally distributed and independent across $m$. The primary difference between DACT and HDMT (and Empirical Bayes methods in general) lies in how the estimated proportions are used. DACT generates individual PCH p-values that are weighted sums of the estimated proportions of each null configuration, on which FDR controlling procedures like BH can be applied. HDMT estimates the FDR of the Max-P test at a given rejection threshold using these proportions, then selects the largest rejection threshold such that the estimated FDR is less than or equal to the desired level. Importantly, both HDMT and DACT are designed for causal mediation analysis, a specific case of PCH testing with $r=m=2$, and thus, are only applicable (as implemented) when $m=2$. However, the statistical frameworks used by Empirical Bayes methods for estimating the proportions of the null configurations are not limited to the $m=2$ case and thus, these methods could reasonably be extended to larger $m$. As the computational cost of Empirical Bayes methods can grow exponentially in $m$ \cite[]{adafilt}, fixing $r=m=2$ allows Empirical Bayes methods to produce accurate estimates of the true proportions within reasonable computational limits, as long as $M$ is sufficiently large and the conditions necessary for the consistency of their estimators are met.

AdaFilter is a filtering method that infers a new (ideally less conservative) rejection threshold in a data-adaptive manner. It first reduces the set of total PCH's to the subset that would be rejected by Bonferroni PCH tests for $H_{0}^{(r-1)/m}$, then applies an adjusted version of the Bonferroni PCH test for $H_0^{r/m}$ to each PCH in the reduced set. 
Intuitively, this filtering is effective because the null PCH's that are rejected for $H_{0}^{(r-1)/m}$ are close to the LFN, so the Bonferroni PCH test will be less conservative on the reduced set. Under mild regularity conditions, AdaFilter guarantees FDR control as $M \to \infty$. For finite $M$ and nominal FDR level $q$, AdaFilter guarantees FDR control at level $q C(M)$ where $C(M) = \sum_{j=1}^M \frac{1}{j}$, when all $p_{ij}$ are independent.

DHH filters out all PCH p-values $p^{r/m}$ above some threshold $\tau$, then applies a transformation to the smaller subset so that, under mild conditions on the $p_{ij}$ and $p^{r/m}$, various FDR controlling procedures such as BH and Storey's procedure can be applied to the smaller subset to control FDR on the entire set. Thus, there are many variations of DHH depending on the PCH p-value and multiple testing procedure used. In our analysis, we use PCH p-values from Simes' and Fisher's tests, which are shown in \cite{dickhaus} to satisfy the conditions necessary for the DHH procedure (with fixed $\tau$) to have FDR control when using BH or Storey's procedure.
   \cite{dickhaus} also provides an algorithm for selecting the threshold $\tau$ in a data-adaptive way, which, under certain conditions, guarantees asymptotic FDR control. We select $\tau = 0.1$ since, as shown in the empirical simulations and real data example presented in \cite{dickhaus}, DHH with $\tau = 0.1$ has similar power to DHH with the data-adaptive threshold in many settings.

 For all PCH multiple testing approaches under consideration (DACT, HDMT, AdaFilter, and DHH), their FDR guarantees hold under the assumption that the $p_{ij}$ are independent, or some mild relaxation of it (e.g., AdaFilter allows weak dependence between the base p-values for their asymptotic results). Thus, for all following simulations, we generate data such that the $p_{ij}$ are independent.

  For methods which generate individual p-values such as the cPCH test, we focus on three FDR controlling procedures: Benjamini--Hochberg (BH) \cite[]{Benjamini1995}, Storey's \cite[]{storey}, and AdaPT--GMM \cite[]{adaptgmm}. Note, since DACT and DHH produce individual PCH p-values (though they rely on being in a PCH multiple testing setting to do so), we also combine them with the procedures above. We now discuss the assumptions necessary for FDR control for these procedures. Storey's procedure and AdaPT--GMM share the assumption that the individual PCH p-values are independent, while BH only requires positive regression dependency on a subset. AdaPT--GMM requires the additional assumption that the null p-values have a non-decreasing density. The standard PCH p-values and DHH-adjusted PCH p-values satisfy the non-decreasing density condition, while DACT and cPCH p-values are not guaranteed to do so. However, since null cPCH p-values are \emph{nearly} uniform, the non-decreasing density assumption is at least justified approximately for the cPCH test. Additionally, we find in our empirical simulations that FDR control is maintained when using cPCH p-values with AdaPT--GMM. In all simulations, we include the standard Simes' and Fisher's tests for single PCH testing in combination with BH, Storey's procedure, and AdaPT--GMM (when applicable) as benchmarks for comparison.

Table~\ref{tab:overview_methods} provides a brief overview of the validity guarantees of all methods under consideration.

      \begin{table}[!htb]
      \centering
        \caption{Overview of Methods under Consideration}\label{tab:overview_methods}
      \begin{threeparttable}
     \begin{tabular}{llp{0.7\linewidth}}
     Method  & Validity Guarantee\\
  \midrule
\textbf{Empirical Bayes} &   \\ 
DACT & Asymptotic FDR control$^{*}$, only considers $m=2$\\ 
HDMT &  Asymptotic FDR control$^{\dagger}$, only considers $m=2$ \\
\noalign{\vskip 2mm} 
\textbf{Filtering} &  \\ 
AdaFilter &  Asymptotic FDR control, upper bound on finite sample FDR\\
DHH  &  Finite sample FDR control \\
\noalign{\vskip 2mm} 
\textbf{Single PCH}$^{\ddagger}$ & \\
cPCH &  Finite sample Type I error control\\
Standard & Finite sample Type I error control\\
     \end{tabular}
    \begin{tablenotes}
    \item ${}^*$: Requires that the regularity conditions of \cite{jinandcai} hold.
     \item  ${}^{\dagger}$: Requires estimated null proportions \cite[]{storey2002, storey} are consistent. 
     \item ${}^{\ddagger}$: With multiple testing procedures BH, Storey's Procedure, and AdaPT--GMM.
    \end{tablenotes}  
      \end{threeparttable}
  \end{table} 

\subsection{Plot of All Methods for Simulations of Section~\ref{section:mt}}\label{section:sm_mt}

\begin{figure}
    \centering
    \includegraphics[width=\textwidth]{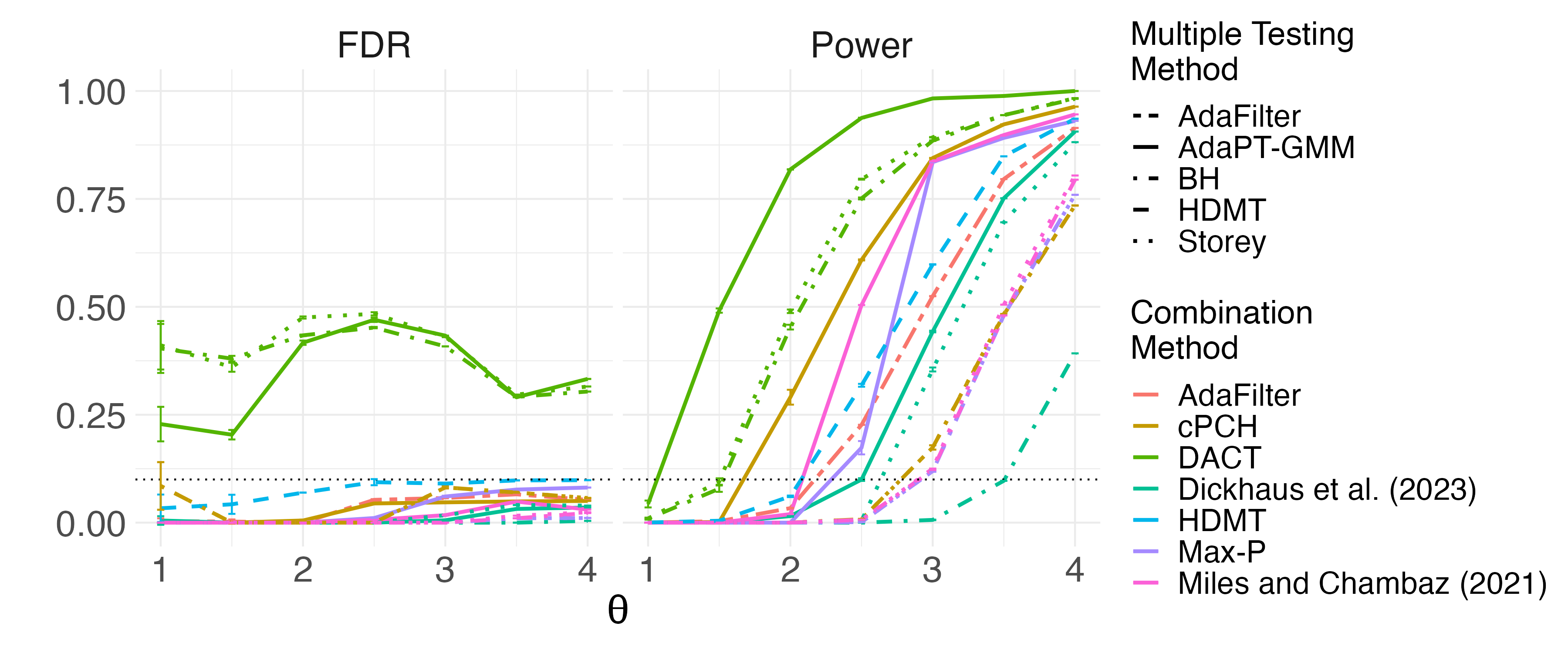}
\caption{FDR and power of various PCH multiple testing methods at nominal FDR level $q = 0.1$ (dotted black line) for testing $H_0^{2/2}$. Each point represents $100$ independent replicates of the data generating procedure described in Section~\ref{section:mt} for a given $\theta$. Error bars depict $\pm 2$ standard errors.} 
    \label{fig:full_adapt}
\end{figure}

\subsection{PCH Multiple Testing for $m = 4$}\label{section:mt_m_4}
To approximate the real-data example in the following section, we conduct a simulation study to assess the empirical FDR and power of various PCH multiple testing approaches when $m = 4$ and $M = 2000$. As we explore a few different testing scenarios in our real data example, we choose a data generating procedure that allows us to flexibly vary the proportion of null and alternative configurations, thus emulating a wide range of possible testing scenarios for $m=4$ and $M = 2000$.
We generate the data $T_{ij}$, $i = 1, ..., 4$, $j = 1, .., 2000$ by sampling independently from the following model:
\begin{align*}
B_j &\sim \text{Bern}(\pi_1)\\
\gamma_{ij}|B_j &\sim  \begin{cases} 
      B_j & \text{w/p } w \\
      \text{Bern}(\pi_1) & \text{w/p } 1-w\\
      \end{cases}\\
T_{ij}|\gamma_{ij}, B_j &\sim \mathcal{N}(\theta \gamma_{ij}, 1)
\end{align*}

 $\pi_1$ tunes the expected proportion of the null and alternative PCH's and $p$ tunes the likelihood of each null and alternative configuration. For example, when $w = 1$, we expect $\pi_1$ of the total PCH's to be alternatives, all of which are ``full alternatives'' (i.e., $r^\star = m$), and all remaining PCH's to be global nulls. 

Discoveries are generated using AdaFilter, the cPCH test ($N=10,000$), the standard  Simes and Fisher tests, and DHH ($\tau = 0.1$), each with BH and Storey's procedures. All procedures are applied at nominal FDR level $q=0.1$. The Empirical Bayes methods under consideration focus on the $m=2$ case and hence cannot be applied in this setting.

      \begin{figure}[!htb]
        \centering
        \includegraphics[width = \textwidth]{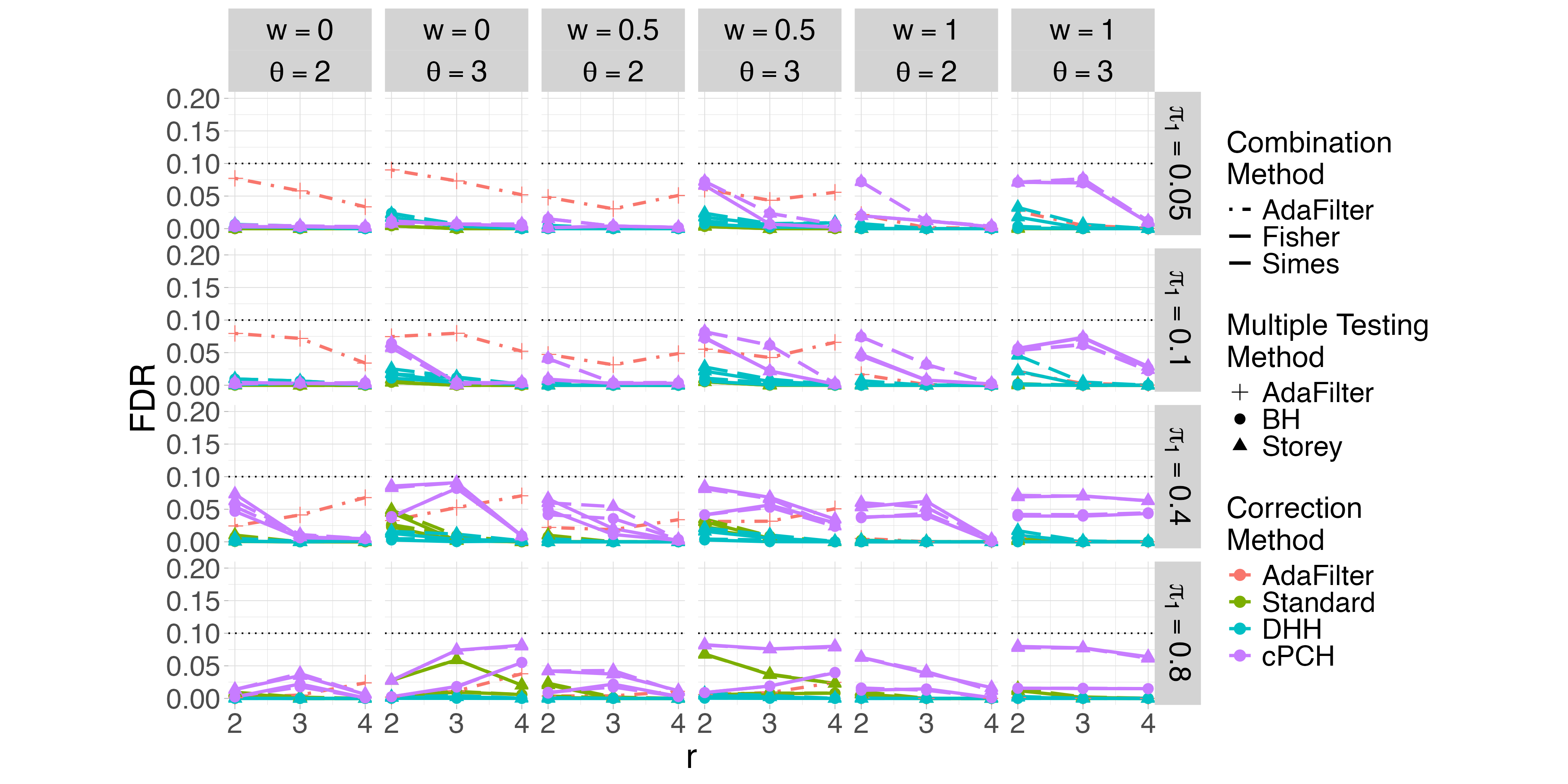}
 \caption{False Discovery Rate (FDR) results corresponding to Section~\ref{section:mt_m_4} where all methods are implemented at nominal level $q = 0.1$ (dotted black line). Each point represents the average proportion of rejected PCH's which are null over $1000$ independent replicates of the data generating procedure described in Section~\ref{section:mt_m_4} for a given $\theta$. Standard errors were all less than $0.008$.} 
        \label{fig:fdr_plot_n4}
    \end{figure}

\begin{figure}[!htb]
        \centering
        \includegraphics[width = \linewidth]{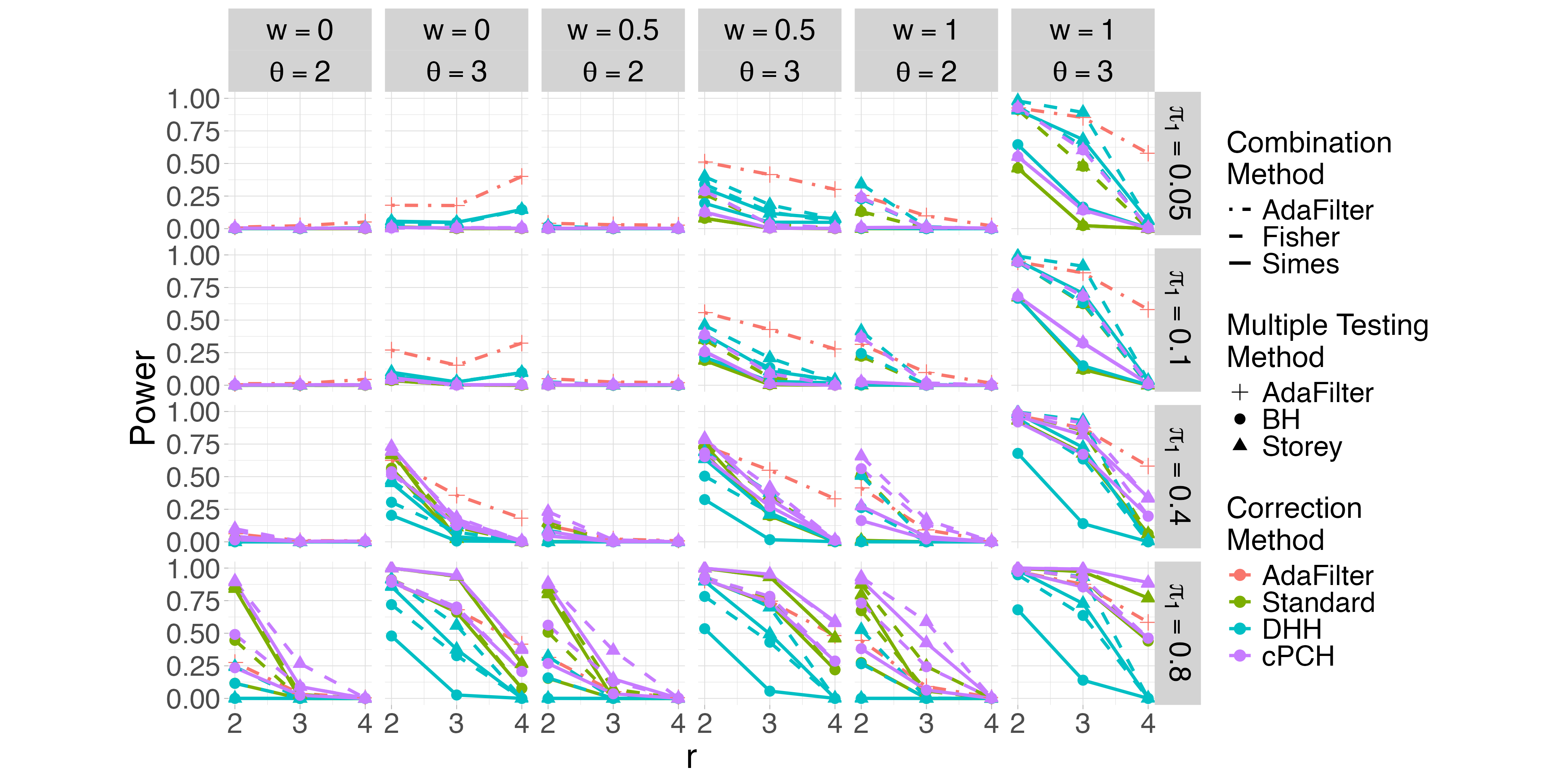}
  \caption{Power of various methods for PCH multiple testing at nominal FDR level $q = 0.1$. Each point represents the average proportion of non-null PCH's which are rejected over $1000$ independent replicates of the data generating procedure described in Section~\ref{section:mt_m_4} for a given $\theta$. All standard errors were less than $0.015$.}
        \label{fig:mt_power_plot}
    \end{figure}

We show that all methods under consideration empirically control the FDR at the nominal level for every combination of $\theta, \pi_1$, and $w$ tested; see Figure~\ref{fig:fdr_plot_n4}. Note, finite-sample FDR control is guaranteed with the Max-P test (with BH and Storey's procedure) and DHH (with BH and Storey's procedure). These results, along with those of the previous section, suggest that AdaFilter and the cPCH--based approaches, neither of which guaranteed FDR control at level $q$ in this setting, have robust FDR control across a wide array of testing settings. As shown in Figure~\ref{fig:mt_power_plot}, we find that the relative power of these methods is highly sensitive to the underlying data generating distribution, so no method dominates all others across all settings. However, the cPCH test is more powerful than its standard counterparts in all data generating settings. As expected, we see that the AdaFilter and DHH methods have the highest power when $\pi_1$ is small, as the construction of filtering methods allows them to be especially effective in this setting. For instance, DHH is likely to significantly reduce the multiplicity of PCH's being tested as a vast majority of the PCH p-values will be above $\tau$. 

However, in settings where $\pi_1$ is large, the cPCH test with Storey's procedure outperforms all other methods. As we alluded to in Section~\ref{section:mt_intro}, even the standard single PCH tests with BH and Storey's procedure can outperform DHH and AdaFilter in this setting. There are various important applications where $\pi_1$ is expected to be large. For instance, in epigenetics, promising sets of candidate genes from pre-screening studies or prior knowledge are often re-evaluated in follow-up studies \cite[]{ odonovan, BHY2009, Rietveld}. In this setting, researchers expect that a relatively large proportion of genes will show replicating results as they have already been identified as promising. We explore this follow-up setting through a real-data example in Section~\ref{section:dmd}.
\clearpage

\section{Additional Duchenne Muscular Dystrophy Analysis Results}\label{section:genes}

\begin{table}[!htb]
      \centering
     \begin{tabular}{lrrrr}
      $r$ & cPCH-AdaPT--GMM &  AdaFilter & Dickhaus et al. (2023) & Simes-AdaPT--GMM\\
  \midrule
2 & 350 & 268 & 147 & 311 \\
3 & 73 & 79 & 13 & 69\\
     \end{tabular}
      \caption{Rejections made in the follow-up study testing scenario of Section~\ref{section:dmd}, which uses the earliest performed microarray as the holdout study. Simes's combining function is used throughout. \re{Note, we only compare the above methods because the other approaches (e.g., HDMT, DACT, and the test of \cite{miles2021optimal}) are only designed for the $r=m=2$ case.}
      }
      \label{tab:follow_dmd_simes}
  \end{table} 

 \begin{table}[ht]
      \centering
     \begin{tabular}{lccc}
    Holdout Study & Method & $r=2$  & $r=3$  \\
     \midrule
  GDS218 &cPCH-Fisher-AdaPT--GMM & 510 & 170 \\
 &cPCH-Simes-AdaPT--GMM & 436   & 176  \\
 &Fisher-AdaPT--GMM & 358 & 100 \\
 &Simes-AdaPT--GMM & 300  & 100 \\
 & AdaFilter & 410  & 225\\
 & Dickhaus et al. (2023)  & 349   & 84 \\
 \midrule
   GDS3027 &cPCH-Fisher-AdaPT--GMM & 308 & 60 \\
 &cPCH-Simes-AdaPT--GMM & 312   & 61 \\
 &Fisher-AdaPT--GMM & 288 & 7 \\
 &Simes-AdaPT--GMM & 251   & 7\\
 & AdaFilter & 218   & 76 \\
 & Dickhaus et al. (2023)  & 144 & 10 \\
  \midrule
    GDS1956 &cPCH-Fisher-AdaPT--GMM & 346 & 75 \\
 &cPCH-Simes-AdaPT--GMM & 348   & 56\\
 &Fisher-AdaPT--GMM & 315 & 63\\
 &Simes-AdaPT--GMM & 277  & 63\\
 & AdaFilter & 355   & 101  \\
 & Dickhaus et al. (2023)  & 244   & 50 \\
 \midrule
     \end{tabular}
      \caption{Rejections made in the follow-up study testing scenario of Section~\ref{section:dmd} using different microarray studies as the holdout study.
      }
      \label{tab:additional_dmd}
  \end{table} 

      \begin{table}[ht]
      \centering
 \begin{tabular}{lcp{0.6\linewidth}}%
    Gene Symbol & cPCH p-value & Gene Function\\
    \midrule
    \begin{filecontents*}{follow_up_3_short.csv}
genesymbol,cpchfpval,function
ART3,0.00011,protein ADP-ribosylation
CFHR1,0.00075,complement activation
CHRNA1,2e-05,cation transmembrane transport/muscle cell cellular homeostasis
DAB2,1e-05,Wnt signaling pathway/apoptotic process
EEF1A1,2e-05,cellular response to epidermal growth factor stimulus/regulation of chaperone-mediated autophagy
EZR,3e-05,actin cytoskeleton reorganization/actin filament bundle assembly
HLA-DMB,0.00045,MHC class II protein complex assembly/antigen processing and presentation of exogenous peptide antigen via MHC class II
HLA-DRA,0.00062,T cell costimulation/T cell receptor signaling pathway
LAMB2,0.00034,Schwann cell development/astrocyte development
LAPTM5,5e-05,transport
MYH3,1e-05,ATP metabolic process/actin filament-based movement
MYH8,1e-05,ATP metabolic process/muscle contraction
MYL4,1e-05,cardiac muscle contraction/muscle filament sliding
MYL5,3e-05,muscle contraction/regulation of muscle contraction
S100A10,1e-05,establishment of protein localization to plasma membrane/membrane budding
S100A11,2e-05,cell-cell adhesion/negative regulation of DNA replication
S100A13,2e-05,cytokine secretion/interleukin-1 alpha secretion
S100A4,0.00041,epithelial to mesenchymal transition/positive regulation of I-kappaB kinase/NF-kappaB signaling
TMSB10,0.0001,actin filament organization
\end{filecontents*}
\csvreader[late after line=\\,head to column names]{follow_up_3_short.csv}{}
    {\csvcoli&\csvcolii& \csvcoliii}
    \end{tabular}
          \caption{A subset of selected genes from applying cPCH using Fisher's combining function with the covariate-assisted multiple testing procedure AdaPT-GMM \cite[]{adaptgmm} at $r=m=3$ and nominal level $q = 0.1$ on the follow-up study design described in Section~\ref{section:dmd}. As desired, many of the genes discovered all correspond to various muscle functions. Notably, the genes MYH3, MYH8, MYL4, and MYL5 are known genetic markers for DMD.}
      \label{tab:selected_genes_3}
  \end{table}

\end{document}